\documentclass[twocolumn,amsmath,amssymb,nofootinbib,fleqn]{revtex4}

\usepackage{amsmath}
\usepackage{amssymb} 
\usepackage{amscd}
\usepackage{amsfonts}
\usepackage{appendix}
\usepackage{bbold}
\usepackage[footnotesize]{caption2}
\usepackage{graphicx}
\usepackage[center,footnotesize,hang]{subfigure}
\usepackage{url}
\usepackage{dcolumn}
\usepackage{bm}

\newcommand{\Rep}[1]{\underline{\mbox{\textbf{#1}}}}
\newcommand{\MoreRep}[2]{\underline{\mbox{\textbf{#1}}} _{\mbox{\textbf{#2}}}}
\newcommand{\Groupname}[2]{$ {#1} _{#2} $}
\newcommand{\Doub}[2]{$ {#1} _{#2} ^{\prime} $}

\newcommand{\TeV}{\ensuremath{\,\mathrm{TeV}}}

\newcommand{\Eqref}[1]{Eq.\eqref{#1}}

\newcommand{\Tabref}[1]{Table \ref{#1}}

\newcommand{\Secref}[1]{Section \ref{#1}}
\newcommand{\Appref}[1]{Appendix \ref{#1}}

\newcommand{\be}{\begin{equation}}
\newcommand{\ee}{\end{equation}}
\newcommand{\bea}{\begin{eqnarray}}
\newcommand{\eea}{\end{eqnarray}}
\newcommand{\ba}{\begin{array}}
\newcommand{\ea}{\end{array}}
\newcommand{\sprime}{^\prime}

\allowdisplaybreaks[1]

\begin{document}

\title{Fermion Masses and Mixings from Dihedral Flavor Symmetries\\
with Preserved Subgroups}

\author{A.~Blum \footnote{alexander.blum@mpi-hd.mpg.de}}
\author{C.~Hagedorn \footnote{claudia.hagedorn@mpi-hd.mpg.de}}
\author{M.~Lindner \footnote{manfred.lindner@mpi-hd.mpg.de}}

\affiliation{Max-Planck-Institut f\"{u}r Kernphysik\\ 
Postfach 10 39 80, 69029 Heidelberg, Germany}

\begin{abstract}
\noindent We perform a systematic study of dihedral groups used
as flavor symmetry. The key feature here is the fact that 
we do not allow the dihedral
groups to be broken in an arbitrary way, but in all cases some
 (non-trivial) subgroup has to be preserved. In this
way we arrive at only five possible (Dirac) mass matrix structures which can arise, if
we require that the matrix has to have a non-vanishing determinant and that
at least two of the three generations of left-handed (conjugate) fermions
are placed into an irreducible two-dimensional representation of the flavor group.
We show that there is no difference between the mass matrix 
structures for single-
and double-valued dihedral groups. Furthermore, we comment on
possible forms of Majorana mass matrices.
As a first application 
we find a way to express the Cabibbo angle, i.e. the CKM matrix
element $|V_{us}|$, in terms of group theory quantities only,  
the group index $n$, the representation index $\mathrm{j}$ and the 
index $m_{u,d}$ of the different preserved subgroups in the up and down
quark sector: $|V_{us}|=\left| \cos \left(\frac{\pi \, (m_{u}-m_{d}) \, \mathrm{j}}{n} \right) \right|$ which is $|\cos (\frac{3 \, \pi}{7})| \approx 0.2225$ for $n=7$, $\mathrm{j}=1$, $m_{u}=3$ 
and $m_{d}=0$.
We prove 
that two successful models which lead to maximal
atmospheric mixing and vanishing $\theta_{13}$ in the lepton sector are
based on the fact that the flavor symmetry is broken in the charged 
lepton, Dirac neutrino and Majorana neutrino sector down to different
preserved subgroups whose mismatch results in the prediction of these
mixing angles. This also demonstrates the power of preserved subgroups
in connection with the prediction of mixing angles in the quark as well as
in the lepton sector.
\end{abstract}

\maketitle

\setcounter{footnote}{0}

%%%%%%%%%%%%%%%%%%%%%%%%%%%%%%%%%%%%%%%%%%%%%%%%%%
\section{Introduction}
%%%%%%%%%%%%%%%%%%%%%%%%%%%%%%%%%%%%%%%%%%%%%%%%%%
\label{sec:intro}

\noindent Experiments have shown that the quark and
lepton mixing angles are completely different in size and hierarchy. They are  
small in the quark sector 
\[
\theta _{12} ^{q} \approx 13.1 ^{\circ} \; , \;\; \theta _{23} ^{q} \approx 2.4 ^{\circ} \; ,
\;\; \theta _{13} ^{q} \approx 0.23 ^{\circ}
\] 
with the Cabibbo angle $\theta _{12} ^{q}$ being the largest \cite{pdg}, 
while in the lepton sector two of them are large, if not maximal,
\[
\theta _{23} ^{l} \approx 45 ^{\circ} \; , \;\; \theta _{12} ^{l} \approx  33.2 ^{\circ} \; , 
\]
\noindent together with the third angle $\theta _{13} ^{l}$ being smaller than $9.1 ^{\circ}$ at the 
$2 \, \sigma$ level \cite{Maltoni:2004ei}.

\noindent From the viewpoint of model building two special structures for the lepton mixing angles are very 
interesting:
a.) the case of tri-bimaximal (TBM) mixing  \cite{hpspaper} and 
b.) the case of $\mu \tau$ symmetry (MTS) \cite{mutau} in the neutrino sector. For
tri-bimaximal mixing the sines of the mixing angles are given by
\[ 
\sin^{2} (\theta _{12} ^{l})=\frac{1}{3} \; , \;\; \sin ^{2} (\theta_{23} ^{l})=\frac{1}{2} \;\;
\mbox{and} \;\; \sin ^{2} (\theta _{13} ^{l})=0
\]
\noindent while $\mu \tau$ symmetry only enforces 
\[
\sin ^{2} (\theta_{23} ^{l})=\frac{1}{2} \;\;
\mbox{and} \;\; \sin ^{2} (\theta _{13} ^{l})=0
\]
\noindent leaving the angle $\theta_{12} ^{l}$ undetermined.
Both ansaetze for the mixing angles are
compatible with the best fit values at the $2 \, \sigma$ level. 

\noindent The Standard Model (SM) can only accommodate, but not explain
these data. The possible special structure for the lepton mixing matrix 
together with the hierarchy of the quark mixing angles
 is quite a strong hint for a \textit{flavor symmetry} 
$G_{F}$ which is broken in a non-trivial way.

\noindent Requiring that at least two of the three fermion generations can be unified by $G_{F}$ and
avoiding extra Goldstone or gauge bosons from breaking $G_{F}$ spontaneously leads us to the
conclusion that the best choice for $G_{F}$ is a \textit{discrete}, \textit{non-abelian}
symmetry.

\noindent Recently, in a series of papers \cite{a4paper} it has been shown that the discrete symmetry
\Groupname{A}{4}, which is the symmetry group of even permutations of four
distinct objects, is able to explain TBM mixing in the lepton sector,
if it acts on the fermion generations in the following way: the three left-handed
lepton doublets transform as the irreducible three-dimensional representation of 
\Groupname{A}{4}, while the left-handed conjugate charged leptons transform as the
three singlets of \Groupname{A}{4}, $\MoreRep{1}{1}$, $\MoreRep{1}{2}$ and $\MoreRep{1}{3}$ (also
called $1$, $1 ^{\prime}$ and $1^{\prime \prime}$).

\noindent In reference \cite{tprimepaper} this 
phenomenologically successful model has been
extended to the quark sector by using the symmetry group \Doub{T}{}, which is the double-valued group of 
\Groupname{A}{4}. In the quark sector it allows for some non-trivial connections
among the CKM matrix elements and the quark masses. 

\noindent The key point of these studies
lies in the fact that the Higgs fields which necessarily also transform non-trivially under the 
flavor group in order to form \Groupname{A}{4} (\Doub{T}{}) invariant Yukawa couplings \textit{do not
acquire arbitrary vacuum expectation values (VEVs), but VEVs which break the flavor group \Groupname{A}{4}
(\Doub{T}{}) down to certain non-trivial subgroups}, such as \Groupname{Z}{2}, \Groupname{Z}{3} and
\Groupname{Z}{4}. The fields which preserve a \Groupname{Z}{2} (in case of \Groupname{A}{4}, \Groupname{Z}{4}
in case of \Doub{T}{}) only couple to neutrinos at the leading order, while the other fields
breaking the flavor symmetry of the model down to a \Groupname{Z}{3} subgroup only couple to charged
fermions at this level. This exactly leads to TBM mixing and
 explains large lepton mixing angles as the result of two
different subgroups, whereas small quark mixing angles come from mass matrices which are
invariant under the same subgroup. 

\noindent These results trigger the following questions: Are \Groupname{A}{4} and its
double-valued group \Doub{T}{} the only groups in which such results can be
achieved? Or are there other groups among the non-abelian discrete symmetries
which can lead to the same or to a similar result by preserving some
non-trivial subgroup? 

\noindent In order to answer these
questions at least partly, we investigate in this paper the dihedral groups \Groupname{D}{n}
and their double-valued groups \Doub{D}{n} as flavor symmetries
\cite{dngrouptheorypaper} . They are non-abelian for $n>2$ and
$n>1$, respectively. The group \Groupname{D}{n} is the symmetry group of a regular
planar $n$-gon. The dihedral groups and their double-valued groups are well suited for
 a general study, since they form a series of groups with similar properties,
e.g. they all contain only one- and two-dimensional representations.
Among the discrete groups there are several such series of groups: the permutation
groups \Groupname{S}{n} of $n$ distinct objects, the alternating groups \Groupname{A}{n}
of even permutations of $n$ objects, and the two series
of subgroups of $SU(3)$, $\Delta (3 \, n^{2})$ and $\Delta (6 \, n^{2})$ \cite{deltasgrouppaper}. The groups
\Groupname{S}{n} and \Groupname{A}{n} are only interesting for small $n$ (\Groupname{S}{2},
\Groupname{S}{3}, \Groupname{S}{4} and \Groupname{A}{3}, \Groupname{A}{4}, \Groupname{A}{5}), since with
increasing $n$ the dimension of their non-trivial representations increases beyond three.
So they are only appropriate for case by case studies and not for a general study like ours.
$\Delta (3 \, n^{2})$ and $\Delta (6 \, n^{2})$ are more similar to the dihedral groups
and therefore more interesting, but also less known in particle theory \cite{deltasphysicspaper}. 

\noindent In our general study we test whether the groups \Groupname{D}{n} and \Doub{D}{n} can be
used to get a similar result as in the mentioned \Groupname{A}{4} and \Doub{T}{} model, i.e.
whether one can induce certain mixing patterns in the lepton and/or quark sector by
breaking to different subgroups. In order to do so, we first need to establish the
group theory of the groups \Groupname{D}{n} and \Doub{D}{n}, in particular we carefully study all
possible subgroups. We show for each of them all representations and directions which 
leave them invariant. In a next step we calculate all possible mass matrices
which can then arise in such a model. Thereby the gauge group is taken to be the one of the SM
for simplicity.
In order to keep the calculation tractable we restrict ourselves
to mass matrices with a non-vanishing determinant and for
Dirac mass matrices we assume in a first step that all Higgs fields which transform under the 
flavor symmetry are copies
of the SM Higgs doublet. Furthermore we do not discuss cases in which all left-handed and left-handed
conjugate fields transform as singlets under the flavor group, since such structures can be produced
by the use of an abelian group as well. Note that this does not exclude the possibility that the
Majorana mass matrix originates from a coupling which involves only fields transforming as
one-dimensional representation under the flavor group, as the Majorana mass term stems solely
from the coupling of two either left-handed or left-handed conjugate fields.
We present all mass matrices and discuss differences among Dirac and Majorana mass matrix structures.

\noindent As one interesting application we show that a prediction of the Cabibbo angle $\theta _{12} ^{q}$, 
more precisely of one of the elements of the quark mixing matrix $V_{CKM}$,
in terms of $\underline{\mbox{only}}$ group theoretical quantities, i.e. the index $n$ of the group 
\Groupname{D}{n} (\Doub{D}{n}),
the index of the representation and the breaking direction in the flavor space, becomes possible.

\noindent Several smaller dihedral groups \cite{dnpaper} and their double-valued groups 
\cite{dnprimepaper} were already used in
the literature to construct models with flavor symmetries. We comment on some
of them by comparing the mass matrices they use with ours. We show that the prediction
of $\theta_{23}^{l}= 45^{\circ}$ and $\theta_{13}^{l}=0$ of models using a
flavor symmetry \Groupname{D}{4} $\times \mbox{\Groupname{Z}{2}} ^{(aux)}$
\cite{Grimus1} and
 \Groupname{D}{3} $\times \mbox{\Groupname{Z}{2}} ^{(aux)}$ \cite{Grimus2}
results 
from the fact that in these models non-trivial subgroups of \Groupname{D}{4}
and \Groupname{D}{3} are preserved, respectively.

\noindent We only briefly touch the question of the VEV alignment and
the choice and stabilization of the desired vacuum structure in our conclusions, since the study
of the Higgs potentials is beyond the scope of this paper. Nevertheless, we
emphasize that of course only the proof that an advocated VEV structure is
realized in a certain potential and the proof of its stability can make
the theory viable. A detailed study of potentials of
SM Higgs doublets which transform under \Groupname{D}{n} or \Doub{D}{n} will be
presented elsewhere \cite{comingsoon}.

\noindent The paper is structured as follows: in \Secref{sec:grouptheory} we present 
the basic group theory of the dihedral groups \Groupname{D}{n} and
\Doub{D}{n}; \Secref{sec:subs} contains the analysis of the subgroups
of \Groupname{D}{n} and \Doub{D}{n}; in \Secref{sec:breakingchains} we
study the stepwise breaking of the dihedral symmetries and show all possible
breaking chains for the single-valued groups $D_n$. We study in \Secref{sec:massmatrices} all Dirac
as well as Majorana mass matrices with a non-vanishing determinant
which can arise from the distinct breakings found in 
\Secref{sec:subs} and  mention some possible applications in 
\Secref{sec:applications}. A comparison of our findings to the literature 
is given in \Secref{sec:litcompare}. Finally, we conclude in 
\Secref{sec:conclusions} and comment on differences among models using
flavor-charged Higgs doublets and gauge singlets (flavons) as well as on some
simple Higgs potentials. The various
appendices contain further group theoretical results which are needed
for our calculations such as Kronecker products and Clebsch Gordan coefficients
as well as the decomposition of representations of the dihedral groups into
representations of their subgroups and the breaking chains for the groups
\Doub{D}{n}.

%%%%%%%%%%%%%%%%%%%%%%%%%%%%%%%%%%%%%%%%%%%%%
\section{Properties of Dihedral Groups}
%%%%%%%%%%%%%%%%%%%%%%%%%%%%%%%%%%%%%%%%%%%%%
\label{sec:grouptheory}

%%%%%%%%%%%%%%%%%%%%%%%%%%%%%%%%%%%%%%%%%%%%%
\mathversion{bold}
\subsection{Single-valued Groups $D_n$}
\mathversion{normal}
%%%%%%%%%%%%%%%%%%%%%%%%%%%%%%%%%%%%%%%%%%%%
\label{sec:grouptheorydn}

\noindent All \Groupname{D}{n} groups are non-abelian apart from $D_1$ ($\cong Z_2$) and $D_2$ ($\cong Z_2 \times Z_2$). They only contain real one- and
two-dimensional irreducible representations. If its index $n$ is even, the group
\Groupname{D}{n} has four one- and $\frac{n}{2}-1$
two-dimensional representations and for $n$ being odd \Groupname{D}{n} has
two one- and $\frac{n-1}{2}$ two-dimensional representations. In the
following we denote the one-dimensional representations with
$\MoreRep{1}{i}$ and the two-dimensional ones with $\MoreRep{2}{j}$
where the indices $\rm i$ and $\rm j$ are $\rm i=1,2$ and $\mathrm{j}=1, ...,\frac{n-1}{2}$
for \Groupname{D}{n} with $n$ odd and $\rm i=1,...4$ and
$\mathrm{j}=1,..,\frac{n}{2} -1$ for \Groupname{D}{n} with $n$ even. The
representation $\MoreRep{1}{1}$ is always the trivial one, i.e. the
one whose characters are $1$ for all classes \footnote{The character of a representation for 
a certain group element is just the trace of the corresponding representation matrix independent of the
choice of basis. For one-dimensional representations the representation matrix is
only a complex number (unequal zero) which is then also the character of this representation.}. The
order of the group \Groupname{D}{n} is $2 \, n$. The generators A and B of the one-dimensional representations are A=B=1 for $\MoreRep11$ and A=1, B=-1 for $\MoreRep12$. For even $n$ we also have A=-1, B=1 for $\MoreRep13$ and A=B=-1 for $\MoreRep14$. The generators of the two-dimensional representations are \cite{Lomont}:
\begin{equation}
\label{eq:generatorsdn}
\rm A =\left(\ba{cc} 
                           \mathrm{e}^{\left( \frac{2 \pi i}{n} \right) \, \mathrm{j}} & 0 \\
                            0 & \mathrm{e}^{-\left( \frac{2 \pi i}{n} \right)
                              \, \mathrm{j}} 
          \ea\right) \; , \; \rm B=\left(\ba{cc} 
                                       0 & 1 \\
                                       1 & 0 
                  \ea\right) 
\end{equation}
with $\mathrm{j}=1, \dots, \frac{n}{2} -1$ for $n$ even and $\mathrm{j}=1, \dots, \frac{n-1}{2}$ for $n$ odd. They fulfill the relations:
\begin{equation}\label{eq:genrelationsdn}
\mathrm{A}^{n} =\mathbb{1} \;\;\; , \;\;\; \rm B^2=\mathbb{1} \;\;\; ,
\;\;\; \rm ABA=B \; . 
\end{equation}

\noindent Note that we have chosen complex
generators for the two-dimensional representations. Since the representations themselves are real,
there exists a unitary matrix $U$ which links their generators to the complex conjugates:
$U= \left( \begin{array}{cc} 0 & 1  \\ 1 & 0 \end{array} \right)$. 
For any $\left( \begin{array}{c} a_{1}
    \\ a_{2} \end{array} \right) \sim \Rep{2}$ the combination $ U \, \left( \begin{array}{c} a_{1} ^{\star}
  \\ a_{2} ^{\star} \end{array} \right) = \left( \begin{array}{c} a_{2} ^{\star}
  \\ a_{1} ^{\star} \end{array} \right)$ transforms as $\Rep{2}$
instead of $ \left( \begin{array}{c} a_{1} ^{\star}
  \\ a_{2} ^{\star} \end{array} \right)$, as would be the case for real
generators $\rm A$ and $\rm B$.

%%%%%%%%%%%%%%%%%%%%%%%%%%%%%%%%%%%%%%%%%%%%%%%%%
\mathversion{bold}
\subsection{Double-valued groups $D_n\sprime$}
\mathversion{normal}
%%%%%%%%%%%%%%%%%%%%%%%%%%%%%%%%%%%%%%%%%%%%%%%%%
\label{sec:grouptheorydnprime}

\noindent The groups \Doub{D}{n} are the double-valued counterparts of the
groups \Groupname{D}{n}. All groups \Doub{D}{n} with $n>1$ are
non-abelian, $D_1\sprime$ is isomorphic to $Z_4$. The simplest non-abelian double-valued dihedral group, $D_2\sprime$, is also called the quaternion group. Hence, one often uses the notation $Q_{2n}$ instead of $D_n\sprime$. The group \Doub{D}{n} is of order $4 \, n$. Similar to the groups \Groupname{D}{n} 
they only contain one- and two-dimensional irreducible representations. The group \Doub{D}{n}
has four one-dimensional and $n-1$ two-dimensional
representations. For $n$ even, the one-dimensional representations are
real, i.e. their characters are real. Furthermore the two-dimensional
representations $\MoreRep{2}{j}$ with j even are real, i.e. not only
their characters are real, but there also exists a set of real
representation matrices. In contrast to this the representations
$\MoreRep{2}{j}$ with j odd are pseudo-real, i.e. their characters
are real, but one cannot find a set of representation matrices which
are also real. If $n$ is odd the one-dimensional
representations $\MoreRep{1}{1,2}$ are real, while $\MoreRep{1}{3}$ and
$\MoreRep{1}{4}$ are complex conjugated to each other. As for $n$ even,
the representations $\MoreRep{2}{j}$  with j even are real and with
j odd are pseudo-real. Compared to the groups \Groupname{D}{n} one
has to add the pseudo-real and complex representations to get
\Doub{D}{n}. Therefore the real representations are usually called
even, while the pseudo-real and complex ones are named odd
representations \cite{Desmier}. The generators for the one-dimensional representations are the same as for \Groupname{D}{n}, if $n$ is even. If $n$ is odd, the generators of the one-dimensional representations are  A=B=1 for $\MoreRep11$ ,  A=1, B=-1 for $\MoreRep12$, A=-1, B=$-i$ for $\MoreRep13$ and A=-1, B=$i$ for $\MoreRep14$. The generators and their relations for the
two-dimensional representations also have a similar form as in case of 
\Groupname{D}{n} \cite{Desmier,Patera}:
\begin{equation}\label{eq:generatorsdnprime1}
\rm A =\left(\begin{array}{cc} 
                           \mathrm{e}^{\left( \frac{\pi i}{n} \right) \, \mathrm{j}} & 0 \\
                            0 & \mathrm{e}^{-\left( \frac{\pi i}{n} \right)
                              \, \mathrm{j}} 
          \end{array}\right) \;\;\; , \;\;\; \rm B=\left(\begin{array}{cc} 
                                       0 & 1 \\
                                       1 & 0 
                  \end{array}\right)  
\end{equation}
\noindent for j even, and
\begin{equation}\label{eq:generatorsdnprime2}
\rm A =\left(\begin{array}{cc} 
                           \mathrm{e}^{\left( \frac{\pi i}{n} \right) \, \mathrm{j}} & 0 \\
                            0 & \mathrm{e}^{-\left( \frac{\pi i}{n} \right)
                              \, \mathrm{j}} 
          \end{array}\right) \;\;\; , \;\;\; \rm B=\left(\begin{array}{cc} 
                                       0 & i \\
                                       i & 0 
                  \end{array}\right)  
\end{equation}
\noindent for j odd. They fulfill:
\begin{equation}\label{eq:genrelationsdnprime}
\mathrm{A}^{n} =\rm R \;\;\; , \;\;\; \rm B^2=R \;\;\; , R^{2} =
\mathbb{1} \;\;\; , \;\;\; \rm ABA=B \; ,
\end{equation}
\noindent with $\rm R$ being $\mathbb{1}$ in case of an even representation and
$-\mathbb{1}$ for an odd one. Comparing the generators of even and odd
representations one recognizes that the generator $\rm B$ contains an
extra factor $i$ for odd representations.
\noindent Since also for \Doub{D}{n} all two-dimensional representations are
real or pseudo-real, i.e. not complex, there has to exist a similarity
transformation $U$ between the representation matrices and their complex
conjugates. If the index $\rm j$ of the representation $\MoreRep{2}{j}$ is
even, $U$ is the same as for the representations of
\Groupname{D}{n}. For $\rm j$ being odd, $U=\left(
  \begin{array}{cc}
    0 & -1\\
    1 & 0
    \end{array}
\right)$
such that $\left( \begin{array}{c} - a_{2} ^{\star} \\ a_{1} ^{\star}
  \end{array} \right) = U \, \left( \begin{array}{c} a_{1} ^{\star} \\
    a_{2} ^{\star} \end{array} \right)$ transforms in the same way as $\left( \begin{array}{c} a_{1} \\
    a_{2} \end{array} \right) \sim \MoreRep{2}{j}$  with $\rm j$ odd.

%%%%%%%%%%%%%%%%%%%%%%%%%%%%%%%%%%%%%%%%%%%
\section{Nontrivial Subgroups}
%%%%%%%%%%%%%%%%%%%%%%%%%%%%%%%%%%%%%%%%%%%
\label{sec:subs}

\noindent In this section we determine the subgroups of a general $D_n$ or $D_n\sprime$ group, using the generators given in \Secref{sec:grouptheory}. This can be done systematically by determining for each representation the eigenvalues and eigenvectors of the group elements. All group elements, which have the same eigenvector corresponding to the eigenvalue of 1, form a subgroup. We then determine the group structure, which is simple, as all subgroups turn out to be either dihedral or cyclic.\\
For the one-dimensional representations we only need to look at which elements of the group are represented by a 1 - these elements then form a subgroup. For the two-dimensional representations, we need to determine all representation matrices which have an eigenvalue of 1. We only need to consider two general matrices: $\mathrm{A}^x$ and $\mathrm{B}\mathrm{A}^y$ and calculate their eigenvalues as a function of $x$ or $y$ and the index $\mathrm{j}$ of the representation $\MoreRep2j$, and then calculate the corresponding eigenvectors. All eigenvectors with eigenvalue 1 turn out to have the same structure:
\begin{equation}
\label{eq:vev}
 \left(
		\ba{c}
		e^{\frac{-4 \pi i \mathrm{j} m}{g}} \\
		1
		\ea
		\right)
\end{equation}
\noindent where $g$ is the order of $G_F$ ($2n$ for single-valued, $4n$ for double-valued groups) and $m$ is an integer. We thus have for a given two-dimensional representation a class of subgroups, parameterized by $m$,  where one of the generators of the subgroup will be $\mathrm{B}\mathrm{A}^m$. For group elements represented by the unit matrix, which appear in unfaithful representations \footnote{A representation is unfaithful if the number of distinct representation matrices is smaller than the order of the group.}, an arbitrary eigenvector corresponds to an eigenvalue of 1. Hence we have for unfaithful two-dimensional representations an additional subgroup made up of all group elements represented by the unit matrix. To make sure that we have determined all subgroups, we need to consider possible combinations of 2 or more representations. Further subgroups will necessarily be subgroups of the subgroups determined above. Since all of the subgroups encountered so far are either dihedral or cyclic, we know that all further subgroups will also be either dihedral or cyclic. 
 As it turns out, we need at most 2 different representations to reach any possible subgroup of our original $D_n$ or $D_n\sprime$.\\
\noindent We come to a physical interpretation of our results by using them to determine how the VEV of a scalar field transforming non-trivially under $G_F$ will break that symmetry. A VEV of a scalar transforming under a given representation conserves the subgroup of elements which leave the VEV invariant, i.e. the VEV is an eigenvector to the eigenvalue 1 and these we determined above. For two-dimensional representations there can be several subgroups and therefore the structure of the scalar VEV is important. We denote an arbitrary VEV by $<\MoreRep2j>$, while a VEV proportional to the eigenvector of \Eqref{eq:vev} will be denoted by $< \MoreRep2j >\sprime $. Subgroups corresponding to a combination of two representations will be conserved by a combination of VEVs. We get the following results for $D_n$ groups:

\footnotesize
\begin{flushleft}
\begin{tabular}{lll}
	$D_{n}$ & $\stackrel {< \MoreRep11 > } {\longrightarrow}$ 
		&
		$D_{n}$
\end{tabular}\\
\begin{tabular}{lll}
	$D_{n}$ & $\stackrel 
		{ < \MoreRep12 >  } {\longrightarrow}$ 
		& $Z_{n}=  < \mathrm{A} >$ 
\end{tabular}\\
\begin{tabular}{lll}
	$D_{n}$ & $\stackrel 
		{ < \MoreRep13 >  } {\longrightarrow}$ 
		& $D_{\frac{n}{2}}=  < \mathrm{A}^2,\mathrm{B} >$ 
\end{tabular}\\
\begin{tabular}{lll}
	$D_{n}$ & $\stackrel 
		{ < \MoreRep14 >  } {\longrightarrow}$ 
		& $D_{\frac{n}{2}}= < \mathrm{A}^2,\mathrm{B}\mathrm{A} >$
\end{tabular}\\
\begin{tabular}{lllr}
	$D_{n}$ & $\stackrel 
		{ < \MoreRep2j >  } {\longrightarrow} $
		& $Z_{\mathrm{j}}= < \mathrm{A}^{\frac{n}{\mathrm{j}}} >$ &
($\mathrm{j} \mid n$)
\end{tabular}\\
\begin{tabular}{lllr}
	$D_{n}$ & $\stackrel 
		{ < \MoreRep2j >  } {\longrightarrow}$ 
		& nothing & ($\mathrm{j} \nmid n$)
\end{tabular}
\begin{tabular}{lllr}
	$D_{n}$ & $\stackrel 
		{ < \MoreRep2j >\sprime } {\longrightarrow} $
		& $D_{\mathrm{j}} = < \mathrm{A}^{\frac{n}{\mathrm{j}}},\mathrm{B}\mathrm{A}^m >$ & ($\mathrm{j} \mid n ; \, m=0,1,...,\frac{n}{\mathrm{j}}-1$)
\end{tabular}
\begin{tabular}{lllr}
	$D_{n}$ & $\stackrel 
		{ < \MoreRep2j >\sprime } {\longrightarrow} $
		& $Z_2 = < \mathrm{B}\mathrm{A}^m >$ & ($\mathrm{j} \nmid n ; \, m=0,1,...,n-1$)
\end{tabular}
\begin{tabular}{lll}
	$D_{n}$ &
	$\stackrel 
		{ < \MoreRep12 >+< \MoreRep13 >}
		{\longrightarrow}$ & $Z_{\frac{n}{2}} = < \mathrm{A}^2 > $ 
\end{tabular}\\
\noindent (One can also use $< \MoreRep12 >+< \MoreRep14 >$ or $< \MoreRep13 >+< \MoreRep14 >$.)
\begin{tabular}{lllr}
	$D_{n}$ & $\stackrel 
		{< \MoreRep13 > + < \MoreRep2j >\sprime} {\longrightarrow}$ 
		& $Z_2 = < \mathrm{B}\mathrm{A}^m >$ & ($\mathrm{j} \mid n ,\,  \frac{n}{\mathrm{j}}\, odd ;\, 
0\le m \le \frac{n}{\mathrm{j}}-1)$ 
\end{tabular}
\noindent ($m$ even for $\MoreRep13$; for $\MoreRep14$ $m$ odd)
\normalsize

\vspace{0.1cm}
\noindent and for $D_n\sprime$ groups we get:\\
\vspace{0.1cm}

\footnotesize
\begin{tabular}{lll}
	$D_{n}\sprime$ & $\stackrel {< \MoreRep11 > } {\longrightarrow} $
		&
		$D_{n}\sprime$ 
\end{tabular}\\
\begin{tabular}{lll}
	$D_{n}\sprime$ & $\stackrel 
		{ < \MoreRep12 >  } {\longrightarrow}$ 
		& $Z_{2n}= < \mathrm{A} >$ 
\end{tabular}
\begin{tabular}{lllr}
	$D_{n}\sprime$ & $\stackrel 
		{ < \MoreRep13 >\,  or\,  < \MoreRep14 > } {\longrightarrow} $
		& $Z_{n}= < \mathrm{A}^2>$ & ($for\, odd\,  n$)
\end{tabular}
\begin{tabular}{lllr}
	$D_{n}\sprime$ & $\stackrel 
		{ < \MoreRep13 >  } {\longrightarrow} $
		& $D_{\frac{n}{2}}\sprime = < \mathrm{A}^2,\mathrm{B} > $ & ($ for\, even\, n$)
\end{tabular}
\begin{tabular}{lllr}
	$D_{n}\sprime $&$ \stackrel 
		{ < \MoreRep14 >  } {\longrightarrow} 
		$&$ D_{\frac{n}{2}}\sprime = < \mathrm{A}^2,\mathrm{B}\mathrm{A} > $ & ($ for\, even\, n$)
\end{tabular}
\begin{tabular}{lllr}
	$D_{n}\sprime $&$ \stackrel 
		{ < \MoreRep2j >  } {\longrightarrow} 
		$&$ Z_{\mathrm{j}}= < \mathrm{A}^{\frac{2n}{\mathrm{j}}} > $ & ($ \mathrm{j} \mid 2n $)
\end{tabular}\\
\begin{tabular}{lllr}
	$D_{n}\sprime $&$ \stackrel 
		{ < \MoreRep2j >  } {\longrightarrow} 
		$& nothing & ($ \mathrm{j} \nmid 2n$)
\end{tabular}
\begin{tabular}{lllr}
	$D_{n}\sprime $&$ \stackrel 
		{ < \MoreRep2j >\sprime } {\longrightarrow} 
		$&$ D_{\frac{\mathrm{j}}{2}}\sprime = < \mathrm{A}^{\frac{2n}{\mathrm{j}}}, \mathrm{B}\mathrm{A}^m >
                $ & ($ \mathrm{j} \, even, \, \mathrm{j} \mid 2n ; m=0,1,...,\frac{2n}{\mathrm{j}}-1$)
\end{tabular}
\begin{tabular}{lllr}
	$D_{n}\sprime $&$ \stackrel 
		{ < \MoreRep2j >\sprime} {\longrightarrow} 
		$&$ Z_4 = < \mathrm{B}\mathrm{A}^m > 
		$ &($ \mathrm{j}\, even, \, \mathrm{j} \nmid 2n ; m=0,1,...,n-1$)
\end{tabular}
\begin{tabular}{lllr}
	$D_{n}\sprime $&$
	\stackrel 
		{ < \MoreRep12 >+< \MoreRep13 >}
		{\longrightarrow} $&$ Z_n = < \mathrm{A}^2 >  $ & $(for\, even\, n $)
\end{tabular}

\noindent (One can also use $< \MoreRep12 >+< \MoreRep14 >$ or $< \MoreRep13 >+< \MoreRep14 >$.)
\begin{tabular}{lllr}
	$D_{n}\sprime $&$ \stackrel 
		{< \MoreRep13 > + < \MoreRep2j >\sprime} {\longrightarrow} 
		$&$ Z_4 = < \mathrm{B}\mathrm{A}^m > $ &
\end{tabular}\\
\noindent ($n \, even$; $ \mathrm{j} \mid 2n , \frac{2n}{\mathrm{j}}\, odd ;\, 
0\le m\le\frac{2n}{\mathrm{j}}-1)$\\
\noindent ($m$ even for $\MoreRep13$; for $\MoreRep14$ $m$ odd)\\
\begin{tabular}{lllr}
	$D_{n}\sprime $&$ \stackrel 
		{< \MoreRep13 > + < \MoreRep2j >\sprime} {\longrightarrow} 
		$&$ Z_2 = < \mathrm{A}^n > $ &
\end{tabular}\\
\noindent  ($n \, even$; $ \frac{2n}{\mathrm{j}} odd ;
0\le m \le \frac{2n}{\mathrm{j}}-1)$\\
\noindent ($m$ odd for $\MoreRep13$; for $\MoreRep14$ $m$ even)
\end{flushleft}
\normalsize

\noindent Some of these results can also be found in references \cite{solid} and \cite{subgroups}. We now know the minimal VEV structure needed to break $G_F$ down to a given subgroup. The maximal VEV structure preserving that subgroup is then achieved by allowing VEVs for all representations, which have at least one component transforming trivially under the subgroup in question. Therefore we list the transformation properties of the representations of our original dihedral group under a given subgroup. These can be found by expressing the generators of the subgroup in terms of the generators A and B of $G_F$. As a general feature we remark that two-dimensional representations become reducible in several cases. The complete list of decompositions is given in Appendix \ref{app:decomposition} along with the maximal VEV structure.

%%%%%%%%%%%%%%%%%%%%%%%%%%%%%%%%%%%
\section{Breaking Chains}
%%%%%%%%%%%%%%%%%%%%%%%%%%%%%%%%%%%
\label{sec:breakingchains}

\noindent We have seen, that a dihedral group will in general have several nontrivial subgroups. We consider all possible breaking patterns of a dihedral group, where the symmetry breaking happens in an arbitrary number of steps.\\
  \noindent The first step in every chain, will be one of the breakings induced by a single VEV, which we considered in \Secref{sec:subs}. For the next step we need to consider how a second (different) VEV will further break our symmetry group, by considering the intersection of the group elements leaving both VEVs invariant, thereby finding the subgroup which is conserved by both VEVs. We determine the group structure of these elements - if the new subgroup is in fact smaller than the old one, we have found a viable next step in the breaking sequence. This procedure is then iterated until we reach either a $Z_2$ (which has no further subgroups, being the smallest nontrivial group), or until we reach a $Z_{\mathrm{j}}$ with arbitrary index $\mathrm{j}$. $Z_{\mathrm{j}}$  can always be further broken down to a $Z_{\mathrm{k}}$, where $\mathrm{k}$ is a divisor of $\mathrm{j}$, as we discuss below.\\
We use the notation $m_{\mathrm{j}}$ for the phase factor $m$ in the VEV $<\MoreRep2j>\sprime$. This classification according to different phase factors becomes important if a breaking sequence contains VEVs of two distinct two-dimensional representations. \\
\noindent Note the breaking patterns marked with a star. These are not just breaking sequences in their own right, but also can be used as building blocks within or at the end of other sequences. In general, we can reduce the order of a dihedral or cyclic group step by step, until we have reached the subgroup which we want to conserve, as long as the conditions given for the starred breaking sequences are fulfilled at each step.\\
\noindent We find two paths in the breaking sequences, one along the dihedral groups and one along the cyclic groups, in the minimal case eliminating one prime divisor of the order in each step. At any point in the sequence we can step over from the dihedral to the cyclic path (from which there is of course no turning back). The cyclic path ends at $Z_q$, $q$ being the smallest prime factor of $n$, while the dihedral path, will end at $D_1$ or $D_1\sprime$ as these groups are nontrivial. $D_1 \cong Z_2$ is simple, while $D_1\sprime \cong Z_4$, is not just nontrivial but also non-simple, so that we can break one step further down to the $Z_2$ group generated by $\mathrm{A}^n$.\\
\noindent The breaking chains for $D_n$ are:

\footnotesize
\begin{flushleft}
\begin{tabular}{cccccccc}
$D_n $&$ \stackrel {< \MoreRep12 >} {\longrightarrow} $&$ Z_n $&$ 
      \stackrel {< \MoreRep13 >} {\longrightarrow} $&$ Z_{\frac{n}{2}} $&$         
      \stackrel {< \MoreRep2j >} {\longrightarrow} $&$ Z_{\mathrm{j}} $
&($ \mathrm{j} \mid \frac{n}{2}$)
\end{tabular}
\begin{tabular}{ccccccc}
$D_n $&$ \stackrel {< \MoreRep12 >} {\longrightarrow} $&$ Z_n $&$ 
      \stackrel {< \MoreRep2j >} {\longrightarrow} $&$ Z_{\mathrm{j}} $&         
       & 
\end{tabular}
\begin{tabular}{cccccccc}
$D_n $&$ \stackrel {< \MoreRep13 >} {\longrightarrow} $&$ D_{\frac{n}{2}} $&$ 
      \stackrel {< \MoreRep12 >} {\longrightarrow} $&$ Z_{\frac{n}{2}} $&$         
      \stackrel {< \MoreRep2j >} {\longrightarrow} $&$ Z_{\mathrm{j}} $
&($ \mathrm{j} \mid \frac{n}{2}$)
\end{tabular}
\begin{tabular}{cccccccc}
$D_n $&$ \stackrel {< \MoreRep13 >} {\longrightarrow} $&$ D_{\frac{n}{2}} $&$ 
      \stackrel {< \MoreRep14 >} {\longrightarrow} $&$ Z_{\frac{n}{2}} $&$         
      \stackrel {< \MoreRep2j >} {\longrightarrow} $&$ Z_{\mathrm{j}} $
&($ \mathrm{j} \mid \frac{n}{2}
$)
\end{tabular}
\begin{tabular}{cccccccc}
$D_n $&$ \stackrel {< \MoreRep13 >} {\longrightarrow} $&$ D_{\frac{n}{2}} $&$ 
      \stackrel {< \MoreRep2j >} {\longrightarrow} $&$ Z_{\mathrm{j}} $&         
      & 
&($ \mathrm{j} \mid \frac{n}{2}
$)
\end{tabular}
\begin{tabular}{cccccccc}
$D_n $&$ \stackrel {< \MoreRep13 >} {\longrightarrow} $&$ D_{\frac{n}{2}} $&$ 
      \stackrel {< \MoreRep2j >\sprime} {\longrightarrow} $&$ D_{\mathrm{j}} $&$         
      \stackrel {< \MoreRep2k >} {\longrightarrow} $&$ Z_{\mathrm{k}} $
&($ \mathrm{j} \mid \frac{n}{2};\, \mathrm{k} \mid \mathrm{j}$) 
\end{tabular}\\
\noindent ( $m_{\mathrm{j}}$ even for $\MoreRep13$ and for $\MoreRep14$ $m_{\mathrm{j}}$ is odd)
\begin{tabular}{cccccccc}
$D_n $&$ \stackrel {< \MoreRep13 >} {\longrightarrow} $&$ D_{\frac{n}{2}} $&$ 
      \stackrel {< \MoreRep2j >\sprime} {\longrightarrow} $&$ D_{\mathrm{j}} $&$         
      \stackrel {< \MoreRep2k >\sprime} {\longrightarrow} $&$ Z_2 $
& ($ \mathrm{j} \mid \frac{n}{2};\, \mathrm{k} \nmid \mathrm{j}; \, m_{\mathrm{j}}=m_{\mathrm{k}} \, even
$)
\end{tabular}
\begin{tabular}{cccccccc}
$D_n $&$ \stackrel {< \MoreRep13 >} {\longrightarrow} $&$ D_{\frac{n}{2}} $&$ 
      \stackrel {< \MoreRep2j >\sprime} {\longrightarrow} $&$ Z_2 $&         
      &  
&($ \mathrm{j} \nmid \frac{n}{2}$)
\end{tabular}\\
\noindent ($m_{\mathrm{j}}$ even for $\MoreRep13$ and for $\MoreRep14$ $m_{\mathrm{j}}$ is odd)
\begin{tabular}{cccccccc}
$D_n $&$ \stackrel {< \MoreRep2j >\sprime} {\longrightarrow} $&$ D_{\mathrm{j}} $&$ 
      \stackrel {< \MoreRep12 >} {\longrightarrow} $&$ Z_{\mathrm{j}} $&         
      & 
&($ m_{\mathrm{j}} \, arbitrary
$)
\end{tabular}
\begin{tabular}{cccccccc}
$D_n $&$ \stackrel {< \MoreRep2j >\sprime} {\longrightarrow} $&$ D_{\mathrm{j}} $&$ 
      \stackrel {< \MoreRep13 >} {\longrightarrow} $&$ Z_2 $&         
      & 
&($\mathrm{j} \nmid \frac{n}{2}
$)
\end{tabular}\\
\noindent ($m_{\mathrm{j}}$ even for $\MoreRep13$ and for $\MoreRep14$ $m_{\mathrm{j}}$ is odd)
\begin{tabular}{cccccccc}
$D_n $&$ \stackrel {< \MoreRep2j >\sprime} {\longrightarrow} $&$ D_{\mathrm{j}} $&$ 
      \stackrel {< \MoreRep2k >} {\longrightarrow} $&$ Z_{\mathrm{k}} $&         
      & 
&($ \mathrm{k} \mid \mathrm{j}; \, m_{\mathrm{j}} \, arbitrary
$)
\end{tabular}
\begin{tabular}{cccccccc}
$D_n $&$ \stackrel {< \MoreRep2j >\sprime} {\longrightarrow} $&$ D_{\mathrm{j}} $&$ 
      \stackrel {< \MoreRep2k >\sprime} {\longrightarrow} $&$ Z_2 $&         
      & 
&($ \mathrm{k} \nmid \mathrm{j};\,  m_{\mathrm{j}}=m_{\mathrm{k}}
$)
\end{tabular}\\
\begin{tabular}{ccccccc}
$D_n $&$ \stackrel {< \MoreRep2j >\sprime} {\longrightarrow} $&$ Z_2 $& 
      & &         
      & 
\end{tabular}\\
\begin{tabular}{cccccccc}
$*$ & $D_n $&$ \stackrel {< \MoreRep2j >} {\longrightarrow} $&$ Z_{\mathrm{j}} $&$ 
      \stackrel {< \MoreRep2k >} {\longrightarrow} $&$ Z_{\mathrm{k}} $&         
      &($ \mathrm{k} \mid \mathrm{j}
$)
\end{tabular}
\begin{tabular}{cccccccc}
$*$& $D_n $&$ \stackrel {< \MoreRep2j >\sprime} {\longrightarrow} $&$ D_{\mathrm{j}} $&$ 
      \stackrel {< \MoreRep2k >\sprime} {\longrightarrow} $&$ D_{\mathrm{k}} $&         
      &($ \mathrm{k} \mid \mathrm{j};\, m_{\mathrm{j}}=m_{\mathrm{k}}
$)\\ &&&&&&&
\end{tabular}
\end{flushleft}
\normalsize

\noindent The corresponding results for $D_n\sprime$ are given in Appendix \ref{app:breakingchains}. The conditions on the indices can be divided into two types: One concerns the divisibility of indices. These appear, when we need to ensure, that we have not broken too far, i.e. that the subgroup we want to break to is actually contained in the subgroup we have already broken to. The second type of condition concerns restrictions on the phase factors $m_{\mathrm{j}}$. These appear, if the direction in which we have broken is important. Several breaking directions occur, if we deal with distinct subgroups showing the same structure - for example in $D_n$ we have two $D_{\frac{n}{2}}$ subgroups: $< \mathrm{A}^2,\mathrm{B} >$ and $< \mathrm{A}^2,\mathrm{B}\mathrm{A} >$. The $Z_2$ subgroup generated by $\mathrm{B}\mathrm{A}^m$, for example, is only a subgroup of the first in case of $m$ even and only a subgroup of the second, if $m$ is odd. Hence we need to impose restrictions on the phase factor $m$.\\
Finally, note that in any of the chains given above we can interchange $\MoreRep13$ and $\MoreRep14$ and again receive a viable breaking sequence. This may cause some of the requirements to change, as shown along with the breaking chains.

%%%%%%%%%%%%%%%%%%%%%%%%%%%%%%%%%%%%
\section{Mass Matrices}
%%%%%%%%%%%%%%%%%%%%%%%%%%%%%%%%%%%%
\label{sec:massmatrices}

\noindent We can determine the Dirac mass matrices $M$ which are generated when $G_F$ is broken to one of the subgroups we determined in \Secref{sec:subs} by Higgs bosons transforming non-trivially under $G_F$. Majorana mass matrices are discussed in \Secref{sec:Majorana}. For simplicity we assume that the Higgs bosons are $SU(2)_L$ doublets like the one in the SM. The group theoretical tools we need, i.e. the Kronecker products and Clebsch Gordan coefficients, are given in Appendix \ref{app:grouptheory}. As mentioned in the Introduction, we restrict ourselves in a  fairly major way: We do not consider mass matrices with a zero determinant, that is with at least one zero eigenvalue.\\
We need to decide how the (SM) fermions will transform under $G_F$. Their transformation properties are not limited by our choice of subgroup. The only limitation we impose here, is that we do not want all fermions to transform under one-dimensional representations of $G_F$, since the resulting structures could also be obtained by an abelian $G_F$. Therefore we will not allow for this possibility (for Dirac mass terms) \footnote{For the case of Majorana mass terms: see below.}, leaving us with two general options for the transformation properties of the (SM) fermions. The first possibility is
\begin{displaymath}
L \sim (\MoreRep1{$\bf i_1$},\MoreRep1{$\bf i_2$},\MoreRep1{$\bf i_3$}) \; , \; L^c \sim (\MoreRep1j,\MoreRep2k)
\end{displaymath}
\noindent which we call the three singlet structure. The second possibility is
\begin{displaymath}
L \sim (\MoreRep1i,\MoreRep2j) \; , \; L^c \sim (\MoreRep1l,\MoreRep2k)
\end{displaymath}
\noindent which we call the two doublet structure. Note, that we do not discuss explicitly the case where the left-handed fermions transform under one one- and one two-dimensional representation with the left-handed conjugate fermions transforming under three one-dimensional representations, since we only need to transpose the mass matrices of the three singlet structure to switch the transformation properties of left-handed and left-handed conjugate fermions.\\
\noindent The mass matrices arise, when the Higgs bosons acquire a VEV. We do not choose their transformation properties, as we did for the fermion fields. In fact, we do not even limit the number of Higgs bosons. In this way the mass matrix structure is entirely determined by the properties of $G_F$ and its subgroup, and not by our choice of scalar fields. When determining the mass matrix generated by breaking down to a certain subgroup, we reference \Tabref{abeld} to \Tabref{nonabelq} in \Appref{app:decomposition} and determine which representations are allowed a VEV, while keeping that subgroup intact. We then start by assuming that our model contains a Higgs boson for each of these possible representations \footnote{We do not need to consider the case of two Higgs bosons transforming under the same representation of the flavor group: These two Higgs bosons would have identical quantum numbers, and only the linear combination acquiring a VEV would appear in the mass matrices. However, two Higgs fields transforming in the same way can be very important for other aspects of a model, especially when discussing the Higgs potential, as shown in reference \cite{AF2}.} and that all of them acquire a VEV, with a structure conserving the relevant subgroup. We can then easily eliminate Higgs bosons from our model, by setting their VEVs to zero in the mass matrix.\\
\noindent We first give our general results in \Secref{sec:results}. In \Secref{sec:notation} we give the conventions and notation we use in \Secref{sec:3singlet} and \Secref{sec:2doublet}, where we discuss the three singlet and two doublet structures, respectively, for single-valued groups $D_n$. The resulting mass matrices will be discussed subgroup by subgroup. In \Secref{Mass DoubleD} we discuss why no new Dirac mass matrix structures appear for double-valued groups. Finally, we present the possible Majorana mass matrices in 
\Secref{sec:Majorana}.

%%%%%%%%%%%%%%%%%%%%%%%%%%%%%%%%%%%%%%%%%%%
\subsection{General Results}
%%%%%%%%%%%%%%%%%%%%%%%%%%%%%%%%%%%%%%%%%%%
\label{sec:results}

\noindent We encounter a very limited number of distinct Dirac mass matrix structures, i.e. in total only five distinct structures are possible. We display them for down-type fermions (down-type quarks and charged leptons). The changes for up-type fermions, i.e. up-type quarks and neutrinos, are discussed in \Secref{sec:notation} \footnote{We have to distinguish these two cases, since our Higgs fields are always assumed to transform as the SM Higgs doublet. In the SM the field $H$ itself couples to the down-type fermions, while its conjugate $\epsilon \, H^{\star}$ is coupled to up-type
ones. This difference is relevant here, since, for example,  we decided to 
use complex generators for the two-dimensional representations.}. 
The first possible structure is a diagonal mass matrix
\be
\left( \begin{matrix}
A & 0 & 0\\
0 & B & 0\\
0 & 0 & C
\end{matrix}
\right)
\label{eq:diagonal}
\ee
\noindent The second type are semi-diagonal mass matrices, of the form
\be
\left( \begin{matrix}
A & 0 & 0\\
0 & 0 & B\\
0 & C & 0
\end{matrix}
\right)
\label{eq:semi-diagonal}
\ee
\noindent where the squared mass matrix $MM^{\dagger}$ has eigenvalues $\vert A \vert^2$, $\vert B \vert^2$ and $\vert C \vert^2$. We also encounter a block matrix structure
\be
 \left( \begin{matrix}
A & 0 & 0\\
0 & B & C\\
0 & D & E
\end{matrix}
\right)
\label{eq:block}
\ee
\noindent for which the squared mass matrix has the characteristic polynomial $(\lambda-\vert A \vert^2)(\lambda^2 - \lambda (\vert B \vert^2 + \vert C \vert^2 + \vert D \vert^2 + \vert E \vert^2) + (\vert B \vert^2 + \vert C \vert^2)(\vert D \vert^2 + \vert E \vert^2) - \vert B D^{\ast} +C E^{\ast} \vert^2)$. These three structures are so common, that we will not give them explicitly each time. Instead, we will give the type of matrix, followed by a listing of the nonzero entries, where we use the notation and nomenclature introduced here. Note, that in some cases the entries in the 2-3-submatrix are further correlated, e.g. there exist equalities among them.\\
Finally, we find two structures, which appear only once, both times for the subgroups $Z_2=<\mathrm{B}\mathrm{A}^m>$. One has one texture zero and the general structure
\be
\left(
\begin{matrix}
0 & C & -C e^{-i \phi \mathrm{k}}\\
A & D & D e^{-i \phi \mathrm{k}}\\
B & E & E e^{-i \phi \mathrm{k}}
\end{matrix}
\right)
\label{eq:onezero}
\ee
\noindent with the group theoretical phase
\be
\phi = \frac{2 \pi m}{n} \; .
\ee
\noindent In this case the squared mass matrix has the characteristic polynomial $(\lambda - 2 \vert C \vert^2)(\lambda^2 - \lambda (\vert A \vert^2 + \vert B \vert^2 + 2(\vert D \vert^2 + \vert E \vert^2)) + (\vert A \vert^2 + 2 \vert D \vert^2)(\vert B \vert^2 + 2 \vert E \vert^2) - \vert A B^{\ast} + 2 D E^{\ast} \vert^2)$. This mass matrix only shows up for a three singlet structure. 
The other mass matrix structure has no texture zeros and is of the form
\be
\left(
\begin{matrix}
A & C & C e^{-i \phi \mathrm{k}}\\
B & D & E \\
B e^{-i \phi \mathrm{j}} & E e^{i(\mathrm{k}-\mathrm{j})\phi} & D e^{-i (\mathrm{j}+\mathrm{k}) \phi}
\end{matrix}
\label{eq:nozero}
\right)
\ee
\noindent where the squared mass matrix has the characteristic polynomial $(\lambda - \vert D - E e^{i \mathrm{k} \phi} \vert^2) (\lambda^2 - (\vert A \vert^2 +2 \vert B \vert^2 + 2 \vert C \vert^2 + \vert D + E e^{i \mathrm{k} \phi} \vert^2) \lambda + (\vert A \vert^2 + 2 \vert C \vert^2)(2 \vert B \vert^2 + \vert D+E e^{i \mathrm{k} \phi} \vert^2)-2 \vert A B^{\ast} + C (D^{\ast} + E^{\ast} e^{-i \mathrm{k} \phi}) \vert^2)$. This mass matrix only appears with a two doublet structure.\\
\noindent In all cases the characteristic polynomial can be 
factorized into a linear and a quadratic one in $\lambda$.\\ 
\noindent We additionally find one case, where all matrix elements are distinct. We do not consider it, as it corresponds to a smaller symmetry being fully broken: $D_n$ breaks down to $Z_q$, where for all two-dimensional representations $\MoreRep2j$ showing up in the model  $\mathrm{j}$  is a multiple of  $q$ , i.e. $\mathrm{j}=\mathrm{c}_{\mathrm{j}} q$, $\mathrm{c}_{\mathrm{j}}$ an integer. We can then replace the original $D_n$ symmetry by a $D_{\frac{n}{q}}$ and all the representations $\MoreRep2j$ by $\MoreRep2{$\bf c_j$}$, as they are in fact all unfaithful representations of the original $D_n$ symmetry. Breaking the original symmetry down to $Z_q$ then corresponds to fully breaking the smaller symmetry \footnote{To be more precise: In all 
cases the representations of the Higgs fields have the property that their index is
divisible by $q$, but this is not necessarily true for the representations under which
the fermions transform. However, one can then always find some other representations for the 
fermions which reproduce exactly the same matrix structure and which have the property
that also their index is divisible by $q$ such that the case can be reduced to a smaller
symmetry which is fully broken.}. 
As we want to consider conserved subgroups, we dismiss this case, when it shows up.\\
Note, that such a case still allows for some non-trivial correlations among the
mass matrix elements. Nevertheless, they are then not determined by a preserved subgroup,
but only by the fact that we use a non-abelian symmetry.

%%%%%%%%%%%%%%%%%%%%%%%%%%%%%%%%%%%%%%%%%%%%%%
\subsection{Conventions and Notation}
%%%%%%%%%%%%%%%%%%%%%%%%%%%%%%%%%%%%%%%%%%%%%%
\label{sec:notation}

\noindent $\kappa$'s and $\alpha$'s denote Yukawa couplings, $<\phi_{\mathrm{i}}>$ denotes the VEV of the Higgs field transforming as $\MoreRep1i$. For the VEVs of Higgs fields transforming as $\MoreRep2j$ we have two possibilities: Either they are allowed to acquire an arbitrary VEV, in which case we denote the VEV by
\begin{displaymath} 
 	\left(
		\ba{c}
		<\psi_{\mathrm{j}}^1> \\
		<\psi_{\mathrm{j}}^2>
		\ea
		\right)
\end{displaymath}
\noindent or they are allowed only a certain VEV structure. As discussed in \Secref{sec:subs} there is only one such structure. We write the VEV of Higgs fields transforming as $\MoreRep2j$ and acquiring this VEV structure as
\begin{displaymath}
<\psi_{\mathrm{j}}> 	\left(
		\ba{c}
		e^{\frac{-2 \pi i \mathrm{j} m}{n}} \\
		1
		\ea
		\right)
\end{displaymath}
\noindent Note that the VEV structure only determines the relative phase between the two doublet components, so that we are in general free to decide which component we want to include the phase factor in. We will on several occasions want to make use of this freedom, to simplify the appearance of our mass matrix. For example, if we have a left-handed fermion transforming as $\MoreRep13$ or $\MoreRep14$ and the left-handed conjugate fermions transforming as $\MoreRep2j$, we need a Higgs boson transforming as \mathversion{bold} $\MoreRep2{$(\frac{n}{2}$\bf -j)}$ \mathversion{normal} to form an invariant Yukawa coupling. We then write the VEVs of these Higgs fields as
\begin{displaymath}
<\psi_{\frac{n}{2} - \mathrm{j}}> 	\left(
		\ba{c}
		(-1)^m\\
		e^{\frac{-2 \pi i \mathrm{j} m}{n}}
		\ea
		\right)
\end{displaymath}
\noindent In this way, only the phase factor $e^{\frac{-2 \pi i \mathrm{j} m}{n}}$ shows up in the mass matrix and not its complex conjugate. Similarly the VEV of Higgs fields transforming as \mathversion{bold} $\MoreRep2{($n$\bf -(j+k))}$ \mathversion{normal} is written as
\begin{displaymath}
<\psi_{n-(\mathrm{j}+\mathrm{k})}> 	\left(
		\ba{c}
		1\\
		e^{\frac{-2 \pi i (\mathrm{j}+\mathrm{k}) m}{n}}
		\ea
		\right)
\end{displaymath}
\noindent so that it contains the same phase as that of the Higgs fields transforming as $\MoreRep2{j+k}$.\\
\noindent To get the corresponding mass matrices for up-type fermions a few changes have to be implemented. Obviously, they have different Yukawa couplings, i.e.:
\begin{equation}
\kappa \rightarrow \kappa^u \;\;\;\;\; \mbox{and} \;\;\;\;\; \alpha \rightarrow \alpha^u
\end{equation}
\noindent Furthermore all VEVs need to be complex conjugated. This corresponds to the following substitutions:
\begin{eqnarray}
&&<\phi_{\mathrm{i}}> \rightarrow <\phi_{\mathrm{i}}>^{\ast} 
\\ \nonumber
&&<\psi^1_{\mathrm{j}}> \rightarrow <\psi^2_{\mathrm{j}}>^{\ast} \; \mbox{and} \; <\psi^2_{\mathrm{j}}> \rightarrow <\psi^1_{\mathrm{j}}>^{\ast} 
\\ \nonumber
&&\mbox{or} <\psi_{\mathrm{j}}> \rightarrow <\psi_{\mathrm{j}}>^{\ast} e^{\frac{2 \pi i \mathrm{j} m}{n}}
\end{eqnarray}
\noindent These changes are only necessary, if up-type fermions couple to Higgs bosons which transform in the same way under $SU(2)_{L} \times U(1)_{Y}$ as the Higgs bosons which couple to the down-type fermions, like it is
in the SM. If they couple to Higgs bosons with other transformation properties under $SU(2)_{L} \times U(1)_{Y}$, as for example in the Minimal Supersymmetric Standard Model (MSSM) with the Higgs doublets $h_u$ and $h_d$, such changes are obsolete and 
obviously then also the VEVs entering the up- and down-type fermion mass matrices can never be the same.\\ 
\noindent We have in general left all minus signs and phases in the mass matrices, even if they can be trivially rotated away.

%%%%%%%%%%%%%%%%%%%%%%%%%%%%%%%%%%%%%%%%%%%
\subsection{Three Singlet Structure}
%%%%%%%%%%%%%%%%%%%%%%%%%%%%%%%%%%%%%%%%%%%
\label{sec:3singlet}

\noindent To get a mass matrix with a nonzero determinant, at least one two-dimensional representation has to get a VEV, otherwise the second and third columns  of the mass matrix will be zero. This means we can ignore all subgroups, where no two-dimensional representation is allowed a VEV, leaving us with three subgroups to consider $Z_2=<\mathrm{B}\mathrm{A}^m>$, $D_q$ and $Z_q$. The two possible two-dimensional representations that can show up are $\MoreRep2k$ and \mathversion{bold} $\MoreRep2{($\frac{n}{2}$\bf -k)}$ \mathversion{normal} 
coming from the product of $\MoreRep2k$ with a one-dimensional representation.\\
 We first consider the subgroup, where all two-dimensional representations acquire a VEV,  that is $Z_2 = <\mathrm{B}\mathrm{A}^m>$.

%%%%%%%%%%%%%%%%%%%%%%%%%%%%%%%%%%%%%%%%%%%%%%%%%%%%%
\subsubsection{$Z_2 = <\mathrm{B}\mathrm{A}^m>$}
%%%%%%%%%%%%%%%%%%%%%%%%%%%%%%%%%%%%%%%%%%%%%%%%%%%%%
\label{sec:3sZ2}

\noindent Due to the large number of possible combinations of one-dimensional representations, we shall only give rows as building blocks for a mass matrix. We group them according to the index $\mathrm{j}$, i.e. according to the transformation properties of the first generation of left-handed conjugate fermions. We give the row vectors for the $p$-th row of the mass matrix, depending on the index $\mathrm{i}_p$, i.e. the transformation properties of the $p$-th generation of left-handed fermions. We assume, that $m$ is even. To get to an odd $m$, we need to switch $\MoreRep13$ and $\MoreRep14$ and $\phi_3$ and $\phi_4$, as can be inferred from \Tabref{abeld}.
\vspace{0.1in}\\
\begin{tabular}{lcr}
 \multicolumn{3}{l}{$ \mathrm{j}=1$ , $\mathrm{i}_p=1 $} \\
($\kappa_p <\phi_1> $,&$ \alpha_p <\psi_{\mathrm{k}}> $,&$ \alpha_p <\psi_{\mathrm{k}}> e^{\frac{-2 \pi i \mathrm{k} m}{n}}$) 
\end{tabular}\\
\begin{tabular}{lcr}
\multicolumn{3}{l}{$ \mathrm{j}=2,4$ , $\mathrm{i}_p=1 $} \\
(0,&$ \alpha_p <\psi_{\mathrm{k}}> $,&$ \alpha_p <\psi_{\mathrm{k}}> e^{\frac{-2 \pi i \mathrm{k} m}{n}}$) 
\end{tabular}\\
\begin{tabular}{lcr}
\multicolumn{3}{l}{$ \mathrm{j}=3$ , $\mathrm{i}_p=1 $} \\
($\kappa_p <\phi_3> $,&$ \alpha_p <\psi_{\mathrm{k}}> $,&$ \alpha_p <\psi_{\mathrm{k}}> e^{\frac{-2 \pi i \mathrm{k} m}{n}}$) 
\end{tabular}\\
\begin{tabular}{lcr}
 \multicolumn{3}{l}{$ \mathrm{j}=1,3$ , $\mathrm{i}_p=2 $} \\
(0,&$ \alpha_p <\psi_{\mathrm{k}}> $,&$ -\alpha_p <\psi_{\mathrm{k}}> e^{\frac{-2 \pi i \mathrm{k} m}{n}}$) 
\end{tabular}\\
\begin{tabular}{lcr}
 \multicolumn{3}{l}{$ \mathrm{j}=2$ , $\mathrm{i}_p=2 $} \\
($\kappa_p <\phi_1> $,&$ \alpha_p <\psi_{\mathrm{k}}> $,&$ -\alpha_p <\psi_{\mathrm{k}}> e^{\frac{-2 \pi i \mathrm{k} m}{n}}$) 
\end{tabular}\\
\begin{tabular}{lcr}
 \multicolumn{3}{l}{$ \mathrm{j}=4$ , $\mathrm{i}_p=2 $} \\
($\kappa_p <\phi_3> $,&$ \alpha_p <\psi_{\mathrm{k}}> $,&$ -\alpha_p <\psi_{\mathrm{k}}> e^{\frac{-2 \pi i \mathrm{k} m}{n}}$) 
\end{tabular}\\
\begin{tabular}{lcr}
 \multicolumn{3}{l}{$ \mathrm{j}=1$ , $\mathrm{i}_p=3 $} \\
($\kappa_p <\phi_3> $,&$ \alpha_p <\psi_{\frac{n}{2}-\mathrm{k}}> $,&$ \alpha_p <\psi_{\frac{n}{2}-\mathrm{k}}> e^{\frac{-2 \pi i \mathrm{k} m}{n}}$) 
\end{tabular}\\
\begin{tabular}{lcr}
 \multicolumn{3}{l}{$ \mathrm{j}=2,4$ , $\mathrm{i}_p=3 $} \\
(0,&$ \alpha_p <\psi_{\frac{n}{2}-\mathrm{k}}> $,&$ \alpha_p <\psi_{\frac{n}{2}-\mathrm{k}}> e^{\frac{-2 \pi i \mathrm{k} m}{n}}$) 
\end{tabular}\\
\begin{tabular}{lcr}
 \multicolumn{3}{l}{$ \mathrm{j}=3$ , $\mathrm{i}_p=3 $} \\
($\kappa_p <\phi_1> $,&$ \alpha_p <\psi_{\frac{n}{2}-\mathrm{k}}> $,&$ \alpha_p <\psi_{\frac{n}{2}-\mathrm{k}}> e^{\frac{-2 \pi i \mathrm{k} m}{n}}$) 
\end{tabular}\\
\begin{tabular}{lcr}
 \multicolumn{3}{l}{$ \mathrm{j}=1,3$ , $\mathrm{i}_p=4 $} \\
(0,&$ -\alpha_p <\psi_{\frac{n}{2}-\mathrm{k}}> $,&$ \alpha_p <\psi_{\frac{n}{2}-\mathrm{k}}> e^{\frac{-2 \pi i \mathrm{k} m}{n}}$) 
\end{tabular}\\
\begin{tabular}{lcr}
 \multicolumn{3}{l}{$ \mathrm{j}=2$ , $\mathrm{i}_p=4 $} \\
($\kappa_p <\phi_3> $,&$- \alpha_p <\psi_{\frac{n}{2}-\mathrm{k}}> $,&$ \alpha_p <\psi_{\frac{n}{2}-\mathrm{k}}> e^{\frac{-2 \pi i \mathrm{k} m}{n}}$) 
\end{tabular}\\
\begin{tabular}{lcr}
 \multicolumn{3}{l}{$ \mathrm{j}=4$ , $\mathrm{i}_p=4 $} \\
($\kappa_p <\phi_1> $,&$ -\alpha_p <\psi_{\frac{n}{2}-\mathrm{k}}> $,&$ \alpha_p <\psi_{\frac{n}{2}-\mathrm{k}}> e^{\frac{-2 \pi i \mathrm{k} m}{n}}$) 
\end{tabular}
\vspace{0.1in}\\
\noindent These building blocks end up giving us only one general option for a mass matrix, that is a mass matrix with one zero entry. Mass matrices with more than one zero entry or no zero entry give a determinant of zero, and so will not be considered here. This is due to the fact, that there is a correlation between the entry in the first column and the relative sign of the entries in the second and third column. A nonzero entry in the first column necessarily implies either a relative sign between the entries in the other two columns (for $\mathrm{j}$=2 and $\mathrm{j}$=4) or no relative sign (for $\mathrm{j}$=1 and $\mathrm{j}$=3) - we can not however have both cases in the same mass matrix. To see what a typical mass matrix looks like, consider an example. Let $\mathrm{j} = \mathrm{i}_2 = 1$,  $\mathrm{i}_1=2$ and $\mathrm{i}_3=3$. The mass matrix then reads:
\be
\left(
\begin{matrix}
0 & \alpha_1 <\psi_{\mathrm{k}}> & - \alpha_1 <\psi_{\mathrm{k}}> e^{\frac{-2 \pi i \mathrm{k} m}{n}} \\
\kappa_2 <\phi_1> & \alpha_2 <\psi_{\mathrm{k}}> & \alpha_2 <\psi_{\mathrm{k}}> e^{\frac{-2 \pi i \mathrm{k} m}{n}} \\
\kappa_3 <\phi_3> & \alpha_3 <\psi_{\frac{n}{2}-\mathrm{k}}> & \alpha_3 <\psi_{\frac{n}{2}-\mathrm{k}}> e^{\frac{-2 \pi i \mathrm{k} m}{n}}
\end{matrix}
\right)
\label{eq:onezeroexpl}
\ee
\noindent The scalar representations and the position of the zero (within the first column) can vary, depending on our assignment of fermion representations, but the general structure will always be the one texture zero structure of \Eqref{eq:onezero}.

%%%%%%%%%%%%%%%%%%%%%%%%%%%%%%%%%%%%%%%%%%%%%%%%%%%%%%%%%%%%%%%%%%%%%%%%%%%%%%
\subsubsection{$D_q=<\mathrm{A}^{\frac{n}{q}},\mathrm{B}\mathrm{A}^m>$}
%%%%%%%%%%%%%%%%%%%%%%%%%%%%%%%%%%%%%%%%%%%%%%%%%%%%%%%%%%%%%%%%%%%%%%%%%%%%%%
\label{sec:3sDq}

\noindent We continue with those cases, where not all two-dimensional representations can acquire a VEV. For the subgroups $D_q$, we only need to consider the case, where $\frac{n}{q}$ is odd. As the only two-dimensional representation for the fermions is $\MoreRep2k$, the only relevant two-dimensional representations for Higgs fields are $\MoreRep2k$ and \mathversion{bold} $\MoreRep2{($\frac{n}{2}$\bf -k)}$. \mathversion{normal} Now we can read off \Tabref{nonabeld} that the Higgs bosons transforming under these representations can only acquire a VEV if  $q$  divides  $\mathrm{k}$  and  $q$  divides $\frac{n}{2}-\mathrm{k}$, respectively. If  $\frac{n}{q}$ is even, these two conditions are equivalent, i.e. either both Higgs fields can receive a VEV and we are effectively dealing with a smaller $G_F$ which is then broken down to $Z_2=<\mathrm{B}\mathrm{A}^m>$, or neither can receive a VEV, and we end up having the second and third columns  equal to zero.\\
\noindent If, however $\frac{n}{q}$ is odd, these two conditions are mutually exclusive (except where $q$=1, but $D_1$ is the same as $Z_2=<\mathrm{B}\mathrm{A}^m>$), and we consider them separately below. We do not need to consider the case where neither of the two conditions are fulfilled, as then no Higgs field transforming under a two-dimensional representation will acquire a VEV and we are again left with a mass matrix containing two zero column vectors.\\
\noindent As in \Secref{sec:3sZ2}, we consider the structure of the $p$-th row, depending on $\mathrm{i}_p$, $\mathrm{j}$ and $\mathrm{k}$. The simplest entry is the $M_{p1}$ element. We have, since only $\phi_1$ can acquire a VEV that $M_{p1}=\kappa_p <\phi_1>$ if  $\mathrm{i}_p=\mathrm{j}$ or $M_{p1}=0$ otherwise.
We give the other entries as building block row vectors $
 	\left(
		\ba{cc}
		M_{p2} \, &
		M_{p3}
		\ea
		\right)$. 
They are, for $q$ dividing k
\vspace{0.1in}\\
\begin{tabular}{clr}
$\mathrm{i}_p=1 $ :&
($\alpha_p <\psi_{\mathrm{k}}> $,& $\alpha_p <\psi_{\mathrm{k}}> e^{\frac{- 2 \pi i \mathrm{k} m}{n}}$) 
\end{tabular}
\begin{tabular}{clr}
$\mathrm{i}_p=2 $ :&
($\alpha_p <\psi_{\mathrm{k}}> $,& $-\alpha_p <\psi_{\mathrm{k}}> e^{\frac{- 2 \pi i \mathrm{k} m}{n}}$) 
\end{tabular}
\begin{tabular}{clr} $\mathrm{i}_p=3,4 $ : &
(0,& 0) 
\end{tabular}
\vspace{0.1in}\\
\noindent and for $q$ dividing $\frac{n}{2}$-k
\vspace{0.1in}\\
\begin{tabular}{clr}
$\mathrm{i}_p=1,2: $ &
(0,&0) 
\end{tabular}\\
\begin{tabular}{lr}
\multicolumn{2}{l}{$\mathrm{i}_p=3 $:}\\
($(-1)^m \alpha_p <\psi_{\frac{n}{2}-\mathrm{k}}> $,& $\alpha_p <\psi_{\frac{n}{2}-\mathrm{k}}> e^{\frac{- 2 \pi i \mathrm{k} m}{n}}$) 
\end{tabular}
\begin{tabular}{lr}
\multicolumn{2}{l}{$\mathrm{i}_p=4 $:}\\
($(-1)^{m+1} \alpha_p <\psi_{\frac{n}{2}-\mathrm{k}}> $,& $\alpha_p <\psi_{\frac{n}{2}-\mathrm{k}}> e^{\frac{- 2 \pi i \mathrm{k} m}{n}}$) 
\end{tabular}
\vspace{0.1in}\\
\noindent As in \Secref{sec:3sZ2} the multiple possibilities can be reduced to one general form. This is again due to the fact, that we have a correlation between the element in the first column and those in the second and third columns. Only those rows, for which $\mathrm{i}_p=\mathrm{j}$ can have a nonzero element in the first column. So if we choose the element in the first column to be nonzero, we have also determined whether the elements in the second and third columns are zero. We must have at least one zero entry in the second and third column, otherwise we are only breaking a smaller $G_F$ down to its subgroup $Z_2=<\mathrm{B}\mathrm{A}^m>$, a case we have already discussed. So, we need to choose j in such a way, that zero elements in the second and third column are not correlated with a zero element in the first column, otherwise we would end up with a zero row vector. This means that $\mathrm{j}=3$ or $\mathrm{j}=4$ for $q$ dividing  $\mathrm{k}$ , while if $q$ divides $\frac{n}{2}-\mathrm{k}$ we must choose $\mathrm{j}=1$ or $\mathrm{j}=2$. We can then not have another row, where $\mathrm{i}_p=\mathrm{j}$, because there again the elements in the second and third column would be zero and two row vectors would be linearly dependent. So exactly one of the $\mathrm{i}_p$ must be equal to $\mathrm{j}$, say $\mathrm{i}_1$. And we can say even more: To ensure a nonzero determinant, $\mathrm{i}_2$ and $\mathrm{i}_3$ must be unequal, since otherwise the second and third row vector will be linearly dependent. We can then write the particle content more exactly as $L \sim (\MoreRep1{$\bf i_1$},\MoreRep1{$\bf i_2$},\MoreRep1{$\bf i_3$})$ and $L^c \sim (\MoreRep1{$\bf i_1$}, \MoreRep2k)$. All mass matrices will then be of the block form of \Eqref{eq:block}. We give as an example the entries for $\mathrm{i}_1=3$, $\mathrm{i}_2=1$, $\mathrm{i}_3=2$ and $q$ dividing $\mathrm{k}$:
\begin{eqnarray}
A&=& \kappa_1 <\phi_1> \\
B&=& \alpha_2 <\psi_{\mathrm{k}}> \nonumber \\
C&=& \alpha_2 <\psi_{\mathrm{k}}> e^{\frac{-2 \pi i \mathrm{k} m}{n}} \nonumber \\
D&=& \alpha_3 <\psi_{\mathrm{k}}> \nonumber \\
E&=& -\alpha_3 <\psi_{\mathrm{k}}> e^{\frac{-2 \pi i \mathrm{k} m}{n}} \nonumber
\end{eqnarray}

%%%%%%%%%%%%%%%%%%%%%%%%%%%%%%%%%%%%%%%%%%%%%%%%%%%%%%%%%
\subsubsection{$Z_q=<\mathrm{A}^{\frac{n}{q}}>$}
%%%%%%%%%%%%%%%%%%%%%%%%%%%%%%%%%%%%%%%%%%%%%%%%%%%%%%%%%

\noindent The major difference between the subgroups $Z_q$ and $D_q$ is that in the former all doublets will acquire arbitrary VEVs, and hence the mass matrices will exhibit less symmetry. We have only, as for the $D_q$ subgroups, to consider $\frac{n}{q}$ odd. The reasons however are slightly different: As can be inferred from \Tabref{abeld}, all one-dimensional representations can acquire a VEV if $\frac{n}{q}$ is even, so either again none of the relevant Higgs bosons that transform under a two-dimensional representation of the flavor group can acquire a VEV (if $q$ does not divide $\mathrm{k}$), or we are faced with the case where all Higgs fields relevant for Yukawa terms (that is both those transforming under one-dimensional and under two-dimensional representations) can acquire a VEV - this corresponds to the case discussed in \Secref{sec:results} of a smaller flavor symmetry being fully broken. \\
\noindent So we can again set $\frac{n}{q}$ to be odd and start by giving the elements in the first column. They are $M_{p1}=\kappa_p <\phi_1>$ if $\mathrm{i}_p=\mathrm{j}$, $M_{p1}= \kappa_p <\phi_2>$ if \mathversion{bold} $\MoreRep1{$\mathrm{i}_p$}$\mathversion{normal}$\times \MoreRep1{j} = \MoreRep12$ or $M_{p1}= 0$ otherwise, whereas the elements in the second and third columns are, for q dividing k:
\vspace{0.1in}\\
\begin{tabular}{clr}
$\mathrm{i}_p=1$: & ($\alpha_p <\psi_{\mathrm{k}}^2> $,& $\alpha_p <\psi_{\mathrm{k}}^1> $)
\end{tabular}
\begin{tabular}{clr}
$\mathrm{i}_p=2$: &
($\alpha_p <\psi_{\mathrm{k}}^2> $,& $-\alpha_p <\psi_{\mathrm{k}}^1> $)
\end{tabular}\\
\begin{tabular}{clr}
$\mathrm{i}_p=3,4$: &
(0,& 0)
\end{tabular}
\vspace{0.1in}\\
\noindent and for q dividing $\frac{n}{2}$-k
\vspace{0.1in}\\
\begin{tabular}{clr}
$\mathrm{i}_p=1,2$: & (0,& 0)
\end{tabular}\\
\begin{tabular}{clr}
$\mathrm{i}_p=3$: &
($\alpha_p <\psi_{\frac{n}{2}-\mathrm{k}}^1> $,& $\alpha_p <\psi_{\frac{n}{2}-\mathrm{k}}^2> $	)
\end{tabular}
\begin{tabular}{clr}
$\mathrm{i}_p=4$: &
($\alpha_p <\psi_{\frac{n}{2}-\mathrm{k}}^1> $,& $-\alpha_p <\psi_{\frac{n}{2}-\mathrm{k}}^2> $	)
\end{tabular}
\vspace{0.1in}\\
\noindent We can reduce the large number of possibilities. First we set $\mathrm{j}$, which must either be in $\{1,2 \}$ or in $\{3,4 \}$. If we want the $M_{p1}$ element to be nonzero, then $\mathrm{i}_p$ must be in the same set as $\mathrm{j}$ and we also know whether the elements in the second and third column are zero or not. If we now choose $\mathrm{j}$ in such a way, that a nonzero element in the first column implies nonzero elements in the second and third column (and thereby a zero element in the first column implies a zero row vector), we are left to choose between a mass matrix with 9 distinct nonzero elements and a mass matrix with at least one zero row vector and thereby a zero determinant. We therefore need to choose $\mathrm{j}$ in such a way, that a nonzero element in the first column implies a zero in the second and third column, i.e. $\mathrm{j}=3$ or $\mathrm{j}=4$ if $q$ divides  $\mathrm{k}$ , $\mathrm{j}=1$ or $\mathrm{j}=2$ if $q$ divides $\frac{n}{2}-\mathrm{k}$. If we however choose two elements in the first column to be nonzero, then those two row vectors will be linearly dependent, and we will have a zero determinant. So, we need to choose one $\mathrm{i}_p$ in the same set as j, while the other two must lie outside that set - and, again, they cannot be equal, to ensure linear independence of the corresponding row vectors. The general structure will then always be a block matrix. We give as an example the entries for the case where $\mathrm{j}=3$, $\mathrm{i}_1=4$, $\mathrm{i}_2=1$, $\mathrm{i}_3=2$ and $q$ divides $\mathrm{k}$:
\begin{eqnarray}
A&=&\kappa_1 <\phi_2> \\
B&=& \alpha_2 <\psi_{\mathrm{k}}^2> \nonumber \\
C&=& \alpha_2 <\psi_{\mathrm{k}}^1> \nonumber \\
D&=& \alpha_3 <\psi_{\mathrm{k}}^2> \nonumber \\
E&=& -\alpha_3 <\psi_{\mathrm{k}}^1> \nonumber 
\end{eqnarray}

%%%%%%%%%%%%%%%%%%%%%%%%%%%%%%%%%%%%%%%%%%%%%
\subsection{Two Doublet Structure}
%%%%%%%%%%%%%%%%%%%%%%%%%%%%%%%%%%%%%%%%%%%%%
\label{sec:2doublet}

\noindent In our discussion of the two doublet structure we frequently use an additional index, p, given by $\MoreRep1i \times \MoreRep1l = \MoreRep1p$.  Without loss of generality we assume $\mathrm{j} \geq \mathrm{k}$.\\
For the two doublet structure we will discuss all possible subgroups, as they all give viable mass matrices.\noindent Another difference compared to the three singlet structure is that the mass matrices given in this section are also potential candidates for Majorana mass matrices, if we impose the conditions $\mathrm{j}=\mathrm{k}$ and $\mathrm{i}=\mathrm{l}$. If a mass matrix can also be used as a Majorana mass matrix, we will mention this and briefly note, which Yukawa couplings need to be equal in that case and which terms drop out due to anti-symmetry.\\

%%%%%%%%%%%%%%%%%%%%%%%%%%%%%%%%%%%%%%%%%%%%%%%
\subsubsection{$Z_n = <\mathrm{A}> $}
%%%%%%%%%%%%%%%%%%%%%%%%%%%%%%%%%%%%%%%%%%%%%%%
\label{sec:2dZn}

\noindent We read off \Tabref{abeld}, that only $<\phi_1>$ and $<\phi_2>$ can get a VEV when conserving this subgroup. This limits our freedom in choosing representations for the fermions: The two doublets must couple to form a $\MoreRep11$ or a $\MoreRep12$, otherwise the second and third row vectors of the mass matrix will be zero. This imposes the condition $\mathrm{j}=\mathrm{k}$. Also, we need $\mathrm{p}$ = 1 or 2, otherwise the first row vector will turn out to be zero. These restrictions leave us with a semi-diagonal mass matrix structure with entries:
\begin{eqnarray}
A&=& \kappa_1 <\phi_{\mathrm{p}}>\\
B&=& \kappa_2 <\phi_1> + \kappa_3 <\phi_2> \nonumber \\
C&=& \kappa_2 <\phi_1> - \kappa_3 <\phi_2> \nonumber
\end{eqnarray}
\noindent If $\mathrm{p}=1$ this is also a possible structure for a Majorana mass matrix. In this case the anti-symmetric part, i.e. the terms containing $<\phi_2>$ drop out.

%%%%%%%%%%%%%%%%%%%%%%%%%%%%%%%%%%%%%%%%%%%%%%%%%%%%%%%%%
\subsubsection{$Z_{\frac{n}{2}} = <\mathrm{A}^2> $}
%%%%%%%%%%%%%%%%%%%%%%%%%%%%%%%%%%%%%%%%%%%%%%%%%%%%%%%%%

\noindent From the \Tabref{abeld} we infer that this subgroup only allows for one-dimensional representations to acquire a VEV. So, we need the product of the two doublets to contain at least one one-dimensional representation. We are thereby left with three possibilities:

\noindent Case (i): $\mathrm{j}=\mathrm{k}$, $\mathrm{j}+\mathrm{k}$ $\neq \frac{n}{2}$ gives a semi-diagonal matrix.
\begin{eqnarray}
A&=& \kappa_3 <\phi_{\mathrm{p}}> \\
B&=& \kappa_1 <\phi_1> + \kappa_2 <\phi_2> \nonumber \\
C&=& \kappa_1 <\phi_1> - \kappa_2 <\phi_2> \nonumber
\end{eqnarray}
Case (ii): $\mathrm{j}+\mathrm{k} = \frac{n}{2}$, $\mathrm{j}=\mathrm{k}=\frac{n}{4}$  gives a block structure.
\begin{eqnarray}
A&=& \kappa_5 <\phi_{\mathrm{p}}> \\
B&=& \kappa_3 <\phi_3> + \kappa_4 <\phi_4> \nonumber \\
C&=& \kappa_1 <\phi_1> + \kappa_2 <\phi_2> \nonumber \\
D&=& \kappa_1 <\phi_1> - \kappa_2 <\phi_2> \nonumber \\
E&=& \kappa_3 <\phi_3> - \kappa_4 <\phi_4> \nonumber
\end{eqnarray}
Case (iii): j$\neq$ k, $\mathrm{j}+\mathrm{k}$ = $\frac{n}{2}$ gives a diagonal structure.
\begin{eqnarray}
A&=& \kappa_3 <\phi_{\mathrm{p}}> \\
B&=& \kappa_1 <\phi_3> + \kappa_2 <\phi_4> \nonumber \\
C&=& \kappa_1 <\phi_3> - \kappa_2 <\phi_4> \nonumber
\end{eqnarray}
\noindent Cases (i) and (ii) are also a possibility for Majorana mass matrices. In this case the anti-symmetric terms containing $<\phi_2>$ drop out.

%%%%%%%%%%%%%%%%%%%%%%%%%%%%%%%%%%%%%%%%%%%%%%%%%%%%%%%%%%%%%%
\subsubsection{$Z_{q}=<\mathrm{A}^{\frac{n}{q}}>$}
%%%%%%%%%%%%%%%%%%%%%%%%%%%%%%%%%%%%%%%%%%%%%%%%%%%%%%%%%%%%%%
\label{sec:2dZq}

\noindent This subgroup requires quite an amount of case differentiation, since we want to make the discussion as general as possible, and allow all possible relations between $q$ and the other indices of the model. We will first discuss the case where $\frac{n}{q}$ is even, and then, at the end of this subsection, discuss the slight changes induced by $\frac{n}{q}$ being odd.\\
As an ordering principle in our discussion, we have taken the structure of the resulting mass matrix, as only the three characteristic types discussed in \Secref{sec:results} will show up.\\
Most of the conditions deal with the question, which two-dimensional representations are allowed a VEV. This translates directly into deciding whether $q$ divides the index of that representation. The two-dimensional representations which can show up are $\MoreRep2j$ and $\MoreRep2k$,  \mathversion{bold} $\MoreRep2{($\frac{n}{2}$\bf -j)}$ and $\MoreRep2{($\frac{n}{2}$\bf -k)}$  from the coupling of two-dimensional with one-dimensional representations and $\MoreRep2{j+k}$,$\MoreRep2{($n$\bf -(j+k))}$ and $\MoreRep2{j-k}$  \mathversion{normal} from the coupling of the two two-dimensional representations. We will only give mass matrices for the case where $\MoreRep2j \times \MoreRep2k$ contains $\MoreRep2{j+k}$. Mass matrices for the case where it contains \mathversion{bold} $\MoreRep2{($n$\bf -(j+k))}$ \mathversion{normal} can be obtained by replacing $\psi_{\mathrm{j}+\mathrm{k}}$ by $\psi_{n-(\mathrm{j}+\mathrm{k})}$ and then switching the components of the doublet (see Appendix \ref{app:CGsDn}).\\
As $q$ must divide  $n$  for $Z_q$ to be a subgroup of $D_n$, $q$ dividing $\mathrm{j}+\mathrm{k}$ and $q$ dividing $n-(\mathrm{j}+\mathrm{k})$ are equivalent. As already noted, $q$ dividing  $\mathrm{j}$  is equivalent to $q$ dividing $\frac{n}{2}-\mathrm{j}$, if $\frac{n}{q}$ is even, which we assume for this discussion.\\
To ensure a nonzero determinant, $q$ must at least divide either $\mathrm{j}-\mathrm{k}$ \footnote{This includes the case j=k, as all numbers divide zero, corresponding to the fact, that $\MoreRep11$ and $\MoreRep12$ can get a VEV for an arbitrary $q$.} or $\mathrm{j}+\mathrm{k}$: If not, the two by two submatrix in the lower right-hand corner of the mass matrix will be zero. This implies directly that $q$ must divide either both  $\mathrm{j}$  and  $\mathrm{k}$ , or neither of the two. If $q$ divides  $\mathrm{j}$  and  $\mathrm{k}$  however, then it also divides $\mathrm{j}+\mathrm{k}$ and $\mathrm{j}-\mathrm{k}$ - hence all relevant two-dimensional representations can acquire an arbitrary VEV and all one-dimensional representations can acquire a VEV anyway. As this leads to the case where the mass matrix contains 9 distinct entries, we disregard this case. We summarize our findings in \Tabref{tab:index} and discuss the different cases below in detail.\\
\begin{table}
\begin{tabular}{ccccc|c}
& j & k & j+k & j-k & Structure \\
\hline
$q$  divides &  &  &  &  & Det[M]=0 \\
$q$  divides &  &  & $\times$ &  & Diagonal \\
$q$  divides &  &  &  & $\times$ & Semi-diagonal \\
$q$  divides & $\times$ &  &  &  & Det[M]=0 \\
$q$  divides &  & $\times$ &  &  & Det[M]=0 \\
$q$ divides &  &  & $\times$ & $\times$ & Block\\
$q$  divides & $\times$ & $\times$ & $\times$ & $\times$ & Full\\
\end{tabular}
\caption{Index relations and corresponding mass matrix structure}
\label{tab:index}
\end{table}
\vspace{0.1in}\\
\noindent Diagonal Matrix
\vspace{0.1in}\\
\noindent This structure appears in the following case: $q$ must divide $\mathrm{j}+\mathrm{k}$ but $q$ does not divide $(\mathrm{j}-\mathrm{k})$,  $\mathrm{j}$  or  $\mathrm{k}$ . Note that this case is not possible for $q$=2, since the sum and the difference of two numbers are either both odd or both even, nor is it possible for $\mathrm{j}=\mathrm{k}$. For $\mathrm{j}+\mathrm{k}$ $\neq \frac{n}{2}$ this gives
\begin{eqnarray}
A&=& \kappa_1 <\phi_{\mathrm{p}}> \\
B&=& \kappa_2 <\psi_{\mathrm{j}+\mathrm{k}}^{2}> \nonumber \\
C&=& \kappa_2 <\psi_{\mathrm{j}+\mathrm{k}}^{1}> \nonumber
\end{eqnarray}
\noindent If $\mathrm{j}+\mathrm{k}$=$\frac{n}{2}$ the mass matrix entries are
\begin{eqnarray}
A&=&\kappa_3 <\phi_{\mathrm{p}}>  \\
B&=&\kappa_1 <\phi_3> + \kappa_2 <\phi_4> \nonumber \\
C&=&\kappa_1 <\phi_3> - \kappa_2 <\phi_4> \nonumber
\end{eqnarray}
\noindent Semi-diagonal Matrix
\vspace{0.1in}\\
\noindent This structure shows up, if $q$ divides $(\mathrm{j}-\mathrm{k})$, but not $\mathrm{j}+\mathrm{k}$,  $\mathrm{j}$  or  $\mathrm{k}$ . This is not possible if $q=2$, nor is it possible for $\mathrm{j}+\mathrm{k}=\frac{n}{2}$, since this contradicts the conditions $q \nmid (\mathrm{j}+\mathrm{k})=\frac{n}{2}$ and $\frac{n}{q}$ being even. This leaves two cases: For $\mathrm{j} \neq \mathrm{k}$ the mass matrix entries are
\begin{eqnarray}
A&=&\kappa_2 <\phi_{\mathrm{p}}>\\
B&=&\kappa_1 <\psi_{\mathrm{j}-\mathrm{k}}^{2}> \nonumber \\
C&=&\kappa_1  <\psi_{\mathrm{j}-\mathrm{k}}^{1}> \nonumber
\end{eqnarray}
\noindent while for $\mathrm{j}=\mathrm{k}$ we get
\begin{eqnarray}
A&=&\kappa_3 <\phi_{\mathrm{p}}> \\
B&=&\kappa_1 <\phi_1> + \kappa_2 <\phi_2> \nonumber \\
C&=&\kappa_1 <\phi_1> - \kappa_2 <\phi_2> \nonumber
\end{eqnarray}
\noindent which is a candidate for a Majorana mass matrix if we omit the anti-symmetric terms.
\vspace{0.1in}\\
\noindent Block Matrix
\vspace{0.1in}\\
\noindent This structure shows up, if $q$ divides $(\mathrm{j}-\mathrm{k})$ and  $\mathrm{j}+\mathrm{k}$, but not  $\mathrm{j}$  and  $\mathrm{k}$ . This forces $q$ to be even, as $q$ must divide $2\mathrm{j}=(\mathrm{j}-\mathrm{k})+(\mathrm{j}+\mathrm{k})$ while not dividing  $\mathrm{j}$ , that is a factor of 2 is relevant for making a number divisible by $q$. In case $\mathrm{j}+\mathrm{k}=\frac{n}{2}$, $\mathrm{j} \neq \mathrm{k}$ we get
\begin{eqnarray}
A&=&\kappa_4 <\phi_{\mathrm{p}}>  \\
B&=&\kappa_1 <\phi_3> + \kappa_2 <\phi_4> \nonumber \\
C&=&\kappa_3 <\psi_{\mathrm{j}-\mathrm{k}}^{2}> \nonumber \\
D&=& \kappa_3  <\psi_{\mathrm{j}-\mathrm{k}}^{1}>  \nonumber \\
E&=& \kappa_1 <\phi_3> - \kappa_2 <\phi_4> \nonumber 
\end{eqnarray}
\noindent In case $\mathrm{j}+\mathrm{k} \neq \frac{n}{2}$ and $\mathrm{j} \neq \mathrm{k}$, we get
\begin{eqnarray}
A&=&\kappa_3 <\phi_{\mathrm{p}}> \\
B&=&\kappa_2 <\psi_{\mathrm{j}+\mathrm{k}}^{2}> \nonumber \\
C&=&\kappa_1 <\psi_{\mathrm{j}-\mathrm{k}}^{2}> \nonumber \\
D&=& \kappa_1  <\psi_{\mathrm{j}-\mathrm{k}}^{1}>  \nonumber \\
E&=& \kappa_2 <\psi_{\mathrm{j}+\mathrm{k}}^{1}> \nonumber
\end{eqnarray}
\noindent and if j$+$k $\neq \frac{n}{2}$ and $\mathrm{j}=\mathrm{k}$, we get
\begin{eqnarray}
A&=&\kappa_4 <\phi_{\mathrm{p}}> \\
B&=&\kappa_3 <\psi_{2\mathrm{j}}^{2}> \nonumber \\
C&=&\kappa_1 <\phi_1> + \kappa_2 <\phi_2> \nonumber \\
D&=&\kappa_1 <\phi_1> - \kappa_2 <\phi_2> \nonumber \\
E&=& \kappa_3 <\psi_{2\mathrm{j}}^{1}> \nonumber 
\end{eqnarray}
\noindent Finally, concerning the case $\mathrm{j}=\mathrm{k}$, $\mathrm{j}+\mathrm{k}= \frac{n}{2}$: this leads us to the conditions $q \mid \frac{n}{2}$ and $q \nmid \frac{n}{4}=\mathrm{j}=\mathrm{k}$, which implies either $q=\frac{n}{2}$ (already covered) or $q$ even and $\frac{n}{4}$ odd. The mass matrices are the same in both cases, with entries:
\begin{eqnarray}
A&=&\kappa_5 <\phi_{\mathrm{p}}> \\
B&=&\kappa_3 <\phi_3> + \kappa_4 <\phi_4> \nonumber \\
C&=& \kappa_1 <\phi_1> + \kappa_2 <\phi_2> \nonumber \\
D&=&\kappa_1 <\phi_1> - \kappa_2 <\phi_2> \nonumber \\
E&=&\kappa_3 <\phi_3> - \kappa_4 <\phi_4> \nonumber
\end{eqnarray}
\noindent For $\frac{n}{q}$ even $q$ either divides both  $\mathrm{j}$  and ($\frac{n}{2}-\mathrm{j}$) or neither, and we need no further case differentiation.\\
 If however $\frac{n}{q}$ is odd, we need to pay closer attention. In case $\mathrm{i}$ and $\mathrm{l}$ are both in $\{ 1,2 \}$, the discussion is as above, since  \mathversion{bold} $\MoreRep2{($\frac{n}{2}$\bf -j)}$ and $\MoreRep2{($\frac{n}{2}$\bf -k)}$  \mathversion{normal} will not show up as the representation of a Higgs field. In case $\mathrm{i}$ and $\mathrm{l}$ are both in $\{ 3,4 \}$ we can also use the above discussion if we substitute $\mathrm{j}$ by ($\frac{n}{2}-\mathrm{j}$) and $\mathrm{k}$ by ($\frac{n}{2}-\mathrm{k}$) in the conditions, which changes nothing concerning the conditions regarding sums and differences, as $q$ will always divide $n$.\\ 
\noindent Further changes occur due
to the fact that $\phi_{3,4} \sim \MoreRep{1}{3,4}$ are not allowed a VEV
anymore, see \Tabref{abeld}.\\
\noindent Finally, we are left with the following: We drop for a moment the condition that $\mathrm{j} \geq \mathrm{k}$ and instead impose the condition $\mathrm{i} \in \{ 1,2 \}$ and $\mathrm{l} \in \{3,4 \}$. This means that $M_{11}$ will be zero, since $\mathrm{p}$ = 3 or 4 and $\phi_{\mathrm{p}}$ is then not allowed a VEV due to $\frac{n}{q}$ being odd. If we now want to avoid having a zero column or row vector in our mass matrix, both $\psi_{\frac{n}{2}-\mathrm{j}}$ and $\psi_{\mathrm{k}}$ must acquire a VEV, i.e. $q$ must divide $\mathrm{k}$ and ($\frac{n}{2}-\mathrm{j}$). By assumption however, $q$ does not divide $\frac{n}{2}$, which leads us straight to the conclusion that $q$ does not divide  $\mathrm{j}+\mathrm{k}$ or $\vert \mathrm{j}-\mathrm{k} \vert$, thereby leaving the two-by-two matrix in the lower right-hand corner of the mass matrix zero, which we have excluded.

%%%%%%%%%%%%%%%%%%%%%%%%%%%%%%%%%%%%%%%%%%%%%%%%%%%%%%%
\subsubsection{$Z_2 = <\mathrm{B}\mathrm{A}^m>$}
%%%%%%%%%%%%%%%%%%%%%%%%%%%%%%%%%%%%%%%%%%%%%%%%%%%%%%%
\label{sec:2dZ2}

\noindent As this structure also strongly depends on the one-dimensional representations under which the fermions transform, we have reduced it entirely to building blocks, to avoid having to deal with too many subcases. As we can read off \Tabref{abeld}, all Higgs bosons transforming under a two-dimensional representation will acquire a structured VEV. The only constraints that arise are therefore due to the Higgs bosons transforming under one-dimensional representations. This means that the $M_{11}$ entry is of special interest. We will first write down the general structure and then use this to explain, why $M_{11}$ has to be nonzero to ensure a nonzero determinant. \\
\be
\left(
\begin{matrix}
\kappa_3 w & \kappa_4 X_1 & \kappa_4 Y_1 e^{- \frac{2 \pi i \mathrm{k} m}{n}}\\
\kappa_5 X_2 & \kappa_1 u & \kappa_2 v \\
\kappa_5 Y_2 e^{- \frac{2 \pi i \mathrm{j} m}{n}}& \kappa_2 v e^{- \frac{2 \pi i (\mathrm{j}-\mathrm{k}) m}{n}}  & \kappa_1 u e^{- \frac{2 \pi i (\mathrm{j}+\mathrm{k}) m}{n}}
\end{matrix}
\right)
\ee
\noindent where:\\
\noindent $u= <\phi_3>$ if $\mathrm{j}+\mathrm{k}=\frac{n}{2}$, $m$ even \\
$u= <\phi_4>$ if $\mathrm{j}+\mathrm{k}=\frac{n}{2}$, $m$ odd \\
$u= <\psi_{\min{[\mathrm{j}+\mathrm{k},n-\mathrm{j}-\mathrm{k}]}}>$ if $\mathrm{j}+\mathrm{k} \neq \frac{n}{2}$\\
\\
$v=<\phi_1>$ if $\mathrm{j}=\mathrm{k}$\\
$v=<\psi_{\mathrm{j}-\mathrm{k}}>$ if $\mathrm{j}$ $\neq$ k\\
\\
$w= <\phi_{\mathrm{p}}>$ where $\mathrm{p}$=1,3 if $m$ even, $\mathrm{p}$=1,4 if $m$ odd
\vspace{0.1in}\\
\noindent Note that some of the phase factors degenerate to signs, in case of j=k or j+k=$\frac{n}{2}$. 
$X_i$ and $Y_i$ depend on the transformation properties of the doublets and singlets involved, i.e. $X_1$ and $Y_1$ depend on $\mathrm{i}$ and $\mathrm{k}$, $X_2$ and $Y_2$ depend on $\mathrm{j}$ and $\mathrm{l}$. Building blocks for $X_1$ and $Y_1$ are given in \Tabref{XY}. The same table can be used for $X_2$ and $Y_2$, substituting $\mathrm{l}$ for $\mathrm{i}$ and $\mathrm{j}$ for $\mathrm{k}$.
\begin{table*}
\begin{tabular}{c|cccc}
 & $\mathrm{i}=1$ & $\mathrm{i}=2$ & $\mathrm{i}=3$ & $\mathrm{i}=4$ \\
\hline
	$\left(
		\ba{c}
		X_1 \\
		Y_1
		\ea
		\right) = $&
	$\left(
		\ba{c}
		<\psi_{\mathrm{k}}> \\
		<\psi_{\mathrm{k}}>
		\ea
		\right) $&
	$\left(
		\ba{c}
		<\psi_{\mathrm{k}}> \\
		-<\psi_{\mathrm{k}}>
		\ea
		\right) $&
	$\left(
		\ba{c}
		(-1)^m <\psi_{\frac{n}{2}-\mathrm{k}}> \\
		<\psi_{\frac{n}{2}-\mathrm{k}}>
		\ea
		\right) $&
	$\left(
		\ba{c}
		(-1)^{m+1} <\psi_{\frac{n}{2}-\mathrm{k}}> \\
		<\psi_{\frac{n}{2}-\mathrm{k}}>
		\ea
		\right)$\\
\end{tabular}
\caption{Building blocks for matrix of two doublet structure under the subgroups $Z_2=<B\mathrm{A}^m>$}
\label{XY}
\end{table*}\\
\noindent For the discussion of the $M_{11}$ element let us assume that $m$ is even; the reasoning for an odd $m$ will be analogous. $w=0$ implies that $\mathrm{p}$=2 or 4, that is $\mathrm{i}$=1 or 3 while $\mathrm{l}$=2 or 4. Switching $\mathrm{i}$ and $\mathrm{l}$ is also possible - the reasoning is again analogous. If we now consult the table for the $X$ and $Y$ entries, we see that this implies a relative minus sign between the $X_2$ and $Y_2$ entries, while there is no relative minus sign between the $X_1$ and $Y_1$ entries, so the sum of the second and third row vector is proportional to the first row vector. This is true also with nontrivial phases. Therefore $w\neq0$ must hold.\\
\noindent If $\mathrm{j}=\mathrm{k}$ and $\mathrm{i}=\mathrm{l}$, the matrix can be made symmetric by imposing $\kappa_4 = \kappa_5$ and can then also show up as a Majorana mass matrix.

%%%%%%%%%%%%%%%%%%%%%%%%%%%%%%%%%%%%%%%%%%%%%%%%%%%%%%%%%%%%%%%%%%%%%%%%%%%%%%%%%%%%%%%%%%%%%%%%%%%%%%%%%%%%%%%
\subsubsection{$D_{\frac{n}{2}}=<\mathrm{A}^2,\mathrm{B}>$ and $D_{\frac{n}{2}}=<\mathrm{A}^2,\mathrm{B}\mathrm{A}>$}
%%%%%%%%%%%%%%%%%%%%%%%%%%%%%%%%%%%%%%%%%%%%%%%%%%%%%%%%%%%%%%%%%%%%%%%%%%%%%%%%%%%%%%%%%%%%%%%%%%%%%%%%%%%%%%%

\noindent The discussion for these subgroups is very similar to that of the subgroup $Z_{\frac{n}{2}}$, as only one-dimensional representations can receive a VEV, however in this case not all of them (see \Tabref{nonabeld}). We receive the mass matrices for $D_{\frac{n}{2}}=<\mathrm{A}^2,\mathrm{B}>$ by simply eliminating all terms containing $\phi_2$ and $\phi_4$ from the mass matrices for $Z_{\frac{n}{2}}$, making sure that we do not end up with a zero determinant. To prevent this $\MoreRep1p$ must be allowed a VEV, that is $\mathrm{p}$ must be 1 or 3. We can then distinguish between the following three sub-cases as for $Z_{\frac{n}{2}}$:
\vspace{0.1in}\\
\noindent Case (i): $\mathrm{j}=\mathrm{k}$ gives a semi-diagonal structure.
\begin{eqnarray}
A&=&\kappa_2 <\phi_{\mathrm{p}}> \\
B&=&\kappa_1 <\phi_1> \nonumber \\
C&=& \kappa_1 <\phi_1> \nonumber
\end{eqnarray}
\noindent Case (ii):$\mathrm{j}=\mathrm{k}$, $\mathrm{j}+\mathrm{k}$ $= \frac{n}{2}$ gives a block structure.
\begin{eqnarray}
A&=&\kappa_3 <\phi_{\mathrm{p}}>  \\
B&=&\kappa_2 <\phi_3> \nonumber \\
C&=& \kappa_1 <\phi_1> \nonumber \\
D&=&\kappa_1 <\phi_1> \nonumber \\
E&=& \kappa_2 <\phi_3> \nonumber 
\end{eqnarray}
\noindent Case (iii): $\mathrm{j} \neq \mathrm{k}$, $\mathrm{j}+\mathrm{k}=\frac{n}{2}$ gives a diagonal structure.
\begin{eqnarray}
A&=&\kappa_2 <\phi_{\mathrm{p}}>  \\
B&=&\kappa_1 <\phi_3> \nonumber \\
C&=&\kappa_1 <\phi_3> \nonumber
\end{eqnarray}
\noindent Note that in cases (i) and (iii) we get two degenerate eigenvalues. We can get the mass matrices generated by breaking to $D_{\frac{n}{2}} = <\mathrm{A}^2,\mathrm{B}\mathrm{A}>$ by substituting $\phi_3$ by $\phi_4$ in the matrices above. In this case $\mathrm{p}$ must either be 1 or 4 and there is also a relative minus sign between the two occurrences of $<\phi_4>$. Case (i) and (ii) are applicable for Majorana mass matrices - in this case $\mathrm{p}$=1 since $\mathrm{i}$ must be equal to $\mathrm{l}$.

%%%%%%%%%%%%%%%%%%%%%%%%%%%%%%%%%%%%%%%%%%%%%%%%%%%%%%%%%%%%%%%%%%%%%%%%%%%%%%
\subsubsection{$D_q=<\mathrm{A}^{\frac{n}{q}},\mathrm{B}\mathrm{A}^m>$}
%%%%%%%%%%%%%%%%%%%%%%%%%%%%%%%%%%%%%%%%%%%%%%%%%%%%%%%%%%%%%%%%%%%%%%%%%%%%%%
\label{sec:2dDq}

\noindent This case is very similar to the $Z_q$ case. The two differences are: a) Less one-dimensional representations are allowed a VEV and b) the two-dimensional representations are only allowed a structured VEV (see \Tabref{nonabeld}). The first difference implies restrictions on the index $\mathrm{p}$ - if the entry in the upper left-hand corner is zero, the determinant is zero too. So, we need to impose the condition that $\MoreRep1p$ is allowed a VEV which depends on the evenness and oddness of $m$ and $\frac{n}{q}$, see \Tabref{nonabeld}.\\
\noindent The second difference means replacing the arbitrary doublet VEV components by one with same absolute value and fixed relative phase. The case where we have 9 independent entries thereby does not show up here - instead the full matrix structure now corresponds to breaking a smaller $G_F$ down to $Z_2=<\mathrm{B}\mathrm{A}^m>$. We briefly comment on this at the end of this subsection.\\
\noindent We will do the entire discussion assuming an even $m$. This means $\mathrm{p}$ has to be either 1 or 3 ($\frac{n}{q}$ even). To get the mass matrices for an odd $m$ exchange $\phi_3$ and $\phi_4$, taking care to also switch indices in the conditions. The relative signs that occur in this case are encoded in factors of $(-1)^m$, which are therefore left in, even though the rest of the discussion concerns only even $m$.\\
\noindent As for the case of $Z_q$ there are no major differences between $\frac{n}{q}$ being even or odd, except that we need to impose the condition $\mathrm{p}=1$, if $\frac{n}{q}$ is odd, as then only $\MoreRep11$ is allowed a VEV. 
Furthermore, if i and l are in $\{ 3, 4 \}$, one must also replace  $\mathrm{j}$  and  $\mathrm{k}$ by $\frac{n}{2}-\mathrm{j}$ and $\frac{n}{2}-\mathrm{k}$ in the conditions, respectively. The structuring we use is the same as in \Secref{sec:2dZq}, that is the mass matrices are classified according to their structure.
\vspace{0.1in}\\
\noindent Diagonal Matrix
\vspace{0.1in}\\
\noindent For this structure we must have $q$ dividing $\mathrm{j}+\mathrm{k}$ but $q$ not dividing $(\mathrm{j}-\mathrm{k})$,  $\mathrm{j}$  or  $\mathrm{k}$ .This is not possible for $q$=2 nor for $\mathrm{j}=\mathrm{k}$. For j$+$ k $\neq \frac{n}{2}$ this gives
\begin{eqnarray}
A&=&\kappa_1 <\phi_{\mathrm{p}}> \\
B&=&\kappa_2 <\psi_{\mathrm{j}+\mathrm{k}}> \nonumber \\
C&=&\kappa_2 <\psi_{\mathrm{j}+\mathrm{k}}> e^{- \frac{2 \pi i m (\mathrm{j}+\mathrm{k})}{n}} \nonumber
\end{eqnarray}
\noindent If j$+$ k $=\frac{n}{2}$ the mass matrix entries are
\begin{eqnarray}
A&=&\kappa_2 <\phi_{\mathrm{p}}> \\
B&=&\kappa_1 <\phi_3> \nonumber \\
C&=&(-1)^m \kappa_1 <\phi_3> \nonumber
\end{eqnarray}
\noindent Note, that for both of these matrices the squared mass matrix $MM^{\dagger}$ has two degenerate eigenvalues.
\vspace{0.1in}\\
\noindent Semi-diagonal Matrix
\vspace{0.1in}\\
\noindent For this structure $q$ must divide $(\mathrm{j}-\mathrm{k})$, but not $\mathrm{j}+\mathrm{k}$,  $\mathrm{j}$  or  $\mathrm{k}$ . This is not possible if $q=2$, nor is it possible for $\mathrm{j}+\mathrm{k}=\frac{n}{2}$. We are left with two cases: If $\mathrm{j} \neq \mathrm{k}$ we end up with
\begin{eqnarray}
A&=&\kappa_2 <\phi_{\mathrm{p}}>  \\
B&=&\kappa_1 <\psi_{\mathrm{j}-\mathrm{k}}> \nonumber \\
C&=&\kappa_1  <\psi_{\mathrm{j}-\mathrm{k}}> e^{-\frac{2 \pi i m (\mathrm{j}-\mathrm{k})}{n}} \nonumber
\end{eqnarray}
\noindent while for $\mathrm{j}=\mathrm{k}$ we get
\begin{eqnarray}
A&=&\kappa_3 <\phi_{\mathrm{p}}> \\
B&=&\kappa_1 <\phi_1>  \nonumber \\
C&=&\kappa_1 <\phi_1> \nonumber
\end{eqnarray}
\noindent which is a candidate for a Majorana mass matrix if $\mathrm{i}=\mathrm{l}$. Both these matrices give degenerate eigenvalues in the squared mass matrix.
\vspace{0.1in}\\
\noindent Block Matrix
\vspace{0.1in}\\
\noindent This structure shows up, for $q$ dividing $(\mathrm{j}-\mathrm{k})$ and  $\mathrm{j}+\mathrm{k}$, but not  $\mathrm{j}$  and  $\mathrm{k}$ . This forces $q$ to be even. For $\mathrm{j}+\mathrm{k}=\frac{n}{2}$, $\mathrm{j} \neq \mathrm{k}$ we get
\begin{eqnarray}
A&=&\kappa_2 <\phi_{\mathrm{p}}>  \\
B&=&\kappa_1 <\phi_3>  \nonumber \\
C&=& \kappa_3 <\psi_{\mathrm{j}-\mathrm{k}}> \nonumber \\
D&=& \kappa_3  <\psi_{\mathrm{j}-\mathrm{k}}> e^{-\frac{2 \pi i m (\mathrm{j}-\mathrm{k})}{n}}  \nonumber \\
E&=& (-1)^m \kappa_1 <\phi_3> \nonumber
\end{eqnarray}
\noindent In case $\mathrm{j}+\mathrm{k} \neq \frac{n}{2}$ and $\mathrm{j} \neq \mathrm{k}$, we get
\begin{eqnarray}
A&=&\kappa_3 <\phi_{\mathrm{p}}> \\
B&=&\kappa_2 <\psi_{\mathrm{j}+\mathrm{k}}> \nonumber \\
C&=&\kappa_1 <\psi_{\mathrm{j}-\mathrm{k}}> \nonumber \\
D&=&\kappa_1  <\psi_{\mathrm{j}-\mathrm{k}}> e^{-\frac{2 \pi i m (\mathrm{j}-\mathrm{k})}{n}}  \nonumber \\
E&=& \kappa_2 <\psi_{\mathrm{j}+\mathrm{k}}> e^{-\frac{2 \pi i m (\mathrm{j}+\mathrm{k})}{n}} \nonumber
\end{eqnarray}
\noindent If $\mathrm{j}+\mathrm{k} \neq \frac{n}{2}$ and $\mathrm{j}=\mathrm{k}$, we get
\begin{eqnarray}
A&=&\kappa_2 <\phi_{\mathrm{p}}> \\
B&=&\kappa_3 <\psi_{2\mathrm{j}}> \nonumber \\
C&=& \kappa_1 <\phi_1> \nonumber \\
D&=&\kappa_1 <\phi_1> \nonumber \\
E&=& \kappa_3 <\psi_{2\mathrm{j}}> e^{-\frac{4 \pi i m \mathrm{j}}{n}} \nonumber
\end{eqnarray}
\noindent and if $\mathrm{j}+\mathrm{k}=\frac{n}{2}$ and $\mathrm{j}=\mathrm{k}=\frac{n}{4}$, that is for $q$ even and $\frac{n}{4}$ odd, we get
\begin{eqnarray}
A&=&\kappa_3 <\phi_{\mathrm{p}}> \\
B&=&\kappa_2 <\phi_3> \nonumber \\
C&=& \kappa_1 <\phi_1> \nonumber \\
D&=&\kappa_1 <\phi_1> \nonumber \\
E&=&  (-1)^m \kappa_2 <\phi_3> \nonumber
\end{eqnarray}
\noindent Full Matrix 
\vspace{0.1in}\\
\noindent If $q$ divides all relevant indices, that is if it divides $\mathrm{j}$ and $\mathrm{k}$ and thereby automatically divides ($\mathrm{j}-\mathrm{k}$) and $\mathrm{j}+\mathrm{k}$, we are actually breaking a smaller $G_F$ down to its subgroup $Z_2=<\mathrm{B}\mathrm{A}^m>$, as discussed in \Secref{sec:3sDq}. This is because $Z_2=<\mathrm{B}\mathrm{A}^m>$ is equivalent to $D_1$, where the above conditions are automatically fulfilled, as 1 divides any integer. We therefore do not need to consider this case.

%%%%%%%%%%%%%%%%%%%%%%%%%%%%%%%%%%%%%%%%%%%%%%%%%%%
\subsection{Mass Matrices in $D_n\sprime$}
%%%%%%%%%%%%%%%%%%%%%%%%%%%%%%%%%%%%%%%%%%%%%%%%%%%
\label{Mass DoubleD}

\noindent One can immediately see, that the Dirac mass matrix structures for double-valued dihedral flavor symmetries are very similar to those generated by single-valued groups. A general correspondence between subgroups of $D_n$ and $D_n\sprime$ can be established, by looking at the allowed VEVs in \Tabref{abeld} to \Tabref{nonabelq}:
\begin{displaymath}
\ba{c|cccccccc}
D_n & Z_n & Z_{\frac{n}{2}} & Z_{\mathrm{q}} & D_2 & D_{\frac{n}{2}} & D_{\mathrm{q}} & 
Z_{(\mathrm{q}=2)} & Z_2\\
\hline
D_n\sprime & Z_{2n} & Z_n & Z_{\mathrm{q}} & Z_4 & D_{\frac{n}{2}}\sprime & D_{\frac{\mathrm{q}}{2}}\sprime & Z_2 & -
\ea
\end{displaymath}
\noindent As one can see $n$ always has to be replaced by $2n$ in the discussion. The only relevant difference between single- and double-valued groups is the existence of odd representations. Therefore, at least one of the fermion generations should transform as an 
odd representation in order to find possible new mass matrix structures.\\  
Odd two-dimensional representations do not get a structured VEV and in case they get a VEV, all VEVs will be arbitrary. This is only possible for the subgroup $Z_{q}$. Similarly,
for the odd one-dimensional representations, i.e. $\MoreRep{1}{3}$ and
$\MoreRep{1}{4}$ of \Doub{D}{n} with $n$ odd,
only two subgroups allow non-vanishing VEVs, namely $Z_{n}$ and $Z_{q}$ for 
$\frac{2 \, n}{q}$ being even. Note that in both cases $\MoreRep{1}{3}$ and $\MoreRep{1}{4}$ simultaneously
get a VEV.\\
\noindent The only changes that appear are additional signs due to differences in the Clebsch Gordan coefficients (see \Appref{app:CGsDnprime}). If the Higgs fields transform as odd 
representations, such additional signs are not relevant, since they can be absorbed into the 
VEVs, as these are arbitrary anyway.\\
Furthermore one sees, that if at least one of the fermion representations is an odd two-dimensional representation, this results in the 3rd column being multiplied by -1 for the mass matrix structures of \Eqref{eq:diagonal}, \Eqref{eq:semi-diagonal} and \Eqref{eq:block}.
These additional signs can be rotated away by redefining the left-handed conjugate fermions. We would need the mass matrix structure of \Eqref{eq:onezero} or \Eqref{eq:nozero} for this sign change to have phenomenological consequences. These mass matrix structures however only appear for the subgroup $Z_2=<\mathrm{B}\mathrm{A}^m>$, which has no counterpart for double-valued groups, so that these two structures do not arise for double-valued groups.\\
Strictly speaking, we encounter fermion mass matrix structures like \Eqref{eq:onezero}
and \Eqref{eq:nozero} also here, but all such cases can be reduced to a single-valued
group. In the simplest case we have used only even representations for the fermions. Then
it is clear that one could also have used a single-valued group right from the beginning.
Another case occurs, if all the fermions transform as odd representations.
Then the Higgs fields have to transform as even ones. One cannot simply reduce
this case to a single-valued group, as one does not find odd representations in 
\Groupname{D}{n} groups. However, one always finds some equivalent assignment for the
fermions using only even representations which leads to the same mass matrix structure
and which can then be reduced to the case of a single-valued group. Furthermore, one finds
cases in which the mass matrix is allowed to have arbitrary entries, but there exists no
smaller symmetry of the original group which is fully broken. This is the same case as already
mentioned for the \Groupname{D}{n} groups, if the subgroup is \Groupname{Z}{q}. Similarly to there, 
we can find equivalent assignments for the fermions which result in the same matrix structure
and which indeed correspond to a smaller group being fully broken.\\
\noindent We thus conclude that no new Dirac mass matrix structures appear for double-valued 
groups.\\
\noindent For an odd $n$, we need to take into account that $\MoreRep13$ and $\MoreRep14$ are complex (conjugated). This implies that $\MoreRep13 \times \MoreRep14 = \MoreRep11$, so we have to replace $<\phi_3>$ in the mass matrices by $<\phi_4>$, and vice versa, even where they show up only implicitly as $<\phi_{\mathrm{p}}>$. This also leads to differences, when switching to up-type mass matrices: $\phi_4^{\ast}$ transforms as $\MoreRep13$. So if we encounter a $<\phi_3>$ in the down-type mass matrix, we need a $<\phi_4>^{\ast}$ for the up-type mass matrix. For odd two-dimensional representations an additional minus sign is introduced along with the second component of the VEV when switching to the up-type mass matrix, due to the matrix $U$ introduced in \Secref{sec:grouptheorydnprime}. All these changes do not lead to new structures.

%%%%%%%%%%%%%%%%%%%%%%%%%%%%%%%%%%%%%%%%%%%%%%%
\subsection{Majorana Mass Matrices}
%%%%%%%%%%%%%%%%%%%%%%%%%%%%%%%%%%%%%%%%%%%%%%%
\label{sec:Majorana}

\noindent Majorana mass matrices correspond to Yukawa couplings involving two identical fermions, either $L$ or $L^c$. The relevant Higgs fields are then $SU(2)_L$ triplets or gauge singlets, respectively, whereby VEVs of total singlets can be replaced by direct mass terms. The fact that we couple identical fermions, forces Majorana mass matrices to be symmetric.\\
\noindent One comment is in order concerning our exclusion criterion of demanding a nonzero determinant: 
This is no longer phenomenologically motivated in this case, since Majorana masses are only allowed for neutrinos 
and the data still allow one neutrino to be massless. Nevertheless, we restrict ourselves in this way in order to 
keep the discussion manageable.\\
\noindent We have already mentioned when and how Majorana mass terms can show up if the Majorana fermions transform under one two- and one one-dimensional representation of $G_F$, as these correspond to mass matrices of the two doublet type. For \Doub{D}{n}, we have to mention that in case of odd two-dimensional representations the terms containing $<\phi_1>$ are anti-symmetric. So, whenever we remark in \Secref{sec:2doublet} that the anti-symmetric terms containing $<\phi_2>$ drop out, it is instead the terms containing $<\phi_1>$ that will drop out. This does not lead to any new structures compared to the Dirac case.\\
However, other structures can appear, since we allow either $L$ or $L^c$ to transform under three one-dimensional representations of $G_F$. Thereby we also need to consider the case where all fermions in the Yukawa term transform under one-dimensional representations.\\ 
\noindent For $D_n$ with $n$ arbitrary and for $D_n\sprime$ with $n$ even we know that, in addition to the Majorana mass matrix being symmetric, all diagonal entries will be nonzero, as two identical one-dimensional representations always couple to form a trivial representation.\\
\noindent We first discuss the case of $n$ even for \Groupname{D}{n} and \Doub{D}{n}. 
Looking at \Tabref{abeld} to \Tabref{nonabelq} we see that depending on the preserved subgroup
the following one-dimensional representations can get a VEV: only $\MoreRep{1}{1}$, 
$\MoreRep{1}{1}$ and $\MoreRep{1}{i}$ with $\mathrm{i} \neq 1$ 
 or all one-dimensional representations. Especially, the
case in which three representations get a VEV is excluded.
 Concerning the assignment of the fermions 
 we can distinguish the following three cases: either all three generations transform in the same way or two of them transform as the same representation or all three transform as different representations. If the
three generations are assigned to $(\MoreRep{1}{i},\MoreRep{1}{i},\MoreRep{1}{i})$, all mass
matrix entries are non-zero, i.e. the Majorana mass matrix is a general symmetric matrix with
6 independent parameters, as $\MoreRep{1}{i} \times \MoreRep{1}{i} = \MoreRep{1}{1}$ holds. 
For the assignment $(\MoreRep{1}{$\mathrm{i}_{1}$},\MoreRep{1}{$\mathrm{i}_{1}$},
\MoreRep{1}{$\mathrm{i}_{2}$})$ there exist two possible structures: either the matrix has a block structure or it has 6 independent entries. In the first case one has to ensure that $\MoreRep{1}{$\mathrm{i}_{1}$} \times \MoreRep{1}{$\mathrm{i}_{2}$}$ is not allowed a VEV by the preserved subgroup, while in the
second case $\MoreRep{1}{$\mathrm{i}_{1}$} \times \MoreRep{1}{$\mathrm{i}_{2}$}$ should 
also acquire a VEV. In the last case, all fermions transform under different one-dimensional 
representations, which
allows apart from the block and the arbitrary structure the possibility of having a matrix with
non-vanishing entries on the diagonal only. The case only occurs, if the preserved subgroup
only allows $\MoreRep{1}{1}$ to acquire a VEV and therefore the flavor symmetry is not broken 
in the Majorana mass sector.\\
\noindent For the case of $D_n$ with $n$ odd, we only have two one-dimensional representations to choose from. Therefore at least two generations of fermions have to transform under the same representation, forbidding the structure with non-vanishing entries only on the diagonal.\\
\noindent For \Doub{D}{n} with $n$ odd, structures not found above could
only arise from fermion assignments involving the representations 
$\MoreRep{1}{3}$ and $\MoreRep{1}{4}$,
 as $\MoreRep13$ and $\MoreRep14$ are complex and hence 
$\MoreRep13 \times \MoreRep13$ and $\MoreRep14 \times \MoreRep14 
= \MoreRep12$. This means if $\MoreRep12$ is not allowed a VEV, 
we can have zero elements on the diagonal. 
However, we find that if $\MoreRep{1}{2}$ is not allowed a VEV, 
only the trivial
representation $\MoreRep{1}{1}$ is allowed a VEV for $n$ odd, 
i.e. the dihedral symmetry is unbroken in the Majorana mass sector. 
The structure arising from $L^{(c)} \sim (\MoreRep{1}{$\mathrm{i}_{1}$},
\MoreRep{1}{$\mathrm{i}_{2}$},\MoreRep{1}{$\mathrm{i}_{3}$})$ with $\mathrm{i}_{1} \in 
\{ 1,2 \}$ and $\mathrm{i}_{2}, \mathrm{i}_{3} \in \{3,4 \}$, $\mathrm{i}_{2} \neq 
\mathrm{i}_{3}$ is then the same as in case of the two doublet assignment, if
the two odd one-dimensional representations $\MoreRep{1}{3}$ and $\MoreRep{1}{4}$
are replaced by the doublet, 
i.e. it is a semi-diagonal matrix
with the obvious restriction that $C$ equals $B$.\\
\noindent As already mentioned, if $\MoreRep{1}{1}$ is the only
representation which gets a VEV, the flavor symmetry is unbroken in
the Majorana sector. This happens in models
in which the flavor symmetry is only spontaneously broken at a low energy
scale, for example at the electroweak scale. All the possible
mass matrix structures have been enumerated above.

%%%%%%%%%%%%%%%%%%%%%%%%%%%%%%%%%%%%%%%%%
\section{Applications}
%%%%%%%%%%%%%%%%%%%%%%%%%%%%%%%%%%%%%%%%%
\label{sec:applications}

\noindent In this section we show that we can predict the Cabibbo angle, i.e. $|V_{us}|$, in the quark sector 
in terms of group theoretical quantities, i.e. the index $n$ of the group, the index of the 
representations $\rm i$, $\rm j$, $\rm k$ and $\rm l$
under which the generations of left- and left-handed conjugate fields transform, 
\[
Q \sim ( \MoreRep{1}{i}, \MoreRep{2}{j} ) \; , \;\; 
d^{c},u^{c} \sim ( \MoreRep{1}{l}, \MoreRep{2}{k}),
\]
\noindent and the breaking direction in flavor space, $m_{u}$ and $m_{d}$.
In \Secref{sec:2dZ2} we showed that in case of a preserved subgroup 
$\mbox{\Groupname{Z}{2}}= \langle \mathrm{B} \mathrm{A}^{m} \rangle$ of \Groupname{D}{n} the
resulting mass matrix $M_{d}$ for the down quark sector is of the form

\begin{equation}
\label{eq:exampleMd}
M_{d}= \left( \begin{array}{ccc}
	  A_{d} & C_{d} & C_{d} \, \mathrm{e} ^{-i \, \phi_{d} \, \mathrm{k}}\\
	  B_{d} & D_{d} & E_{d}\\
	  B_{d} \, \mathrm{e} ^{-i \, \phi_{d} \, \mathrm{j}} & E_{d} \, \mathrm{e} ^{-i \, \phi_{d} \, 
(\mathrm{j} -\mathrm{k})} & D_{d} \, \mathrm{e} ^{-i \, \phi_{d} \, (\mathrm{j} + \mathrm{k})}
	\end{array}
	\right)
\end{equation}

\noindent where we defined:
\[
\phi_{d} = \frac{2 \, \pi}{n} \, m_{d} \; , \;\; m_{d}=0,1,2,... \; .
\]
such that we break down to a subgroup $\mbox{\Groupname{Z}{2}}= \langle \mathrm{B} \mathrm{A}^{m_{d}} \rangle$
of \Groupname{D}{n}. $A_{d}, B_{d}, ...$ are in general independent complex numbers which are
products of VEVs and Yukawa couplings.
$M_{d} \, M_{d} ^{\dagger}$ can be diagonalized by the unitary matrix

\begin{equation}
\label{eq:exampleUd}
U_{d}= \left( \begin{array}{ccc}
	\cos (\theta_{d}) \, \mathrm{e} ^{i \, \beta_{d}} & 0 & \sin (\theta_{d}) \, 
\mathrm{e} ^{i \, \beta_{d}}\\
	-\frac{\sin (\theta_{d})}{\sqrt{2}} & \frac{\mathrm{e} ^{i \, \phi_{d} \, \mathrm{j}}}{\sqrt{2}}
	& \frac{\cos (\theta_{d})}{\sqrt{2}}\\	
	-\frac{\sin (\theta_{d})}{\sqrt{2}} \, \mathrm{e} ^{-i \, \phi_{d} \, \mathrm{j}} & -\frac{1}{\sqrt{2}}
	& \frac{\cos (\theta_{d})}{\sqrt{2}} \, \mathrm{e} ^{-i \, \phi_{d} \, \mathrm{j}}
	\end{array}
	\right)
\end{equation}

\noindent with $\beta _{d}= \arg \left[A_{d} \, B_{d} ^{\star} + C_{d} \, (D_{d} + E_{d} \, \mathrm{e} ^{i \, \phi_{d} \, 
\mathrm{k}}) ^{\star} \right]$ and $\theta_{d}$ depending on $A_{d}, B_{d}, ...$ in a non-trivial way.
\noindent The mass eigenvalues for the first, second and third generation are then:
\small
\begin{eqnarray}\label{eq:examplemds}
&&\mathrm{m}_{d} ^2=\frac{1}{2} \, [|A_{d}|^2 + 2 \, (|B_{d}|^2 + |C_{d}|^2) + |D_{d} + E_{d} \, \mathrm{e} ^{i \, \phi_{d} \, \mathrm{k}}|^2 \\ \nonumber
&&- (-|A_{d}|^2 + 2 \, (|B_{d}|^2 - |C_{d}|^2) + |D_{d} + E_{d} \, \mathrm{e} ^{i \, \phi_{d} \, \mathrm{k}}|^2) \, \sec (2 \, \theta_{d})] \; , \\ \nonumber
&& \mathrm{m}_{s}=|D_{d}-E_{d} \, \mathrm{e} ^{i \, \phi_{d} \, \mathrm{k}}| \; ,\\ \nonumber
&& \mathrm{m}_{b} ^2=\frac{1}{2} \, [|A_{d}|^2 + 2 \, (|B_{d}|^2 + |C_{d}|^2) + |D_{d} + E_{d} \, \mathrm{e} ^{i \, \phi_{d} \, \mathrm{k}}|^2 \\ \nonumber
&&+ (-|A_{d}|^2 + 2 \, (|B_{d}|^2 - |C_{d}|^2) + |D_{d} + E_{d} \, \mathrm{e} ^{i \, \phi_{d} \, \mathrm{k}}|^2) \, \sec (2 \, \theta_{d})] \; .
\end{eqnarray}
\normalsize

\noindent The mass matrix $M_{u}$ for the up-type quarks has the form:

\begin{equation}
\label{eq:exampleMu}
M_{u}= \left( \begin{array}{ccc}
	  A_{u} & C_{u} \,  \mathrm{e} ^{i \, \phi_{u} \, \mathrm{k}} & C_{u}\\
	  B_{u}  \, \mathrm{e} ^{i \, \phi_{u} \, \mathrm{j}} & D_{u} 
\, \mathrm{e} ^{i \, \phi_{u} \, (\mathrm{j} + \mathrm{k})} & E_{u} 
\, \mathrm{e} ^{i \, \phi_{u} \, (\mathrm{j} -\mathrm{k})} \\
	  B_{u} & E_{u} & D_{u}
	\end{array}
	\right)
\end{equation}
\noindent with
\[
\phi_{u} =\frac{2 \, \pi}{n} \, m_{u} \; , \;\; m_{u}=0,1,2,... \; .
\]
such that $\mbox{\Groupname{Z}{2}}= \langle \mathrm{B} \mathrm{A}^{m_{u}} \rangle$ is the subgroup of 
\Groupname{D}{n} which is preserved by the Higgs fields coupling to
the up-type quarks. $A_{u}, B_{u}, ....$ are in general complex quantities like $A_{d}, B_{d}, ...$.

\noindent The unitary transformation of the left-handed fields is given by:

\begin{equation}
\label{eq:exampleUu}
U_{u}= \left( \begin{array}{ccc}
	0 & \cos (\theta _{u}) \, \mathrm{e} ^{i \, \beta_{u}} & \sin (\theta _{u}) \, 
\mathrm{e} ^{i \, \beta_{u}}\\
	-\frac{\mathrm{e}^{i \, \phi_{u} \, \mathrm{j}}}{\sqrt{2}} & -\frac{\sin (\theta_{u})}{\sqrt{2}}&
	\frac{\cos (\theta_{u})}{\sqrt{2}}\\
	\frac{1}{\sqrt{2}} & -\frac{\sin (\theta _{u})}{\sqrt{2}} \, 
\mathrm{e} ^{- i \, \phi_{u} \, \mathrm{j}} & \frac{\cos (\theta_{u})}{\sqrt{2}} \, 
\mathrm{e} ^{- i \, \phi_{u} \, \mathrm{j}}
	\end{array}
	\right)
\end{equation}

\noindent where $\beta _{u}= \arg \left[A_{u} \, B_{u} ^{\star} + C_{u} \, (D_{u} + E_{u} \, \mathrm{e} ^{-i \, \phi_{u} 
\, \mathrm{k}}) ^{\star}\right] - \phi_{u} \, \mathrm{j}$ and $\theta_{u}$ is a function of the parameters
$A_{u}, B_{u}, ...$.

\noindent The masses for the first, second and third generation read:
\small
\begin{eqnarray}\label{eq:examplemus}
&& \mathrm{m}_{u} = |D_{u}-E_{u} \, \mathrm{e} ^{-i \, \phi _{u} \, \mathrm{k}}| \; , \\ \nonumber
&& \mathrm{m}_{c}^2 =\frac{1}{2} \, [ |A_{u}|^2 + 2 \, (|B_{u}|^2 + |C_{u}|^2) + |D_{u} + E_{u} 
\, \mathrm{e} ^{- i \, \phi_{u} \, \mathrm{k}}|^2 
\\ \nonumber
&&- (- |A_{u}|^2 + 2 \, (|B_{u}|^2 -|C_{u}|^2) 
+ |D_{u} + E_{u} \, \mathrm{e} ^{-i \, \phi_{u} \, \mathrm{k}}|^2) \, \sec (2 \, \theta_{u})] \; ,
\\ \nonumber
&& \mathrm{m}_{t}^2 =\frac{1}{2} \, [ |A_{u}|^2 + 2 \, (|B_{u}|^2 + |C_{u}|^2) + |D_{u} + E_{u} 
\, \mathrm{e} ^{- i \, \phi_{u} \, \mathrm{k}}|^2 
\\ \nonumber
&&+ (- |A_{u}|^2 + 2 \, (|B_{u}|^2 -|C_{u}|^2) 
+ |D_{u} + E_{u} \, \mathrm{e} ^{-i \, \phi_{u} \, \mathrm{k}}|^2) \, \sec (2 \, \theta_{u})] \; .
\end{eqnarray}
\normalsize

\noindent Taking \Eqref{eq:exampleUd} and \Eqref{eq:exampleUu} we arrive at the following
form for the CKM mixing matrix $V_{CKM}$:

\begin{widetext} 
\begin{equation}
\label{eq:exampleVCKM}
V_{CKM}= U_{u} ^{T} \, U_{d} ^{\star}= 
 \frac{1}{2} \, \left( \begin{array}{ccc}
	-\mathrm{e}^{i \, \phi_{d} \, \mathrm{j}} \, x_{-} \, s_{d} & - x_{+} & \mathrm{e} ^{i \, \phi_{d} \, \mathrm{j}} \, x_{-} \, c_{d}\\
 	2 \, \mathrm{e} ^{i \, \alpha} \, c_{d} \, c_{u} + y_{+} \, s_{d} \, s_{u}
& - \mathrm{e} ^{-i \, \phi_{d} \, \mathrm{j}} \, y_{-} \,
s_{u} & 2 \, \mathrm{e}^{i \, \alpha} \, c_{u} \, s_{d} - y_{+} \, c_{d} \, s_{u}\\
	- y_{+} \, c_{u} \,  
s_{d} + 2 \, \mathrm{e} ^{i \, \alpha} \, c_{d} \, s_{u}
& \mathrm{e} ^{- i \, \phi_{d} \, \mathrm{j}} \, y_{-}
\, c_{u} & y_{+} \, c_{d}
\, c_{u} + 2 \, \mathrm{e} ^{i \, \alpha} \, s_{d} \, s_{u} 
\end{array}
\right)
\end{equation}
\end{widetext} 

\noindent where we defined: $x_{\pm}= (1 \pm
\mathrm{e} ^{i \, \delta \phi \, \mathrm{j}})$, $y_{\pm}= (1 \pm
\mathrm{e} ^{-i \, \delta \phi \, \mathrm{j}})$, $\delta \phi = \phi_{u}-\phi_{d}$, $\alpha= \beta_{u}-\beta_{d}$ and
used the abbreviations: $s_{d,u}=\sin (\theta_{d,u})$ and $c_{d,u}=\cos (\theta_{d,u})$.

\noindent As one can see the element $|V_{us}|$ \textit{solely} depends on the group theoretical quantities 
$n$, $\mathrm{j}$, $m_{u}$ and $m_{d}$: 

\begin{equation}
\label{eq:exampleVus}
|V_{us}|= \frac{1}{2} \, \left| 1+ \mathrm{e} ^{i \, \delta \phi \, \mathrm{j}} \right|= 
\left| \cos \left(\frac{\pi \, (m_{u}-m_{d}) \, \mathrm{j}}{n} \right) \right|
\end{equation}

\noindent Note further that only the transformation properties of the left-handed fields which form a 
doublet under the flavor group \Groupname{D}{n} are relevant, since only their representation
index $\mathrm{j}$ appears in the expressions \Eqref{eq:exampleVCKM} and \Eqref{eq:exampleVus} \footnote{To be correct, the phases $\beta_{u,d}$ also depend on the index $\mathrm{k}$ of the doublet 
representation 
under which the left-handed conjugate fields transform. 
Nevertheless these phases depend on many other parameters $A_{u,d}, 
B_{u,d}, ...$ such that they are primarily not determined by the group theory of the flavor symmetry.}.

\noindent The other two mixing angles $\theta_{13} ^{q}$ and $\theta _{23} ^{q}$ can be 
tuned by the use of the two unconstrained angles $\theta _{u}$ and $\theta _{d}$. The Jarlskog
invariant $\mathcal{J} _{CP}$ \cite{Jarlskog} depends on the phase $\alpha$. 
In this way the experimental value of $\mathcal{J} _{CP}$ of $3.08 \times 10^{-5}$ can be 
reproduced.

\noindent In case of $n=7$ and $(m_{u}-m_{d}) \, \mathrm{j}=3$, e.g. $m_{d}=0$, $m_{u}=1$ and $\mathrm{j}=3$, 
we arrive at a value of $|\cos (\frac{3 \, \pi}{7})| \approx 0.2225$ for
$|V_{us}|$ which is only $2 \%$ smaller than the measured value $0.2272 ^{+ 0.0010} _{-0.0010}$
\cite{pdg}. In other words the size of the Cabibbo angle can be explained by group theoretical means
derived from a flavor symmetry \Groupname{D}{n} which is broken down to two \textit{distinct} subgroups, 
$\mbox{\Groupname{Z}{2}}= \langle \mathrm{B} \mathrm{A}^{m_{u}} \rangle$ and 
$\mbox{\Groupname{Z}{2}}= \langle \mathrm{B} \mathrm{A}^{m_{d}} \rangle$ with $m_{u} \neq m_{d}$,
in the up and down quark sector. This obviously requires that up-type and down-type quarks do not
couple to the same Higgs fields, i.e. a further separation mechanism is needed here. In the SM one can
simply assume an extra \Groupname{Z}{2} symmetry:
\[
Q_{i} \;\; \rightarrow Q_{i} \; , u_{i} ^{c} \;\; \rightarrow \;\; u_{i} ^{c} \;\; \mbox{and} \;\; 
d_{i} ^{c} \;\; \rightarrow \;\; - d_{i} ^{c}
\]
for the fermions and for the Higgs fields:
\[
\varphi_{u \, i} \;\; \rightarrow \;\; \varphi_{u \, i} \;\; \mbox{and} \;\; 
\varphi_{d \, i} \;\; \rightarrow \;\; - \varphi_{d \, i}
\]
where $\varphi_{u \, i}$ denote all Higgs fields which can couple to the up-type quarks, while
$\varphi_{d \, i}$ only couple to down-type quarks. In order to obtain two distinct VEV configurations
for the fields $\varphi_{u \, i}$ and $\varphi_{d \, i}$ which either break to $\mbox{\Groupname{Z}{2}}= 
\langle \mathrm{B} \mathrm{A}^{m_{u}} \rangle$ or
$\mbox{\Groupname{Z}{2}}= \langle \mathrm{B} \mathrm{A}^{m_{d}} \rangle$ we also need a separation 
mechanism in the Higgs sector, but this issue will not be discussed here.

\noindent \Eqref{eq:examplemds} and \Eqref{eq:examplemus} for the masses of the up-type and down-type quarks already show some possible source
of fine-tuning, since the smallness of the up quark mass $\mathrm{m}_{u}$ as well as 
of the strange quark mass $\mathrm{m} _{s}$ must
be due to the smallness of the expression $|D_{u}-E_{u} \, \mathrm{e} ^{-i \, \phi _{u} \, \mathrm{k}}|$
and $|D_{d}-E_{d} \, \mathrm{e} ^{i \, \phi_{d} \, \mathrm{k}}|$,
respectively. A further study of this issue is however beyond the scope of this paper and will be
treated in detail elsewhere \cite{comingsoonAH}.

\noindent There is a second way to generate such a mixing matrix: If we change the transformation properties of the 
left-handed conjugate fields to 
\[
d^{c}, u^{c} \sim ( \MoreRep{1}{$\mathrm{i}_{1}$}, \MoreRep{1}{$\mathrm{i}_{2}$},  \MoreRep{1}{$\mathrm{i}_{3}$}) 
\; ,
\]
we arrive at a mass matrix which is of the form of the transpose of the one shown in \Eqref{eq:onezero}.
The preserved subgroup is then also of the form $\mbox{\Groupname{Z}{2}}= \langle \mathrm{B} \mathrm{A}^{m} \rangle$, as explained in \Secref{sec:3sZ2}.
The number of free parameters is again five and therefore the chances to find a numerical solution 
which can reproduce all the masses and mixing parameters correctly apart from the predicted value 
for one
of the elements of $V_{CKM}$ are similar as in the case above.
However, in a complete model one or the other mass matrix may be more
easily accommodated with the suitable number of Higgs fields, VEV alignments, etc.. 
From the theoretical viewpoint the model in which the left- as well as the left-handed conjugate 
fields are partially unified into a doublet and a singlet representation of the discrete group 
might be favorable.

\noindent In all cases a thorough check whether one can accommodate all the quark masses and the rest of the
mixing angles is necessary, although we believe that this can be done due to the number of free parameters.
Furthermore a Higgs potential which allows for the breaking into the two different directions needs to 
be constructed and proven to be stable against corrections. And also an extension of this idea to 
include leptons is desirable. All these issues combined with a numerical study are
delegated to a future publication \cite{comingsoonAH}.

%%%%%%%%%%%%%%%%%%%%%%%%%%%%%%%%%%%%%%%%%%%%%
\section{Comparison with the literature}
%%%%%%%%%%%%%%%%%%%%%%%%%%%%%%%%%%%%%%%%%%%%%
\label{sec:litcompare}

\noindent The fact that preserved subgroups of a dihedral group can play an important role in obtaining certain mixing patterns is often not explicitly discussed. Several results from the literature can however easily be obtained from the group theoretical considerations we have performed in this paper. Realizing this can help to understand how and why flavor symmetry models work, as the following exemplary discussion shows.\\
In reference \cite{Grimus1} a $D_4$ flavor symmetry is used, with an additional $Z_2^{(aux)}$ for separating the three different Yukawa sectors: the charged lepton sector, the Dirac neutrino sector and the Majorana neutrino sector. The transformation properties of the fermions are chosen to give a two doublet structure in all three sectors. Then one breaks down to a different subgroup in each sector: $D_2$ is conserved in the charged lepton sector. $D_4$ is conserved in the Dirac neutrino sector, while it is broken down to $Z_2$ in the Majorana neutrino sector. As the authors use a different basis, the actual mass matrices differ from ours. The resulting physical mixing angles however are unaffected, i.e independent of the basis. Here we just describe the result for the mass matrices in their basis. The mass matrix for the charged leptons as well as the Dirac neutrino mass matrix are
diagonal, i.e. they do not contribute to the lepton mixing.
Two texture zeros in the right-handed neutrino mass matrix are obtained, which only appear because a gauge singlet transforming as a further non-trivial \Groupname{D}{4} singlet is absent from the model. They are removed when the seesaw mechanism is applied. The resulting mass matrix for the light neutrinos is $\mu\tau$ symmetric, i.e it has a maximal mixing angle $\theta_{23}^l$, a zero mixing angle $\theta_{13}^l$ and a free mixing angle $\theta_{12}^l$. The texture zeros in the right-handed neutrino mass matrix effect the relations between the eigenvalues of the light neutrino mass matrix, i.e. they enforce a normal hierarchy. Nevertheless, this model is not complete, as quarks are not discussed in this context. \cite{Grimus3} goes on to discuss soft breaking of the $D_4$ in the scalar potential. More precisely, the \Groupname{Z}{2}
subgroup is broken explicitly. Thus, 
the resulting doublet VEV no longer conserves $Z_2$, which can lead to sizable deviations from a maximal $\theta_{23}^l$, while leaving $\theta_{13}^l=0$.\\
A similar result is reproduced by a $D_3$ flavor symmetry in reference \cite{Grimus2}. In this case, the authors work in the same group basis we have used, so a comparison is more straightforward.
Nevertheless, they present their Dirac mass matrices in the basis $\bar{L} R$, while our
results are always given in the basis of $L L^{c}$.
 The $D_3 (\cong S_3)$ flavor symmetry is again joined by a $Z_2$ symmetry, which creates the same three sectors. $D_3$ is then broken to $Z_3$ in the charged lepton sector. In the Dirac neutrino sector the entire flavor symmetry is conserved, while it is broken down to a $Z_2$ in the Majorana neutrino sector - in this case all three equivalent subgroups $<\mathrm{B}>$, $\mathrm{<BA>}$ and $<\mathrm{B}\mathrm{A}^2>$ are mentioned. The charged lepton mass matrix and the Dirac mass matrix for the neutrinos are again diagonal, i.e. the non-trivial
mixing stems from the right-handed neutrino mass matrix. 
Here the maximal scalar field content is allowed, resulting in the same light neutrino mass matrix as for the $D_4$ model. The additional parameter which arises from having the full scalar field content does not affect the structure of the neutrino mass matrix obtained from the Type I seesaw formula, can however affect the neutrino mass hierarchy, as in this case the normal hierarchy is not necessarily predicted.\\
\noindent We finally mention some further examples, where nontrivial subgroups of dihedral flavor symmetries are conserved and explicitly mentioned. \cite{D3toZ2} and \cite{D5paper} mention the possibility of breaking their dihedral flavor symmetry down to a nontrivial subgroup. \cite{D3toZ2} uses a $D_3$ flavor symmetry, which is then broken down to $Z_2$ to achieve maximal atmospheric mixing and $\theta_{13}^l=0$. \cite{D5paper} uses a $D_5$ symmetry. If this is broken down to $Z_2$, maximal mixing can be achieved. However,
 without separating the Yukawa sectors or explicitly breaking the residual symmetry, the resulting maximal mixing angles in the neutrino and charged lepton sector will cancel in the leptonic mixing matrix.\\
The use of a dihedral symmetry is not only limited to the determination of the leptonic mixing angles. The breaking chains of \Secref{sec:breakingchains} can also be employed to explain the fermion mass hierarchy, by giving mass to different generations at different breaking scales. \cite{breakingchains1} breaks $D_3\sprime$ down to $Z_6$ and then further down, while \cite{breakingchains2} breaks a $D_6$ flavor symmetry down to $Z_2$ and then fully breaks it.\\

%%%%%%%%%%%%%%%%%%%%%%%%%%%%%%%%%
\section{Conclusions}
%%%%%%%%%%%%%%%%%%%%%%%%%%%%%%%%%
\label{sec:conclusions}

\noindent We have determined the possible Dirac and Majorana mass matrix structures that arise if a dihedral flavor group $D_n$ is broken down to a non-trivial subgroup by VEVs of some scalar fields. In order to exploit the non-abelian structure of the dihedral groups we only 
considered assignments where at least two of the left-handed or left-handed conjugate (SM)
fermions are unified into an irreducible two-dimensional representation, i.e.
\begin{equation}
\label{eq:threesinglet}
L \sim (\MoreRep1{$\bf i_1$},\MoreRep1{$\bf i_2$},\MoreRep1{$\bf i_3$}) \; , \; L^c \sim (\MoreRep1j,\MoreRep2k)
\end{equation}
or 
\begin{equation}
\label{eq:twodoublet}
L \sim (\MoreRep1i,\MoreRep2j) \; , \; L^c \sim (\MoreRep1l,\MoreRep2k)
\end{equation}
\noindent We constrained ourselves by the requirement that the mass matrices have to have a non-vanishing
determinant to reduce the number of cases so that a general discussion becomes possible. The number of different mass matrix structures we encounter is then limited.
We find the following mass matrix structures:
\be
\left(
\begin{matrix}
0 & C & -C e^{-i \phi \mathrm{k}} \\
A & D & D  e^{-i \phi \mathrm{k}}\\
B & E & E e^{-i \phi \mathrm{k}}
\end{matrix}
\right)
\ee
\noindent and
\be
\label{eq:mutaumat}
\left(
\begin{matrix}
A & C & C e^{-i \phi \mathrm{k}}\\
B & D & E \\
B e^{-i \phi \mathrm{j}} & E e^{i(\mathrm{k}-\mathrm{j})\phi} & D e^{-i (\mathrm{j}+\mathrm{k}) \phi}
\end{matrix}
\right)
\ee
Both of these matrix structures are associated with a conserved subgroup $Z_2$. The other mass matrices which can arise from a conserved subgroup are either diagonal, semi-diagonal or of block form.\\
\noindent All these mass matrices can result from a Dirac mass term and several ones also
from a Majorana mass.\\
\noindent For Majorana mass terms we additionally allow the assignment that all involved fermions
can transform as one-dimensional representations. Furthermore, we also studied the case of
an unbroken dihedral group.\\
\noindent It turns out that the double-valued groups \Doub{D}{n} do not lead to additional structures. We observe one slight difference in case of a Majorana mass term for \Doub{D}{n}
with $n$ odd, if the fermions
transform as three one-dimensional representations. Due to the odd one-dimensional representations a semi-diagonal mass matrix can arise. Although this structure cannot be deduced with
one-dimensional representations in the groups \Groupname{D}{n} ($n$ arbitrary)
and \Doub{D}{n} ($n$ even), this structure is not new, as it can be reproduced, if the
fermions which transform as the two odd one-dimensional representations are unified into a
two-dimensional representation.\\
\noindent Note that all these matrix structures base on the assumption that for each 
representation $\mu$ which is/contains a trivial representation under the residual subgroup of the 
flavor symmetry there is a Higgs field present in the theory which transforms as $\mu$. In this
way the structures of the mass matrices only depend on the choice of the original group,
the subgroup as well as the transformation properties of the fermions, 
but not on the choice of the
Higgs content. Therefore our results are less arbitrary. Reducing the Higgs content  
is also possible by simply setting the corresponding VEV to zero in the mass matrix. In some
cases this just reduces the number of parameters and therefore increases the predictive power
of the model \cite{Grimus1}, while in other cases this is essential in order to get the desired mixing 
pattern \cite{tprimepaper,AF2,AF1}.\\
\noindent In general one observes that for several subgroups the mass matrices (can) exhibit $2-3$-interchange symmetry, as for example \Eqref{eq:mutaumat}, if the group theoretical phase $\phi$ is set to zero.
This can be used to obtain maximal atmospheric mixing in the leptonic sector.\\
\noindent Interestingly, it is not only possible to predict such special mixing angles as $45 ^{\circ}$
or zero with a dihedral flavor symmetry, as we showed in the very first application of our general 
study. There we were able to describe one element of the CKM mixing matrix, namely $|V_{us}|$, which
corresponds to the Cabibbo angle, only in terms of group theoretical quantities, like the index $n$
of the dihedral group \Groupname{D}{n}, the index $\mathrm{j}$ of the representation under which 
the (left-handed) quarks transform as well as the indices of the preserved subgroups of 
\Groupname{D}{n}, $m_{u}$ and $m_{d}$:

\small
\begin{equation}
\label{eq:Vus}
|V_{us}|= \left| \cos \left(\frac{\pi \, (m_{u}-m_{d}) \, \mathrm{j}}{n} \right) \right|= |\cos (\frac{3 \,\pi}{7})|\approx 0.2225
\end{equation}
\normalsize
\noindent Since these dihedral symmetries have already been used several times in the literature,
it was interesting to see whether the successful models, like the ones by Grimus and Lavoura which
can predict $\mu\tau$ symmetry with the help of the flavor symmetry \Groupname{D}{4} $\times$ $Z_{2} ^{(aux)}$ \cite{Grimus1} or \Groupname{D}{3} $\times$ $Z_{2} ^{(aux)}$
\cite{Grimus2}, already make use of our described idea.
And indeed these two models are examples in which the three sectors represented by the charged 
lepton mass matrix, the Dirac neutrino mass matrix and the Majorana mass matrix for the right-handed
neutrinos preserve different non-trivial subgroups of the employed flavor symmetry. This 
mismatch of the different preserved subgroups then leads to the non-trivial mixing pattern.\\
\noindent The notation $L$ and $L^{c}$ of the fermions in our discussion 
always implied 
that they transform under the SM gauge group. 
However, all structures remain the same, if we
consider the MSSM as framework. Concerning a GUT like $SU(5)$ 
the fifteen fermions
of one generation are unified into $\Rep{10}$ and
$\bar{\Rep{5}}$. As a consequence, for example, the (Dirac) mass matrix 
for the up-type
quarks stems from the couplings of two $\Rep{10}$s. Therefore only the
two doublet structure with $\rm i = l$ and $\rm j = k$ applies here.
Similar changes occur, if we consider other (unified) gauge groups.
As our study includes all the possible mass matrix structures arising
from the two assignments \Eqref{eq:threesinglet} and \Eqref{eq:twodoublet}, 
choosing a gauge group other than the one of the SM under which the mass terms are 
invariant does not give rise to new cases.\\
\noindent In our study we always assumed that the involved Higgs fields (in case of the Dirac mass matrix) are copies of the SM Higgs $SU(2)_{L}$ doublet. This might cause problems
in a complete model, since it is well-known that multi-Higgs doublet models suffer from FCNCs
whose bounds usually demand that the Higgs fields have masses above $10 \, \TeV$. However, all
structures we have presented can also be achieved in models in which only the SM Higgs doublet
(or in the MSSM the Higgs fields $h_{u}$ and $h_{d}$) exists and the flavor symmetry is broken
by some gauge singlets, usually called flavons, mostly at a high energy scale. 
Then one has to deal in general with non-renormalizable operators 
consisting of two fermions, the usual Higgs field and appropriate
combinations of the flavon fields. According to their mass dimension 
the operators then have different suppression factors compared to the
multi-Higgs doublet case where all operators arise at the renormalizable
level. In an explicit realization one therefore has to check whether all
results which are produced in a multi-Higgs doublet model can be 
reproduced by the usage of non-renormalizable operators in a natural way.
For example, the top Yukawa coupling has to be large and should come
from a renormalizable term.\\ 
\noindent We mainly concentrated on the explanation of the mixing pattern of the fermions
by the flavor symmetry and its breaking to subgroups. However it is also interesting to
ask whether the hierarchy among the fermion masses, 
especially among the up-type quarks, could also be accommodated in this
framework, for example by some stepwise breaking of the
symmetry. Obviously, it is an advantage, if this could also be done.
If not, the Froggatt-Nielsen mechanism \cite{FN}
could be used as an explanation. For this an extra $U(1)$
flavor symmetry is needed. It is in general non-trivial to combine a certain assignment
of the fermions under the discrete flavor group with a certain set of $U(1)$ charges being
necessary for the mass hierarchy. As mentioned above
the usage of flavon fields is already a kind of implementation of the 
Froggatt-Nielsen mechanism, as their
existence generally leads to non-renormalizable operators,
 whose suppression according to the number of insertions of flavon fields might be the origin of 
the fermion mass hierarchy.\\ 

\noindent And finally we want to comment on the VEV alignment.
As we saw, there exists a specific VEV structure for scalar fields transforming under a two-dimensional representation which is necessary for conserving certain subgroups. This  VEV structure
\begin{equation}
\label{eq:VEVstructure}
<\psi_{\mathrm{j}}> 	\left(
		\ba{c}
		e^{\frac{-2 \pi i \mathrm{j} m}{n}} \\
		1
		\ea
		\right)
\end{equation}
can in fact arise naturally from the extremization of a scalar potential. For example: 
\begin{itemize}
\item a $D_3$-invariant two Higgs potential, with the two real scalar fields transforming as SM gauge singlets and as $\Rep2$ under $D_3$,
\item  the equivalent potential for $D_4$, with two real scalar fields transforming as SM gauge singlets and as $\Rep2$ under $D_4$,
\item  the most simple phenomenologically viable  $D_3$ invariant potential for $SU(2)_L$ doublet scalars, which is now a three Higgs potential with the three scalar fields transforming as $SU(2)_L$ doublets and as $\Rep2$ and $\MoreRep11$ under $D_3$ \footnote{Without the additional singlet the potential exhibits an accidental $U(1)$ symmetry, apart from the gauge symmetries \cite{comingsoon}.} \footnote{Here we preferred to use real representation matrices which then
have real eigenvectors belonging to the eigenvalue 1 such that we could solve
the extremization conditions under the assumption of real parameters and
also real VEVs. In this way all allowed VEV configurations correspond to the
specific VEV structure.}
\end{itemize}
\noindent One finds in all cases that the VEV structure given above is the only one that can extremize the potential while breaking the flavor symmetry, if one assumes real parameters in the potential and possibly also no correlations
among the parameters of the potential.\\
\noindent In contrast to this, one finds that for the simplest $SU(2)_L$ doublet potential in $D_4$ no such explicit statement can be made. Nevertheless, the above VEV structure is still allowed.\\
\noindent These statements make us confident that there is some way to 
get the announced VEV structure, even if more scalar fields exist in the model.\\ 
\noindent Regarding the success of models like \cite{AF2,tprimepaper} the 
crucial issue of the VEV alignment might be solved in a more efficient way 
in the context of supersymmetric
models with flavon fields. The main reason is that the (super-)potential itself is
significantly simplified by using gauge singlet scalars which transform non-trivially under $G_F$.
Furthermore, as these fields are gauge singlets which break the flavor symmetry at a high energy scale 
$\Lambda > m_{SUSY}$, the condition that the F-terms of some flavored gauge singlet fields vanish
determine the equations for the VEVs. These can then be often solved 
analytically.
\vspace{0.1in}\\
\textit{Note added:} After completing this work the paper \cite{Lam} by C.S. Lam  appeared in
which he also deals with the fact that non-trivial subgroups of some discrete flavor symmetry
can help to explain a certain mixing pattern and also very briefly mentions that the Cabibbo angle 
might be the result of some dihedral group.

%%%%%%%%%%%%%%%%%%%%%%%%%%%%%%%%%%%%%%%%%%%%%%%%%%
\begin{acknowledgments} 
\noindent A.B. acknowledges support from the Studienstiftung des Deutschen Volkes.
C.H. was supported by the ``Sonderforschungsbereich'' TR27.
\end{acknowledgments}
%%%%%%%%%%%%%%%%%%%%%%%%%%%%%%%%%%%%%%%%%%%%%%%%%%

\begin{appendix}
\footnotesize

%%%%%%%%%%%%%%%%%%%%%%%%%%%%%%%%%%%%%%%%%%%
\section{More Group Theory}
%%%%%%%%%%%%%%%%%%%%%%%%%%%%%%%%%%%%%%%%%%%
\label{app:grouptheory}

%%%%%%%%%%%%%%%%%%%%%%%%%%%%%%%%%%%%%%%%%%%
\subsection{Kronecker Products}
%%%%%%%%%%%%%%%%%%%%%%%%%%%%%%%%%%%%%%%%%%%
\label{app:kronprods}

\noindent The products of the one-dimensional representations of \Groupname{D}{n} are:
\begin{center}
\begin{tabular}{c|cccc}
$\times$&$\MoreRep{1}{1}$&$\MoreRep{1}{2}$&$\MoreRep{1}{3}$&$\MoreRep{1}{4}$\\
\hline
$\MoreRep{1}{1}$        &$\MoreRep{1}{1}$       &$\MoreRep{1}{2}$       &$\MoreRep{1}{3}$       &$\MoreRep{1}{4}$\\
$\MoreRep{1}{2}$        &$\MoreRep{1}{2}$       &$\MoreRep{1}{1}$       &$\MoreRep{1}{4}$       &$\MoreRep{1}{3}$\\
$\MoreRep{1}{3}$        &$\MoreRep{1}{3}$       &$\MoreRep{1}{4}$       &$\MoreRep{1}{1}$       &$\MoreRep{1}{2}$\\
$\MoreRep{1}{4}$        &$\MoreRep{1}{4}$       &$\MoreRep{1}{3}$       &$\MoreRep{1}{2}$       &$\MoreRep{1}{1}$\\
\end{tabular}
\end{center}
\noindent where the representations $\MoreRep{1}{3,4}$ only exist in groups \Groupname{D}{n} with an even index $n$. The products $\MoreRep{1}{i} \times \MoreRep{2}{j}$ transform as \[ \MoreRep{1}{1,2} \times \MoreRep{2}{j} = \MoreRep{2}{j} \] and for $n$ even there are also \[ \MoreRep{1}{3,4} \times \MoreRep{2}{j} = \MoreRep{2}{k} \;\;\; \mbox{with} \;\;\; \mathrm{k}= \frac{n}{2} - \mathrm{j} \].\\
If 4 is a divisor of $n$, the product of the representation $\MoreRep{2}{j}$ with  $\mathrm{j}= \frac{n}{4}$ with any one-dimensional representation of the group also transforms as $\MoreRep{2}{j}$.\\
\noindent The products $\MoreRep{2}{i} \times \MoreRep{2}{i}$ are of the form $\MoreRep{1}{1} + \MoreRep{1}{2} + \MoreRep{2}{j}$ with $\mathrm{j}= \min (2 \, \mathrm{i}, n - 2 \, \mathrm{i})$. In case that the group \Groupname{D}{n} has an index $n$ which is divisible by four one also finds the structure $\MoreRep{2}{i} \times \MoreRep{2}{i} = \sum \limits _{\mathrm{j}=1} ^{4} \MoreRep{1}{j}$ for $\mathrm{i}= \frac{n}{4}$. This shows that there is at most one representation in each group \Groupname{D}{n} with this property. The mixed products $\MoreRep{2}{i} \times \MoreRep{2}{j}$ can have two structures: $a.)$ $\MoreRep{2}{i} \times \MoreRep{2}{j} = \MoreRep{2}{k} + \MoreRep{2}{l}$ with $\rm k= |i-j|$ and $\mathrm{l}=\min (\mathrm{i}+\mathrm{j}, n-(\mathrm{i}+\mathrm{j}))$  and $b.)$ $\MoreRep{2}{i} \times \MoreRep{2}{j} = \MoreRep{1}{3} + \MoreRep{1}{4} + \MoreRep{2}{k}$ with $\rm k=|i-j|$ for $\mathrm{i}+\mathrm{j} = \frac{n}{2}$.\\
\noindent For \Doub{D}{n} with $n$ even the one-dimensional representations have the same product structure as for \Groupname{D}{n} while for $n$ being odd they are: 
\begin{center}
\begin{tabular}{c|cccc}
$\times$&$\MoreRep{1}{1}$&$\MoreRep{1}{2}$&$\MoreRep{1}{3}$&$\MoreRep{1}{4}$\\
\hline
$\MoreRep{1}{1}$        &$\MoreRep{1}{1}$       &$\MoreRep{1}{2}$       &$\MoreRep{1}{3}$       &$\MoreRep{1}{4}$\\
$\MoreRep{1}{2}$        &$\MoreRep{1}{2}$       &$\MoreRep{1}{1}$       &$\MoreRep{1}{4}$       &$\MoreRep{1}{3}$\\
$\MoreRep{1}{3}$        &$\MoreRep{1}{3}$       &$\MoreRep{1}{4}$       &$\MoreRep{1}{2}$       &$\MoreRep{1}{1}$\\
$\MoreRep{1}{4}$        &$\MoreRep{1}{4}$       &$\MoreRep{1}{3}$       &$\MoreRep{1}{1}$       &$\MoreRep{1}{2}$\\
\end{tabular}
\end{center}
\noindent due to the fact that the two one-dimensional representations $\MoreRep{1}{3}$ and $\MoreRep{1}{4}$ are complex conjugated to each other.
\noindent The rest of the formulae for the different product structures are the same as in the case of \Groupname{D}{2 \, n},i.e. in each formula above which is given for \Groupname{D}{n} one has to replace $n$ by $2 \, n$.

\noindent The Kronecker products can also be found in reference \cite{kronprods}.

%%%%%%%%%%%%%%%%%%%%%%%%%%%%%%%%%%%%%%%%%%%%%%%%%%%
\subsection{Clebsch Gordan Coefficients}
%%%%%%%%%%%%%%%%%%%%%%%%%%%%%%%%%%%%%%%%%%%%%%%%%%%
\label{app:CGsDn}

%%%%%%%%%%%%%%%%%%%%%%%%%%%%%%%%%%%%%%%%%
\mathversion{bold}
\subsubsection{for $D_n$}
\mathversion{normal}
%%%%%%%%%%%%%%%%%%%%%%%%%%%%%%%%%%%%%%%%%

\noindent For $\MoreRep{1}{i} \times \MoreRep{1}{j}= \MoreRep{1}{k}$ the Clebsch Gordan coefficient is trivially one. For $\MoreRep{1}{i} \times \MoreRep{2}{j}$ the Clebsch Gordan coefficients are:
\[ 
\text{ for $\rm i=1$:}\left( \begin{array}{c} \left( \begin{array}{cc} 1 & 0 \end{array} \right) \\
  \left( \begin{array}{cc} 0 & 1 \end{array} \right) \end{array} \right) \sim \MoreRep{2}{j} \text{ and for $\rm i=2$:} \left( 
\begin{array}{c}
\left( \begin{array}{cc} 1 & 0 \end{array} \right) \\
\left( \begin{array}{cc} 0 & -1 \end{array} \right) \end{array} \right) \sim \MoreRep{2}{j} 
\]
\noindent If the index $n$ of \Groupname{D}{n} is even, the group has two further one-dimensional representations $\MoreRep{1}{3,4}$ whose products with $\MoreRep{2}{j}$ are of the form:
\[
\text{for $\rm i=3$:} \left( 
\begin{array}{c} \left( \begin{array}{cc} 0 & 1 \end{array} \right) \\
  \left( \begin{array}{cc} 1 & 0 \end{array} \right) \end{array} \right) \sim \mathversion{bold} \MoreRep{2}{$\frac{n}{2}$\bf -j} \mathversion{normal} \text{ and for $\rm i=4$:} \left( 
\begin{array}{c} \left( \begin{array}{cc} 0 & 1 \end{array} \right) \\
  \left( \begin{array}{cc} -1 & 0 \end{array} \right) \end{array}
\right) \sim \mathversion{normal} \MoreRep{2}{$\frac{n}{2}$\bf -j} \mathversion{normal}
\]
\noindent For the products $\MoreRep{2}{i} \times \MoreRep{2}{i}$ the covariant combinations are:
\[
\left( \begin{array}{cc}
    0 & 1\\
    1 & 0
\end{array}
\right) \sim \MoreRep{1}{1} \;\;\; , \;\;\; \left( \begin{array}{cc}
    0 & 1\\
    -1 & 0
\end{array}
\right) \sim \MoreRep{1}{2}
\]
\noindent and
\[
\left( \begin{array}{c}
\left( \begin{array}{cc} 
    1 & 0 \\
    0 & 0
\end{array} \right)\\
\left( \begin{array}{cc} 
    0 & 0 \\
    0 & 1
\end{array} \right)
\end{array}
\right) \sim \MoreRep{2}{2i} \; \; \mbox{or} \; \;
\left( \begin{array}{c}
\left( \begin{array}{cc} 
    0 & 0 \\
    0 & 1
\end{array} \right)\\
\left( \begin{array}{cc} 
    1 & 0 \\
    0 & 0
\end{array} \right)
\end{array}
\right) \sim \mathversion{bold} \MoreRep{2}{$n$ \bf -2i} \mathversion{normal}
\]
\noindent If the index $n$ of \Groupname{D}{n} is even and $\mathrm{i}=\frac{n}{4}$ (4 has to be a divisor of $n$), there is a second possibility: $\MoreRep{2}{i} \times \MoreRep{2}{i} = \sum \limits _{\mathrm{j}=1} ^{4}
\MoreRep{1}{j}$. The Clebsch Gordan coefficients are
\[
\left( \begin{array}{cc}
    0 & 1\\
    1 & 0
\end{array}
\right) \sim \MoreRep{1}{1} \;\;\; , \;\;\; \left( \begin{array}{cc}
    0 & 1\\
    -1 & 0
\end{array}
\right) \sim \MoreRep{1}{2} \;\;\; ,
\]
\[
\left( \begin{array}{cc}
    1 & 0\\
    0 & 1
\end{array}
\right) \sim \MoreRep{1}{3} \;\;\; , \;\;\; \left( \begin{array}{cc}
    1 & 0\\
    0 & -1
\end{array}
\right) \sim \MoreRep{1}{4}
\]
\noindent For the products $\MoreRep{2}{i} \times \MoreRep{2}{j}$ with $\rm i \neq
j$ there are the two structures $\MoreRep{2}{i} \times \MoreRep{2}{j} = \MoreRep{2}{k} + \MoreRep{2}{l}$ with $\rm k= |i-j|$ and $\mathrm{l} = \min (\mathrm{i}+\mathrm{j}, n- (\mathrm{i}+\mathrm{j}))$ or $\MoreRep{2}{i} \times \MoreRep{2}{j} = \MoreRep{1}{3} + \MoreRep{1}{4} + \MoreRep{2}{k}$ with $\rm k=|i-j|$, if $\mathrm{i} +\mathrm{j} = \frac{n}{2}$ (obviously $n$ has to be even). The Clebsch Gordan coefficients for  $\MoreRep{2}{i} \times \MoreRep{2}{j}= \MoreRep{2}{k} + \MoreRep{2}{l}$  are:
\[
\left( \begin{array}{c}
\left( \begin{array}{cc} 
    0 & 1 \\
    0 & 0
\end{array} \right)\\
\left( \begin{array}{cc} 
    0 & 0 \\
    1 & 0
\end{array} \right)
\end{array}
\right) \sim \MoreRep{2}{i-j} \;\; \mbox{or} \; \;
\left( \begin{array}{c}
\left( \begin{array}{cc} 
    0 & 0 \\
    1 & 0
\end{array} \right)\\
\left( \begin{array}{cc} 
    0 & 1 \\
    0 & 0
\end{array} \right)
\end{array}
\right) \sim \MoreRep{2}{j-i} 
\]
\noindent and
\[
\left( \begin{array}{c}
\left( \begin{array}{cc} 
    1 & 0 \\
    0 & 0
\end{array} \right)\\
\left( \begin{array}{cc} 
    0 & 0 \\
    0 & 1
\end{array} \right)
\end{array}
\right) \sim \MoreRep{2}{i+j} \;\; \mbox{or} \; \; 
\left( \begin{array}{c}
\left( \begin{array}{cc} 
    0 & 0 \\
    0 & 1
\end{array} \right)\\
\left( \begin{array}{cc} 
    1 & 0 \\
    0 & 0
\end{array} \right)
\end{array}
\right) \sim \mathversion{bold} \MoreRep{2}{$n$ \bf-(i+j)} \mathversion{normal}
\]
\noindent For the structure $\MoreRep{2}{i} \times \MoreRep{2}{j} = \MoreRep{1}{3} + \MoreRep{1}{4} + \MoreRep{2}{k}$ with $\mathrm{i}+\mathrm{j} = \frac{n}{2}$ the Clebsch Gordan coefficients are
\[
\left( \begin{array}{cc}
    1 & 0 \\
    0 & 1
\end{array}
\right) \sim \MoreRep{1}{3} \;\;\; , \;\;\; \left( \begin{array}{cc}
    1 & 0 \\
    0 & -1
\end{array}
\right) \sim \MoreRep{1}{4}
\]
and
\[
\left( \begin{array}{c}
\left( \begin{array}{cc} 
    0 & 1 \\
    0 & 0
\end{array} \right)\\
\left( \begin{array}{cc} 
    0 & 0 \\
    1 & 0
\end{array} \right)
\end{array}
\right) \sim \MoreRep{2}{i-j} \;\; \mbox{or} \; \;
\left( \begin{array}{c}
\left( \begin{array}{cc} 
    0 & 0 \\
    1 & 0
\end{array} \right)\\
\left( \begin{array}{cc} 
    0 & 1 \\
    0 & 0
\end{array} \right)
\end{array}
\right) \sim \MoreRep{2}{j-i} 
\]

%%%%%%%%%%%%%%%%%%%%%%%%%%%%%%%%%%%%%%%%%%%%
\mathversion{bold}
\subsubsection{for \Doub{D}{n}}
\mathversion{normal}
%%%%%%%%%%%%%%%%%%%%%%%%%%%%%%%%%%%%%%%%%%%%
\label{app:CGsDnprime}

\noindent For $n$ even the Clebsch Gordan coefficients for the products $\MoreRep{1}{i} \times \MoreRep{2}{j} = \MoreRep{2}{k}$ are the same as in the case of \Groupname{D}{2 \, n}, i.e. for $\rm i=3,4$ the condition for $\rm k$ is $\mathrm{j} +\mathrm{k}=n$ instead of $\frac{n}{2}$.\\ 
If $n$ is odd, the same holds for $\rm j$ odd whereas for $\rm j$ even the Clebsch Gordan coefficients of the products $\MoreRep{1}{3} \times \MoreRep{2}{j}$ and $\MoreRep{1}{4} \times \MoreRep{2}{j}$ have to be interchanged.\\
The Clebsch Gordan coefficients for the products $\MoreRep{2}{i} \times \MoreRep{2}{i}$ are the same as for \Groupname{D}{2 \, n}, if $\rm i$ is even. Similarly, the ones of $\MoreRep{2}{i} \times \MoreRep{2}{j}$ with $\rm i \neq j$ are the same, if $\rm i$, $\rm j$ are both even or one is even and one is odd, if $n$ is even. For $n$ being odd the only difference is that in the case that the product is of the form $\MoreRep{2}{i} \times \MoreRep{2}{j} = \MoreRep{1}{3} + \MoreRep{1}{4} + \MoreRep{2}{k}$ the Clebsch Gordan coefficients for the covariant combination transforming as $\MoreRep{1}{3}$ and $\MoreRep{1}{4}$ are interchanged.\\
Concerning the structure of the products $\MoreRep{2}{i} \times \MoreRep{2}{i} = \MoreRep{1}{1} + \MoreRep{1}{2} + \MoreRep{2}{j}$ with $\mathrm{j}= \min (2 \, \mathrm{i},2 \, n-2 \, \mathrm{i})$ for $\rm i$ odd,  one finds the following:
\[
\left( \begin{array}{cc}
    0 & 1 \\
    -1 & 0
\end{array}
\right) \sim \MoreRep{1}{1} \;\;\; \left( \begin{array}{cc}
    0 & 1 \\
    1 & 0
\end{array}
\right) \sim \MoreRep{1}{2}
\]
\noindent and
\[
\left( \begin{array}{c}
\left( \begin{array}{cc} 
    1 & 0 \\
    0 & 0
\end{array} \right)\\
\left( \begin{array}{cc} 
    0 & 0 \\
    0 & -1
\end{array} \right)
\end{array}
\right) \sim \MoreRep{2}{2i} \;\; \mbox{or} \; \;
\left( \begin{array}{c}
\left( \begin{array}{cc} 
    0 & 0 \\
    0 & 1
\end{array} \right)\\
\left( \begin{array}{cc} 
    -1 & 0 \\
    0 & 0
\end{array} \right)
\end{array}
\right) \sim \mathversion{bold} \MoreRep{2}{2$n$ \bf -2i} \mathversion{normal}
\]
\noindent If $\mathrm{i}= \frac{n}{2}$ ($n$ even), then one has $\MoreRep{2}{i} \times \MoreRep{2}{i} = \sum \limits _{\mathrm{j}=1} ^{4} \MoreRep{1}{j}$. The Clebsch Gordan coefficients are
\[
\left( \begin{array}{cc}
    0 & 1 \\
    -1 & 0
\end{array}
\right) \sim \MoreRep{1}{1} \;\;\; , \;\;\; \left( \begin{array}{cc}
    0 & 1 \\
    1 & 0
\end{array}
\right) \sim \MoreRep{1}{2} \;\;\; ,
\]

\[
 \left( \begin{array}{cc}
    1 & 0 \\
    0 & -1
\end{array}
\right) \sim \MoreRep{1}{3} \;\;\; , \;\;\; \left( \begin{array}{cc}
    1 & 0 \\
    0 & 1
\end{array}
\right) \sim \MoreRep{1}{4}
\]
\noindent $\MoreRep{2}{i} \times \MoreRep{2}{j}$ for $\rm i,j$ being odd is either $\MoreRep{2}{k} + \MoreRep{2}{l}$ with $\rm k= |i-j|$ and $\mathrm{l}= 
\min (\mathrm{i}+\mathrm{j}, 2 \, n- (\mathrm{i}+\mathrm{j}))$  or $\MoreRep{1}{3} + \MoreRep{1}{4} + \MoreRep{2}{k}$ with $\rm k= |i-j|$, if $\mathrm{i}+\mathrm{j}=n$. The Clebsch Gordan coefficients in the first case are:
\[
\left( \begin{array}{c}
\left( \begin{array}{cc} 
    0 & 1 \\
    0 & 0
\end{array} \right)\\
\left( \begin{array}{cc} 
    0 & 0 \\
    -1 & 0
\end{array} \right)
\end{array}
\right) \sim \MoreRep{2}{i-j} \;\; \mbox{or} \; \;
\left( \begin{array}{c}
\left( \begin{array}{cc} 
    0 & 0 \\
    1 & 0
\end{array} \right)\\
\left( \begin{array}{cc} 
    0 & -1 \\
    0 & 0
\end{array} \right)
\end{array}
\right) \sim \MoreRep{2}{j-i} 
\]
\noindent and
\[
\left( \begin{array}{c}
\left( \begin{array}{cc} 
    1 & 0 \\
    0 & 0
\end{array} \right)\\
\left( \begin{array}{cc} 
    0 & 0 \\
    0 & -1
\end{array} \right)
\end{array}
\right) \sim \MoreRep{2}{i+j} \;\; \mbox{or} \; \;
\left( \begin{array}{c}
\left( \begin{array}{cc} 
    0 & 0 \\
    0 & 1
\end{array} \right)\\
\left( \begin{array}{cc} 
    -1 & 0 \\
    0 & 0
\end{array} \right)
\end{array}
\right) \sim \mathversion{bold} \MoreRep{2}{$2n$ \bf -(i+j)} \mathversion{normal}
\]
\noindent In the second one the Clebsch Gordan coefficients are:
\[
\left( \begin{array}{cc}
    1 & 0 \\
    0 & -1
\end{array}
\right) \sim \MoreRep{1}{3} \;\;\; , \;\;\; \left( \begin{array}{cc}
    1 & 0 \\
    0 & 1
\end{array}
\right) \sim \MoreRep{1}{4}
\]
\noindent and
\[
\left( \begin{array}{c}
\left( \begin{array}{cc} 
    0 & 1 \\
    0 & 0
\end{array} \right)\\
\left( \begin{array}{cc} 
    0 & 0 \\
    -1 & 0
\end{array} \right)
\end{array}
\right) \sim \MoreRep{2}{i-j} \;\; \mbox{or} \; \;
\left( \begin{array}{c}
\left( \begin{array}{cc} 
    0 & 0 \\
    1 & 0
\end{array} \right)\\
\left( \begin{array}{cc} 
    0 & -1 \\
    0 & 0
\end{array} \right)
\end{array}
\right) \sim \MoreRep{2}{j-i} 
\]

%%%%%%%%%%%%%%%%%%%%%%%%%%%%%%%%%%%%%%%%%%%%%%%%%
\section{Decomposition under subgroups}
%%%%%%%%%%%%%%%%%%%%%%%%%%%%%%%%%%%%%%%%%%%%%%%%%
\label{app:decomposition}

\noindent The decomposition of representations of $G_F$ under its subgroups is given in \Tabref{abeld}, \Tabref{nonabeld}, \Tabref{abelq} and \Tabref{nonabelq}. We have used the following non-standard convention for the representation of $Z_n$: The representation $\MoreRep1k$ transforms as $e^{\frac{2 \pi i}{n} (\mathrm{k})}$, so that $\MoreRep10$ denotes the trivial representation and \mathversion{bold} $\MoreRep1{($n$\bf + k)} = \MoreRep1k $ \mathversion{normal}.\\
We will denote the components of the two-dimensional representation $\MoreRep2k$ by
\begin{displaymath}
\MoreRep2k \sim
\left(
\ba{c}
		a_{\mathrm{k}} \\
		b_{\mathrm{k}}
		\ea
		\right)
\end{displaymath}
\noindent For two-dimensional representations under dihedral subgroups, we find in the tables the general identification $\MoreRep2k \sim \MoreRep2k$. However, dihedral subgroups will have less two-dimensional representations than the original group, so we need to make the following identifications, if the dihedral subgroup has no representation $\MoreRep2k$:\\
\noindent In $D_{\mathrm{j}}$, $\mathrm{j}$ even:
\be
(e^{\frac{ \pi i m \mathrm{j}}{n}} a_{\frac{\mathrm{j}}{2}} + b_{\frac{\mathrm{j}}{2}}) \sim \MoreRep13 \; , \; (-e^{\frac{ \pi i m \mathrm{j}}{n}} a_{\frac{\mathrm{j}}{2}} + b_{\frac{\mathrm{j}}{2}}) \sim \MoreRep14
\ee
\be
(e^{\frac{2 \pi i m \mathrm{j}}{n}} a_{\mathrm{j}} + b_{\mathrm{j}}) \sim \MoreRep11 \; , \; (-e^{\frac{2 \pi i m \mathrm{j}}{n}} a_{\mathrm{j}} + b_{\mathrm{j}}) \sim \MoreRep12
\ee
\be
\left(
\ba{c}
		e^{\frac{2 \pi i m \mathrm{k}}{n}} a_{\mathrm{k}} \\
		b_{\mathrm{k}}
		\ea
		\right) \sim \MoreRep2k \, (if \, \mathrm{k}<\frac{\mathrm{j}}{2})
\ee
\be
\left(
\ba{c}
		b_{\mathrm{k}} \\
		e^{\frac{2 \pi i m \mathrm{k}}{n}} a_{\mathrm{k}}
		\ea
		\right) \sim \MoreRep2{j-k}\, (if \, \mathrm{j}>\mathrm{k}>\frac{\mathrm{j}}{2})
\ee
\noindent In $D_{\mathrm{j}}$, $\mathrm{j}$ odd:\\
\be
(e^{\frac{2 \pi i m \mathrm{j}}{n}} a_{\mathrm{j}} + b_{\mathrm{j}}) \sim \MoreRep11 \; , \; (-e^{\frac{2 \pi i m {\mathrm{j}}}{n}} a_{\mathrm{j}} + b_{\mathrm{j}}) \sim \MoreRep12
\ee
\be
\left(
\ba{c}
		e^{\frac{2 \pi i m \mathrm{k}}{n}} a_{\mathrm{k}} \\
		b_{\mathrm{k}}
		\ea
		\right) \sim \MoreRep2k \, (if \, \mathrm{k} \leq \frac{{\mathrm{j}}-1}{2})
\ee
\be
\left(
\ba{c}
		b_{\mathrm{k}} \\
		e^{\frac{2 \pi i m \mathrm{k}}{n}} a_{\mathrm{k}}
		\ea
		\right) \sim \MoreRep2{j-k}\, (if \, {\mathrm{j}}>\mathrm{k}>\frac{{\mathrm{j}}-1}{2})
\ee
\noindent If $ \mathrm{k}>{\mathrm{j}}$, $\MoreRep2k \sim \MoreRep2{(k\, mod j)}$ . If ${\mathrm{j}}$ divides $\mathrm{k}$, then $\MoreRep2k$ transforms just as $\MoreRep2j$, i.e. as $\MoreRep11 + \MoreRep12$. For $D_{\mathrm{j}}\sprime$ one can make the same identifications as for $D_{\mathrm{j}}$, j even, if one makes the substitutions j $\rightarrow$ 2j and $n$ $\rightarrow$ $2n$. For j dividing k, $\MoreRep2{2k}$ then transforms as $\MoreRep2{2j}$, i.e. as $\MoreRep11 + \MoreRep12$.\\
\begin{table*}
\begin{center}
\begin{tabular}{ccc|c}
$D_n $&$ \rightarrow $&$ Subgroup $&$ VEV\, allowed?$
\\
\hline
\hline
& & $Z_n = < \mathrm{A}>$ & 
\\
\hline
${\bf \MoreRep11} $&$ \rightarrow $&$ \MoreRep10 $& $yes$
\\
${\bf \MoreRep12} $&$ \rightarrow $&$ \MoreRep10 $& $yes$
\\
${\bf \MoreRep13} $&$ \rightarrow $&\mathversion{bold} $ \MoreRep1{$\frac{n}{2}$} $ \mathversion{normal}& 
\\
${\bf \MoreRep14} $&$ \rightarrow $&$ \mathversion{bold} \MoreRep1{$\frac{n}{2}$} $ \mathversion{normal}&
\\
${\bf \MoreRep2k} $&$ \rightarrow $&$ a_{\mathrm{k}} \sim \MoreRep1{k} \, , \, b_{\mathrm{k}} \sim$\mathversion{bold} $\MoreRep1{($n$\bf -k)} $ \mathversion{normal}&
\\
\hline
& & $Z_{\mathrm{j}} = < \mathrm{A}^{\frac{n}{\mathrm{j}}}> $& 
\\
\hline
${\bf \MoreRep11} $&$ \rightarrow $&$ \MoreRep10 $& $yes$
\\
${\bf \MoreRep12} $&$ \rightarrow $&$ \MoreRep10 $& $yes$
\\
${\bf \MoreRep13} $&$ \rightarrow $& \mathversion{bold} $ \MoreRep1{$\frac{n}{2}$} $ \mathversion{normal} &$ if\,  \frac{n}{\mathrm{j}} \, even$ 
\\
${\bf \MoreRep14} $&$ \rightarrow $& \mathversion{bold} $ \MoreRep1{$\frac{n}{2}$} $ \mathversion{normal} &$ if\,  \frac{n}{\mathrm{j}} \, even$
\\
${\bf \MoreRep2k} $&$ \rightarrow $&$  a_{\mathrm{k}} \sim \, \MoreRep1{k} \, , \, b_{\mathrm{k}} \sim \MoreRep1{j-k} $&$ if\,  \mathrm{j} \mid \mathrm{k}$
\\
\hline
& & $Z_2 = < \mathrm{B}\mathrm{A}^m> $& 
\\
\hline
${\bf \MoreRep11} $&$ \rightarrow $&$ \MoreRep10 $&$ yes$
\\
${\bf \MoreRep12} $&$ \rightarrow $&$ \MoreRep11 $& 
\\
${\bf \MoreRep13} $&$ \rightarrow $&$ \mathversion{bold} \MoreRep1{$m$} $ \mathversion{normal} &$ if\,  m\,  even $
\\
${\bf \MoreRep14} $&$ \rightarrow $&\mathversion{bold} $ \MoreRep1{$m$\bf +1} $ \mathversion{normal} &$ if\,  m\,  odd$
\\
${\bf \MoreRep2k} $&$ \rightarrow $&$ 
(e^{\frac{2 \pi i m \mathrm{k}}{n}} a_{\mathrm{k}} + b_{\mathrm{k}}) \sim \MoreRep10 \, , \,
(- e^{\frac{2 \pi i m \mathrm{k}}{n}} a_{\mathrm{k}} + b_{\mathrm{k}}) \sim \MoreRep11 $&$ 	\left(
		\ba{c}
		e^{\frac{-2 \pi i \mathrm{k} m}{n}} \\
		1
		\ea
		\right)$
\\
\hline
& & $Z_{\frac{n}{2}} = < \mathrm{A}^2> $& 
\\
\hline
${\bf \MoreRep11} $&$ \rightarrow $&$ \MoreRep10 $&$ yes$
\\
${\bf \MoreRep12} $&$ \rightarrow $&$ \MoreRep10 $&$ yes$
\\
${\bf \MoreRep13} $&$ \rightarrow $&$ \MoreRep10 $&$ yes $
\\
${\bf \MoreRep14} $&$ \rightarrow $&$ \MoreRep10 $&$ yes$
\\
${\bf \MoreRep2k} $&$ \rightarrow $&$ a_{\mathrm{k}} \sim \MoreRep1k \, , \, b_{\mathrm{k}} \sim$ \mathversion{bold} $\MoreRep1{($\frac{n}{2}$\bf -k)}$ \mathversion{normal}&
\end{tabular}
\caption{Transformation properties of the representations of a dihedral group under its abelian subgroups, as determined in \Secref{sec:subs}. The rightmost column shows whether a representation has a component, which transforms trivially under the subgroup, i.e. if a scalar field transforming under this representation can acquire a VEV, while conserving this subgroup. If only a specific VEV structure is allowed, it is given explicitly, otherwise an arbitrary VEV is allowed.}
\label{abeld}
\end{center}
\end{table*}
\begin{table*}
\footnotesize
\begin{center}
\begin{tabular}{ccc|c}
$D_n $&$ \rightarrow $&$ Subgroup $&$ VEV\,  allowed?$
\\
\hline
\hline
& & $D_{\frac{n}{2}} = < \mathrm{A}^2,\mathrm{B}> $& 
\\
\hline
${\bf \MoreRep11} $&$ \rightarrow $&$ {\bf \MoreRep11} $&$ yes$
\\
${\bf \MoreRep12} $&$ \rightarrow $&$ {\bf \MoreRep12} $& 
\\
${\bf \MoreRep13} $&$ \rightarrow $&$ {\bf \MoreRep11} $&$ yes$
\\
${\bf \MoreRep14} $&$ \rightarrow $&$ {\bf \MoreRep12} $&
\\
${\bf \MoreRep2k} $&$ \rightarrow $&$ {\bf \MoreRep2k} $&
\\
\hline
& & $D_{\frac{n}{2}} = < \mathrm{A}^2,\mathrm{B}\mathrm{A}> $& 
\\
\hline
${\bf \MoreRep11} $&$ \rightarrow $&$ {\bf \MoreRep11} $&$ yes$
\\
${\bf \MoreRep12} $&$ \rightarrow $&$ {\bf \MoreRep12} $& 
\\
${\bf \MoreRep13} $&$ \rightarrow $&$ {\bf \MoreRep12} $& 
\\
${\bf \MoreRep14} $&$ \rightarrow $&$ {\bf \MoreRep11} $&$ yes$
\\
${\bf \MoreRep2k} $&$ \rightarrow $&$ {\bf \MoreRep2k} $&
\\
\hline
& & $D_{\mathrm{j}} = < \mathrm{A}^{\frac{n}{\mathrm{j}}},\mathrm{B}\mathrm{A}^m> $& 
\\
\hline
${\bf \MoreRep11} $&$ \rightarrow $&$ {\bf \MoreRep11} $&$ yes$
\\
${\bf \MoreRep12} $&$ \rightarrow $&$ {\bf \MoreRep12} $& 
\\
${\bf \MoreRep13} $&$ \rightarrow $&$ {\bf \MoreRep11} (\frac{n}{\mathrm{j}}\, even,\, m\, even) $&$ yes$
\\
& &  ${\bf \MoreRep12} (\frac{n}{\mathrm{j}}\, even,\, m\, odd) $& 
\\
& &  ${\bf \MoreRep13} (\frac{n}{\mathrm{j}}\, odd,\, m\, even)$ & 
\\
& &  ${\bf \MoreRep14} (\frac{n}{\mathrm{j}}\, odd,\, m\, odd) $& 
\\
${\bf \MoreRep14} $&$ \rightarrow $&$ {\bf \MoreRep11} (\frac{n}{\mathrm{j}}\, even,\, m\, odd) $&$ yes$
\\
& &  ${\bf \MoreRep12} (\frac{n}{\mathrm{j}}\, even,\, m\, even) $& 
\\
& &  ${\bf \MoreRep13} (\frac{n}{\mathrm{j}}\, odd,\, m\, odd) $& 
\\
& &  ${\bf \MoreRep14} (\frac{n}{\mathrm{j}}\, odd,\, m\, even) $& 
\\
${\bf \MoreRep2k} $&$ \rightarrow $&$ {\bf \MoreRep2k} $&$ 	\left(
		\ba{c}
		e^{\frac{- 2 \pi i \mathrm{k} m}{n}} \\
		1
		\ea
		\right) (if \, \mathrm{j} \mid \mathrm{k})$

\end{tabular}
\caption{Transformation properties of the representations of a dihedral group under its non-abelian subgroups. For further details see caption of \Tabref{abeld}.}
\label{nonabeld}
\end{center}
\end{table*}
\begin{table*}
\footnotesize
\begin{center}
\begin{tabular}{ccc|c}
$D_n\sprime $&$ \rightarrow $&$ Subgroup $&$ VEV\, allowed?$
\\
\hline
\hline
& &$ Z_{2n} = < \mathrm{A}> $& 
\\
\hline
${\bf \MoreRep11} $&$ \rightarrow $&$ \MoreRep10 $&$ yes$
\\
${\bf \MoreRep12} $&$ \rightarrow $&$ \MoreRep10 $&$ yes$
\\
${\bf \MoreRep13} $&$ \rightarrow $& \mathversion{bold} $\MoreRep1{$n$} $ \mathversion{normal}& 
\\
${\bf \MoreRep14} $&$ \rightarrow $&\mathversion{bold} $\MoreRep1{$n$} $ \mathversion{normal}&
\\
${\bf \MoreRep2k} $&$ \rightarrow $&$ a_{\mathrm{k}} \sim \MoreRep1k \, , \, b_{\mathrm{k}} \sim$ \mathversion{bold} $\MoreRep1{(2$n$\bf -k)} $ \mathversion{normal}&
\\
\hline
& &$ Z_{n} = < \mathrm{A}^2> $& 
\\
\hline
${\bf \MoreRep11} $&$ \rightarrow $&$ \MoreRep10 $&$ yes$
\\
${\bf \MoreRep12} $&$ \rightarrow $&$ \MoreRep10 $&$ yes$
\\
${\bf \MoreRep13} $&$ \rightarrow $&$ \MoreRep10 $&$ yes $
\\
${\bf \MoreRep14} $&$ \rightarrow $&$ \MoreRep10 $&$ yes$
\\
${\bf \MoreRep2k} $&$ \rightarrow $&$ a_{\mathrm{k}} \sim \MoreRep1k \, , \, b_{\mathrm{k}} \sim$ \mathversion{bold} $\MoreRep1{($n$\bf -k)} $ \mathversion{normal}&
\\
\hline
& &$ Z_{\mathrm{j}} = < \mathrm{A}^{\frac{2n}{\mathrm{j}}}> $& 
\\
\hline
${\bf \MoreRep11} $&$ \rightarrow $&$ \MoreRep10 $&$ yes$
\\
${\bf \MoreRep12} $&$ \rightarrow $&$ \MoreRep10 $&$ yes$
\\
${\bf \MoreRep13} $&$ \rightarrow $&\mathversion{bold} $\MoreRep1{$n$} $ \mathversion{normal}&$ if\,  \frac{2n}{\mathrm{j}} \, even $
\\
${\bf \MoreRep14} $&$ \rightarrow $&\mathversion{bold} $\MoreRep1{$n$} $ \mathversion{normal}&$ if\,  \frac{2n}{\mathrm{j}} \, even$
\\
${\bf \MoreRep2k} $&$ \rightarrow $&$ a_{\mathrm{k}} \sim \MoreRep1k \, , \, b_{\mathrm{k}} \sim \MoreRep1{j-k} $&$ if\,  \mathrm{j} \mid \mathrm{k} \, (\mathrm{k} \, can \, be \, odd)$
\\
\hline
& &$ Z_4 = < \mathrm{B}\mathrm{A}^m> $& 
\\
\hline
${\bf \MoreRep11} $&$ \rightarrow $&$ \MoreRep10 $&$ yes$
\\
${\bf \MoreRep12} $&$ \rightarrow $&$ \MoreRep12 $& 
\\
${\bf \MoreRep13} $&$ \rightarrow $&$ {\bf \MoreRep10} (n\, even,\, m\, even) $&$ yes$
\\
& &$  {\bf \MoreRep11} (n\, odd,\, m\, odd) $& 
\\
& &$  {\bf \MoreRep12} (n\, even,\, m\, odd) $& 
\\
& &$  {\bf \MoreRep13} (n\, odd,\, m\, even) $& 
\\
${\bf \MoreRep14} $&$ \rightarrow $&$ {\bf \MoreRep12} (n\, even,\, m\, even) $& 
\\
& &$  {\bf \MoreRep13} (n\, odd,\, m\, odd) $& 
\\
& &$  {\bf \MoreRep10} (n\, even,\, m\, odd) $&$ yes$
\\
& &$  {\bf \MoreRep11} (n\, odd,\, m\, even) $& 
\\
${\bf \MoreRep2k} $&$ \rightarrow $&$
(e^{\frac{\pi i m \mathrm{k}}{n}} a_{\mathrm{k}} + b_{\mathrm{k}}) \sim \MoreRep{1}{{0,1}} \, , \,
(-e^{\frac{\pi i m \mathrm{k}}{n}} a_{\mathrm{k}} + b_{\mathrm{k}}) \sim \MoreRep{1}{{2,3}} $&$ 	\left(
		\ba{c}
		e^{\frac{- \pi i \mathrm{k} m}{n}} \\
		1
		\ea
		\right)  \, (if\, \mathrm{k}\, even)$
\\
\hline
& &$ Z_2 = < \mathrm{A}^n> $& 
\\
\hline
${\bf \MoreRep11} $&$ \rightarrow $&$ \MoreRep10 $&$ yes$
\\
${\bf \MoreRep12} $&$ \rightarrow $&$ \MoreRep10 $&$ yes$
\\
${\bf \MoreRep13} $&$ \rightarrow $&\mathversion{bold} $\MoreRep1{$n$} $ \mathversion{normal}&$ if\, n\, even $
\\
${\bf \MoreRep14} $&$ \rightarrow $&\mathversion{bold} $\MoreRep1{$n$} $ \mathversion{normal}&$ if\, n\, even$
\\
${\bf \MoreRep2k} $&$ \rightarrow $&$ a_{\mathrm{k}} \sim \MoreRep1k \, , \, b_{\mathrm{k}} \sim  \MoreRep1k $&$ if\, \mathrm{k}\, even$

\end{tabular}
\caption{Transformation properties of the representations of a double-valued dihedral group under its abelian subgroups. For the decomposition of the
two-dimensional \Doub{D}{n} representations under its subgroup \Groupname{Z}{4}
one has to mention that $\MoreRep{2}{k}$ for $\mathrm{k}$ even splits up
into $\MoreRep{1}{0}$ and $\MoreRep{1}{2}$ under \Groupname{Z}{4}, while for
$\mathrm{k}$ being odd the representations are $\MoreRep{1}{1}$ and $\MoreRep{1}{3}$. For further details see caption of \Tabref{abeld}.}
\label{abelq}
\end{center}
\end{table*}
\begin{table*}
\footnotesize
\begin{center}
\begin{tabular}{ccc|c}
$D_n\sprime $&$ \rightarrow $&$ Subgroup $&$ VEV\,  allowed?$
\\
\hline
\hline
& &$ D_{\frac{n}{2}}\sprime = < \mathrm{A}^2,\mathrm{B}> $& 
\\
\hline
${\bf \MoreRep11} $&$ \rightarrow $&$ {\bf \MoreRep11} $&$ yes$
\\
${\bf \MoreRep12} $&$ \rightarrow $&$ {\bf \MoreRep12} $& 
\\
${\bf \MoreRep13} $&$ \rightarrow $&$ {\bf \MoreRep11} $&$ yes$
\\
${\bf \MoreRep14} $&$ \rightarrow $&$ {\bf \MoreRep12} $&
\\
${\bf \MoreRep2k} $&$ \rightarrow $&$ {\bf \MoreRep2k} $&
\\
\hline
& &$ D_{\frac{n}{2}}\sprime = < \mathrm{A}^2,\mathrm{B}\mathrm{A}> $& 
\\
\hline
${\bf \MoreRep11} $&$ \rightarrow $&$ {\bf \MoreRep11} $&$ yes$
\\
${\bf \MoreRep12} $&$ \rightarrow $&$ {\bf \MoreRep12} $& 
\\
${\bf \MoreRep13} $&$ \rightarrow $&$ {\bf \MoreRep12} $& 
\\
${\bf \MoreRep14} $&$ \rightarrow $&$ {\bf \MoreRep11} $&$ yes$
\\
${\bf \MoreRep2k} $&$ \rightarrow $&$ {\bf \MoreRep2k} $&
\\
\hline
& &$ D_{\frac{\mathrm{j}}{2}}\sprime = < \mathrm{A}^{\frac{2n}{\mathrm{j}}},\mathrm{B}\mathrm{A}^m> $& 
\\
\hline
${\bf \MoreRep11} $&$ \rightarrow $&$ {\bf \MoreRep11} $&$ yes$
\\
${\bf \MoreRep12} $&$ \rightarrow $&$ {\bf \MoreRep12} $& 
\\
${\bf \MoreRep13} $&$ \rightarrow $&$ {\bf \MoreRep11} (\frac{2n}{\mathrm{j}}\, even,\, m\, even) $&$ yes$
\\
& &$  {\bf \MoreRep12} (\frac{2n}{\mathrm{j}}\, even,\, m\, odd) $& 
\\
& &$  {\bf \MoreRep13} (\frac{2n}{\mathrm{j}}\, odd,\, m\, even) $& 
\\
& &$  {\bf \MoreRep14} (\frac{2n}{\mathrm{j}}\, odd,\, m\, odd) $& 
\\
${\bf \MoreRep14} $&$ \rightarrow $&$ {\bf \MoreRep11} (\frac{2n}{\mathrm{j}}\, even,\, m\, odd) $&$ yes$
\\
& &$  {\bf \MoreRep12} (\frac{2n}{\mathrm{j}}\, even,\, m\, even) $& 
\\
& &$  {\bf \MoreRep13} (\frac{2n}{\mathrm{j}}\, odd,\, m\, odd) $& 
\\
& &$  {\bf \MoreRep14} (\frac{2n}{\mathrm{j}}\, odd,\, m\, even) $& 
\\
${\bf \MoreRep2k} $&$ \rightarrow $&$ {\bf \MoreRep2k} $&$ 	\left(
		\ba{c}
		e^{\frac{- \pi i \mathrm{k} m}{n}} \\
		1
		\ea
		\right) (if\, \mathrm{j} \mid \mathrm{k}) \, , \, \mathrm{k} \, even$

\end{tabular}
\caption{Transformation properties of the representations of a double-valued dihedral group under its non-abelian subgroups. For further details see caption of \Tabref{abeld}.}
\label{nonabelq}
\end{center}
\end{table*}

%%%%%%%%%%%%%%%%%%%%%%%%%%%%%%%%%%%%%%%%%%%%%%%%%%%%%%%%
\mathversion{bold}
\section{Breaking Chains for $D_n\sprime$}
\mathversion{normal}
%%%%%%%%%%%%%%%%%%%%%%%%%%%%%%%%%%%%%%%%%%%%%%%%%%%%%%%%%
\label{app:breakingchains}

\noindent We give the possible breaking sequences for a double-valued dihedral group $D_n\sprime$. The breaking sequences for $D_n$ along with a discussion of the conventions used is given in \Secref{sec:breakingchains}.
\begin{flushleft}
\begin{tabular}{cccccccccc}
$D_{n}\sprime $&$ \stackrel {< \MoreRep12 >} {\longrightarrow} $&$ Z_{2n} $&$ 
      \stackrel {< \MoreRep13 >} {\longrightarrow} $&$ Z_n $&$         
      \stackrel {< \MoreRep2j >} {\longrightarrow} $&$ Z_{\mathrm{j}} $& 
      & 
&( $\mathrm{j} \mid n$)
\end{tabular}\\
\begin{tabular}{cccccccccc}
$D_{n}\sprime $&$ \stackrel {< \MoreRep12 >} {\longrightarrow} $&$ Z_{2n} $&$ 
      \stackrel {< \MoreRep2j >} {\longrightarrow} $&$ Z_{\mathrm{j}} $&         
       & &
       & 
\end{tabular}\\
\begin{tabular}{cccccccccc}
$D_{n}\sprime $&$ \stackrel {< \MoreRep13 >} {\longrightarrow} $&$ Z_n $&$ 
      \stackrel {< \MoreRep2j >} {\longrightarrow} $&$ Z_{\mathrm{j}} $&         
      & & 
      & 
&($\mathrm{j} \mid n; \, n\, odd$)
\end{tabular}\\
\begin{tabular}{cccccccccc}
$D_{n}\sprime $&$ \stackrel {< \MoreRep13 >} {\longrightarrow} $&$ D_{\frac{n}{2}}\sprime $&$ 
      \stackrel {< \MoreRep12 >} {\longrightarrow} $&$ Z_n $&$         
      \stackrel {< \MoreRep2j >} {\longrightarrow} $&$ Z_{\mathrm{j}} $& 
      & 
&($\mathrm{j} \mid n; \, n \, even$ ) 
\end{tabular}\\
\begin{tabular}{cccccccccc}
$D_{n}\sprime $&$ \stackrel {< \MoreRep13 >} {\longrightarrow} $&$ D_{\frac{n}{2}}\sprime $&$ 
      \stackrel {< \MoreRep14 >} {\longrightarrow} $&$ Z_n $&$         
      \stackrel {< \MoreRep2j >} {\longrightarrow} $&$ Z_{\mathrm{j}} $&
      & 
&($ \mathrm{j} \mid n; \, n \, even$ ) 
\end{tabular}\\
\begin{tabular}{cccccccccc}
$D_{n}\sprime $&$ \stackrel {< \MoreRep13 >} {\longrightarrow} $&$ D_{\frac{n}{2}}\sprime $&$ 
      \stackrel {< \MoreRep2j >} {\longrightarrow} $&$ Z_{\mathrm{j}} $&          
      & &
      & 
&($\mathrm{j} \mid n; \, n \, even$)
\end{tabular}\\
\begin{tabular}{cccccccccc}
$D_{n}\sprime $&$ \stackrel {< \MoreRep13 >} {\longrightarrow} $&$ D_{\frac{n}{2}}\sprime $&$ 
      \stackrel {< \MoreRep2j >\sprime} {\longrightarrow} $&$ Z_4 $&$         
      \stackrel {< \MoreRep14 >} {\longrightarrow} $&$ Z_2 $&
      & 
&($\mathrm{j} \nmid \frac{n}{2}$)
\end{tabular}\\
\noindent ($m_{\mathrm{j}}$ even and for $\MoreRep13$ and $\MoreRep14$ exchanged $m_{\mathrm{j}}$ is odd)
\begin{tabular}{cccccccccc}
$D_{n}\sprime $&$ \stackrel {< \MoreRep13 >} {\longrightarrow} $&$ D_{\frac{n}{2}}\sprime $&$ 
      \stackrel {< \MoreRep2j >\sprime} {\longrightarrow} $&$ Z_4 $&$         
      \stackrel {< \MoreRep2k >\sprime} {\longrightarrow} $&$ Z_2 $&
      & 
&($\mathrm{j} \nmid \frac{n}{2};\, m_{\mathrm{j}} \not= m_{\mathrm{k}}$)
\end{tabular}\\
\noindent ( $m_{\mathrm{j}}$ even for $\MoreRep13$ and for $\MoreRep14$ $m_{\mathrm{j}}$ is odd)
\begin{tabular}{cccccccccc}
$D_{n}\sprime $&$ \stackrel {< \MoreRep13 >} {\longrightarrow} $&$ D_{\frac{n}{2}}\sprime $&$ 
      \stackrel {< \MoreRep2j >\sprime} {\longrightarrow} $&$ D_{\frac{\mathrm{j}}{2}}\sprime $&$         
      \stackrel {< \MoreRep14 >} {\longrightarrow} $&$ Z_{\mathrm{j}} $&
      & 
&($\mathrm{j} \mid n$) 
\end{tabular}\\
\noindent ($m_{\mathrm{j}}$ even - for $\MoreRep13$ and $\MoreRep14$ exchanged $m_{\mathrm{j}}$ is 
odd)
\begin{tabular}{cccccccccc}
$D_{n}\sprime $&$ \stackrel {< \MoreRep13 >} {\longrightarrow} $&$ D_{\frac{n}{2}}\sprime $&$ 
      \stackrel {< \MoreRep2j >\sprime} {\longrightarrow} $&$ D_{\frac{\mathrm{j}}{2}}\sprime $&$         
      \stackrel {< \MoreRep2k >} {\longrightarrow} $&$ Z_{\mathrm{k}} $&
      & 
&($\mathrm{j} \mid n; \, \mathrm{k} \mid \mathrm{j}$)
\end{tabular}\\
\noindent ( $m_{\mathrm{j}}$ even for $\MoreRep13$ and for $\MoreRep14$ $m_{\mathrm{j}}$ is odd)
\begin{tabular}{cccccccccc}
$D_{n}\sprime $&$ \stackrel {< \MoreRep13 >} {\longrightarrow} $&$ D_{\frac{n}{2}}\sprime $&$ 
      \stackrel {< \MoreRep2j >\sprime} {\longrightarrow} $&$ D_{\frac{\mathrm{j}}{2}}\sprime $&$         
      \stackrel {< \MoreRep2k >\sprime} {\longrightarrow} $&$ Z_4 $&$
      \stackrel {< \MoreRep14 >} {\longrightarrow} $&$ Z_2 $
&
\end{tabular}
\noindent ($ \mathrm{j} \mid n;\,  m_{\mathrm{j}}=m_{\mathrm{k}} ; \, \mathrm{k} \nmid \mathrm{j}$)\\
\noindent ($m_{\mathrm{j}}$ even - for $\MoreRep13$ and $\MoreRep14$ exchanged, $m_{\mathrm{j}}$ is odd)
\begin{tabular}{cccccccccc}
$D_{n}\sprime $&$ \stackrel {< \MoreRep13 >} {\longrightarrow} $&$ D_{\frac{n}{2}}\sprime $&$ 
      \stackrel {< \MoreRep2j >\sprime} {\longrightarrow} $&$ D_{\frac{\mathrm{j}}{2}}\sprime $&$         
      \stackrel {< \MoreRep2k >\sprime} {\longrightarrow} $&$ Z_4 $&$
      \stackrel {< \MoreRep2l >\sprime} {\longrightarrow} $&$ Z_2 $
&
\end{tabular}
\noindent ($ \mathrm{j} \mid n;\,  m_{\mathrm{j}}=m_{\mathrm{k}}\not= m_{\mathrm{l}}; \, \mathrm{k} \nmid \mathrm{j}
$)\\
\noindent ($m_{\mathrm{j}}$ even for $\MoreRep13$ and for $\MoreRep14$ $m_{\mathrm{j}}$ is odd)
\begin{tabular}{cccccccccc}
$D_{n}\sprime $&$ \stackrel {< \MoreRep2j >\sprime} {\longrightarrow} $&$ Z_4 $&$ 
      \stackrel {< \MoreRep13 >} {\longrightarrow} $&$ Z_2 $&         
      & &
      & 
&($ m_{\mathrm{j}} \, arbitrary 
$)
\end{tabular}\\
\begin{tabular}{cccccccccc}
$D_{n}\sprime $&$ \stackrel {< \MoreRep2j >\sprime} {\longrightarrow} $&$ Z_4 $&$ 
      \stackrel {< \MoreRep2k >\sprime} {\longrightarrow} $&$ Z_2 $&         
      & &
      & 
&($  m_{\mathrm{j}} \not= m_{\mathrm{k}}
$)
\end{tabular}
\begin{tabular}{cccccccccc}
$D_{n}\sprime $&$ \stackrel {< \MoreRep2j >\sprime} {\longrightarrow} $&$ D_{\frac{\mathrm{j}}{2}}\sprime $&$ 
      \stackrel {< \MoreRep12 >} {\longrightarrow} $&$ Z_{\mathrm{j}} $&         
      & &
      & 
&($ m_{\mathrm{j}} \, arbitrary
$)
\end{tabular}\\
\begin{tabular}{cccccccccc}
$D_{n}\sprime $&$ \stackrel {< \MoreRep2j >\sprime} {\longrightarrow} $&$ D_{\frac{\mathrm{j}}{2}}\sprime $&$ 
      \stackrel {< \MoreRep13 >} {\longrightarrow} $&$ Z_{\mathrm{j}} $&         
      & &
      & 
&($  \mathrm{j} \mid n$)
\end{tabular}\\
\noindent ($n$ odd or $m_{\mathrm{j}}$ odd - if $\MoreRep13 \rightarrow \MoreRep14$, then $n$ is odd or $m_{\mathrm{j}}$ is even.)
\begin{tabular}{cccccccccc}
$D_{n}\sprime $&$ \stackrel {< \MoreRep2j >\sprime} {\longrightarrow} $&$ D_{\frac{\mathrm{j}}{2}}\sprime $&$ 
      \stackrel {< \MoreRep13 >} {\longrightarrow} $&$ Z_2 $&         
      & &
      & 
&($  \mathrm{j} \nmid n ; \, n \, even$)
\end{tabular}\\
\noindent ( $m_{\mathrm{j}}$ odd for $\MoreRep13$ and for $\MoreRep14$ $m_{\mathrm{j}}$ is even)
\begin{tabular}{cccccccccc}
$D_{n}\sprime $&$ \stackrel {< \MoreRep2j >\sprime} {\longrightarrow} $&$ D_{\frac{\mathrm{j}}{2}}\sprime $&$ 
      \stackrel {< \MoreRep13 >} {\longrightarrow} $&$ Z_4 $&$         
      \stackrel {< \MoreRep14 >} {\longrightarrow} $&$ Z_2 $&
      & 
&($ n \, even$)
\end{tabular}\\
\noindent ($m_{\mathrm{j}}$ even - for $\MoreRep13$ and $\MoreRep14$ exchanged $m_{\mathrm{j}}$ is odd)
\begin{tabular}{cccccccccc}
$D_{n}\sprime $&$ \stackrel {< \MoreRep2j >\sprime} {\longrightarrow} $&$ D_{\frac{\mathrm{j}}{2}}\sprime $&$ 
      \stackrel {< \MoreRep13 >} {\longrightarrow} $&$ Z_4 $&$         
      \stackrel {< \MoreRep2k >\sprime} {\longrightarrow} $&$ Z_2 $&
      & 
&($ n \, even; \, m_{\mathrm{j}} \not= m_{\mathrm{k}} $)
\end{tabular}\\
\noindent ( $m_{\mathrm{j}}$ even for $\MoreRep13$ and for $\MoreRep14$ $m_{\mathrm{j}}$ is odd)
\begin{tabular}{cccccccccc}
$D_{n}\sprime $&$ \stackrel {< \MoreRep2j >\sprime} {\longrightarrow} $&$ D_{\frac{\mathrm{j}}{2}}\sprime $&$ 
      \stackrel {< \MoreRep2k >} {\longrightarrow} $&$ Z_{\mathrm{k}} $&         
      & &
      & 
&($  \mathrm{k} \mid \mathrm{j}; \, m_{\mathrm{j}} \, arbitrary
$)
\end{tabular}\\
\begin{tabular}{cccccccccc}
$D_{n}\sprime $&$ \stackrel {< \MoreRep2j >\sprime} {\longrightarrow} $&$ D_{\frac{\mathrm{j}}{2}}\sprime $&$ 
      \stackrel {< \MoreRep2k >\sprime} {\longrightarrow} $&$ Z_4 $&$         
      \stackrel {< \MoreRep13 >} {\longrightarrow} $&$ Z_2 $&
      & 
&($ n \, even; \, m_{\mathrm{j}}=m_{\mathrm{k}} ; \, \mathrm{k} \nmid \mathrm{j}
$)
\end{tabular}
\noindent ( $m_{\mathrm{j}}$ odd for $\MoreRep13$ and for $\MoreRep14$ $m_{\mathrm{j}}$ is even)
\begin{tabular}{cccccccccc}
$D_{n}\sprime $&$ \stackrel {< \MoreRep2j >\sprime} {\longrightarrow} $&$ D_{\frac{\mathrm{j}}{2}}\sprime $&$ 
      \stackrel {< \MoreRep2k >\sprime} {\longrightarrow} $&$ Z_4 $&$         
      \stackrel {< \MoreRep2l >\sprime} {\longrightarrow} $&$ Z_2 $&
      & 
&($ m_{\mathrm{j}}=m_{\mathrm{k}}(\not=m_{\mathrm{l}}) ; \, \mathrm{k} \nmid \mathrm{j}$)
\end{tabular}
\begin{tabular}{cccccccccc}
$ *  $ &$D_{n}\sprime $&$ \stackrel {< \MoreRep2j >} {\longrightarrow} $&$ Z_{\mathrm{j}} $&$ 
      \stackrel {< \MoreRep2k >} {\longrightarrow} $&$ Z_{\mathrm{k}} $&         
      & &
      &($  \mathrm{k} \mid \mathrm{j}
$)
\end{tabular}
\begin{tabular}{cccccccccc}
$ *  $ & $D_{n}\sprime $&$ \stackrel {< \MoreRep2j >\sprime} {\longrightarrow} $&$ D_{\frac{\mathrm{j}}{2}}\sprime $&$ 
      \stackrel {< \MoreRep2k >\sprime} {\longrightarrow} $&$ D_{\frac{\mathrm{k}}{2}}\sprime $&         
      & &
      &($  \mathrm{k} \mid \mathrm{j}; \, m_{\mathrm{j}}=m_{\mathrm{k}}
$)
\end{tabular}
\end{flushleft}

\end{appendix}

\newpage

%%%%%%%%%%%%%%%%%%%%%%%%%%%%%%%%%%%%%%%%%%%%%%%%%%%%%%

\end{document}